%% file: 01_Manuscript.tex
\documentclass[pdflatex,sn-mathphys-num]{sn-jnl}
\usepackage{graphicx}
\usepackage{multirow}
\usepackage{amsmath,amssymb,amsfonts}
\usepackage{amsthm}
\usepackage{mathrsfs}
\usepackage[title]{appendix}
\usepackage{xcolor}
\usepackage{textcomp}
\usepackage{manyfoot}
\usepackage{booktabs}
\usepackage{algorithm}
\usepackage{algorithmicx}
\usepackage{algpseudocode}
\usepackage{listings}
\usepackage{verbatim}

\theoremstyle{thmstyleone}

\theoremstyle{thmstyletwo}

\theoremstyle{thmstylethree}

\begin{document}

\title[Article Title]{Real-Time Single-Pixel Imaging via Exploiting Temporal Mixing: A Unified Calibration and Composite Readout Framework}

\author{\fnm{Jing-Sen} \sur{Li}}\email{luzi23887@163.com}

\author{\fnm{En-Rui} \sur{Zhang}}\email{2591763411@qq.com}
\author{\fnm{Kan-Xu} \sur{Jia}}\email{jiakanxu@163.com}
\author{\fnm{Yi-Zhu} \sur{Zhang}}\email{1362319116@qq.com}

\author{\fnm{Qian-Qian} \sur{Bao}}\email{baoqianqian@lnu.edu.cn}
\author{\fnm{Shao-Ying} \sur{Meng}}\email{mengshaoying@163.com}
\author*{\fnm{Xi-Hao} \sur{Chen}\textsuperscript{*}}
\email{xi-haochen@163.com}
\makeatletter
\g@addto@macro\corrauthemail{\par Tel.: +86~18704088217.}
\makeatother
\affil{\orgdiv{Key Laboratory of Optoelectronic Devices and Detection Technology, School of Physics}, \orgname{Liaoning University}, \orgaddress{\street{66 Chongshan Middle Road }, \city{Shenyang}, \postcode{110036}, \state{Liaoning}, \country{China}}}

\abstract{In high-speed optical measurements and computational imaging—such as in single-pixel imaging—temporal mixing arising from the mismatch between modulation and the electro-optical acquisition chain is conventionally treated as an unmitigated distortion that corrupts measurement orthogonality, breaks compressive sensing priors, fundamentally limits imaging speed, and induces catastrophic reconstruction failure. Here, we establish a transferable, unified calibration and composite-readout framework for single-pixel imaging that systematically repurposes temporal memory from a readout limitation into a powerful measurement resource. Beyond achieving robust high-fidelity recovery, this engineered approach enables cost-effective, lower-bandwidth configurations to consistently deliver superior image quality compared to higher-bandwidth reference setups across diverse operating conditions—turning a historical hardware bottleneck into an active imaging advantage. Specifically, target scenes are reconstructed directly from mixed observations without the need to recover individual pattern projections, thereby significantly advancing high-speed and video-rate single-pixel imaging capabilities. Crucially, the transferability of our framework allows calibration routines established on one hardware platform (such as digital micromirror device-based setups) to be readily transferred across different optical systems (such as rotating coding disks), paving the way for calibrated temporal memory as a versatile and generalized measurement paradigm for broader computational imaging.}


\maketitle
\section{Introduction}\label{sec1}
High-speed optical measurements and computational imaging lie at the front lines of modern optical science and engineering, offering unprecedented capabilities in dynamic scene visualization~\cite{ref1,ref2}, real-time tracking~\cite{ref3}, transient event sensing~\cite{ref4}, and multidimensional scientific diagnostics~\cite{ref5,ref6}. By circumventing the physical and economic constraints of traditional focal-plane array detectors, computational imaging shifts the burden of image formation from hardware optics to advanced algorithmic recovery~\cite{ref7,ref8}. Among various implementations, single-pixel imaging (SPI), which encodes spatial information onto structured illumination or masks and captures the backscattered or transmitted light using a single-element photodetector, has evolved into a remarkably versatile paradigm~\cite{ref9,ref10}. Benefiting from the superior performance and spectral versatility of single-element detectors across infrared (IR)~\cite{ref11,ref12}, terahertz (THz)~\cite{ref13}, and X-ray~\cite{ref14} regimes, where conventional arrays face severe cost bottlenecks, SPI bypasses traditional spatial-detection limits~\cite{ref15,ref16}. It now serves as a cornerstone for addressing formidable optical challenges in complex and hidden environments, ranging from non-line-of-sight imaging~\cite{ref17} and scattering-media diagnostics~\cite{ref18} to large-scale remote sensing~\cite{ref19}. Driven by this growing technological landscape, modern computational imaging represented by SPI is currently undergoing a paradigm shift toward higher temporal resolutions~\cite{ref20,ref21} and ultrafast frame rates~\cite{ref22,ref23}, aiming to capture transient physical dynamics and video-rate processes~\cite{ref24,ref25} that remain fundamentally inaccessible to conventional multiplexing architectures.

Pushing single-pixel and computational imaging systems toward such high-speed, high-frame-rate regimes exposes a fundamental bottleneck governed by hardware dynamics rather than pure computational constraints~\cite{ref10,ref13,ref26,ref27}. As high modulation speeds, such as those of digital micromirror devices (DMDs) or high-speed coding units, approach or exceed the intrinsic bandwidth of the electro-optical acquisition chain, temporal mixing becomes inevitable~\cite{ref13,ref26,ref27}. This phenomenon, arising from the combined transient responses, RC delays, and finite impulse characteristics of photodetectors, amplifiers, analog-to-digital converters (ADCs), and data acquisition cards, creates a severe mismatch between spatial modulation and signal readout~\cite{ref13,ref26,ref28}. In response to these challenges, prior efforts have largely pursued two distinct pathways: hardware-side mitigation, which attempts to suppress latency or avoid overlap through bandwidth upgrades, active gating, electronic equalization, or hardware-software co-design~\cite{ref22,ref27,ref29,ref30}, and software-side correction, which relies on heuristic deconvolution, degradation compensation, or two-stage inverse filtering algorithms to clean up degraded signals~\cite{ref13,ref28,ref31}. However, both strategies face severe intrinsic ceilings. Hardware upgrades are invariably constrained by fundamental gain-bandwidth and economic trade-offs, sacrificing transimpedance sensitivity and elevating broadband noise while failing to eliminate residual memory effects at ultra-high throughputs~\cite{ref10,ref26,ref29}. Meanwhile, conventional deconvolution and two-stage methods must reconcile intermediate-projection fidelity with final image quality; under strong inter-symbol interference or limited sampling, inverting isolated waveforms prior to reconstruction becomes ill-conditioned, amplifying noise, discarding common-target consistency, and propagating unmodeled residuals~\cite{ref13,ref31,ref32}. Moreover, exact projections do not guarantee optimal images, and in complex environments certain waveform distortions or smoothing can even aid recovery, rendering waveform-level criteria unreliable.
As a result, operating beyond the steady-state settling limit conventionally exacts a heavy toll: corrupted measurement orthogonality, restricted throughput, and structural reconstruction failure~\cite{ref13,ref28,ref32}. To overcome this fundamental dilemma, a pivotal question arises: must temporal mixing in high-speed SPI be strictly treated as a detrimental artifact to be eliminated, or can it be harnessed as a constructive spatiotemporal encoding resource when the physical measurement process and target recovery are jointly modeled?

Here, we move beyond conventional mitigation and inversion paradigms~\cite{ref13,ref27,ref28}, motivated by the core physical insight that temporal mixing does not erase spatial information, but rather redistributes it across the temporal dimension. Rather than treating this temporal memory as a parasitic artifact to be passively tolerated, we establish a new SPI paradigm: by employing a data-driven calibration to precisely characterize the complete-history electro-optical dynamics, we reformulate the forward measurement operator and turn hardware inertia into an active, controllable design variable. Applicable across diverse acquisition chains and operating conditions, this unified calibration lays the foundation for harnessing a series of intrinsic advantages embedded within temporally mixed observations. Beyond unlocking these physical benefits, our approach dramatically expands the system degrees of freedom, allowing temporal dynamics to be freely co-designed alongside spatial coding patterns, modulation sequences, sampling schedules, and composite readouts. By systematically exploiting the temporally redistributed spatial information, our framework transforms the entire imaging chain into a flexibly configurable architecture, one that not only breaks free from traditional steady-state settling constraints, but also enables lower-bandwidth, cost-effective hardware configurations to surpass conventional high-bandwidth SPI systems. In essence, this framework decouples high-performance computational imaging from the prohibitive costs and physical constraints of ultra-high-speed component upgrades. Methodologically, this co-design paradigm is anchored by a direct common-target recovery principle: rather than judging uncertain intermediate waveforms, it unifies the evaluation criterion of all design degrees of freedom directly onto the reconstructed image. Bypassing ill-posed projection separation not only enables target scenes to be retrieved directly from time-integrated mixed observations at video rates, but also provides a single end-to-end standard for jointly tailoring spatial modulation, temporal response, and readout processing.

Furthermore, because our framework structurally decouples the intrinsic impulse response of the receiving chain from the spatial probing sequence, the complete measurement paradigm of reference calibration, composite readout, and common-target recovery can be transferred across diverse operating regimes and independent detection platforms, while conceptually extending to broader modulation architectures ranging from digital micromirror devices and LED arrays to mechanical coding masks~\cite{ref9,ref22,ref23,ref25,ref33}. By transforming hardware inertia from a passive physical ceiling into an active dimension of system design, this work establishes a generalized measurement paradigm that redefines the interplay between optoelectronic dynamics and computational inversion, offering a versatile roadmap for high-speed, high-dimensional imaging well beyond conventional bandwidth limits~\cite{ref5,ref6,ref24}.

\section{Theoretical Model and Co-Design Solution Mechanisms}\label{sec2}
\subsection{Continuous-Time Forward Model and Reference Calibration}\label{sec2.1}

\begin{figure}[th]
    \centering
    \includegraphics[width=0.8\textwidth]{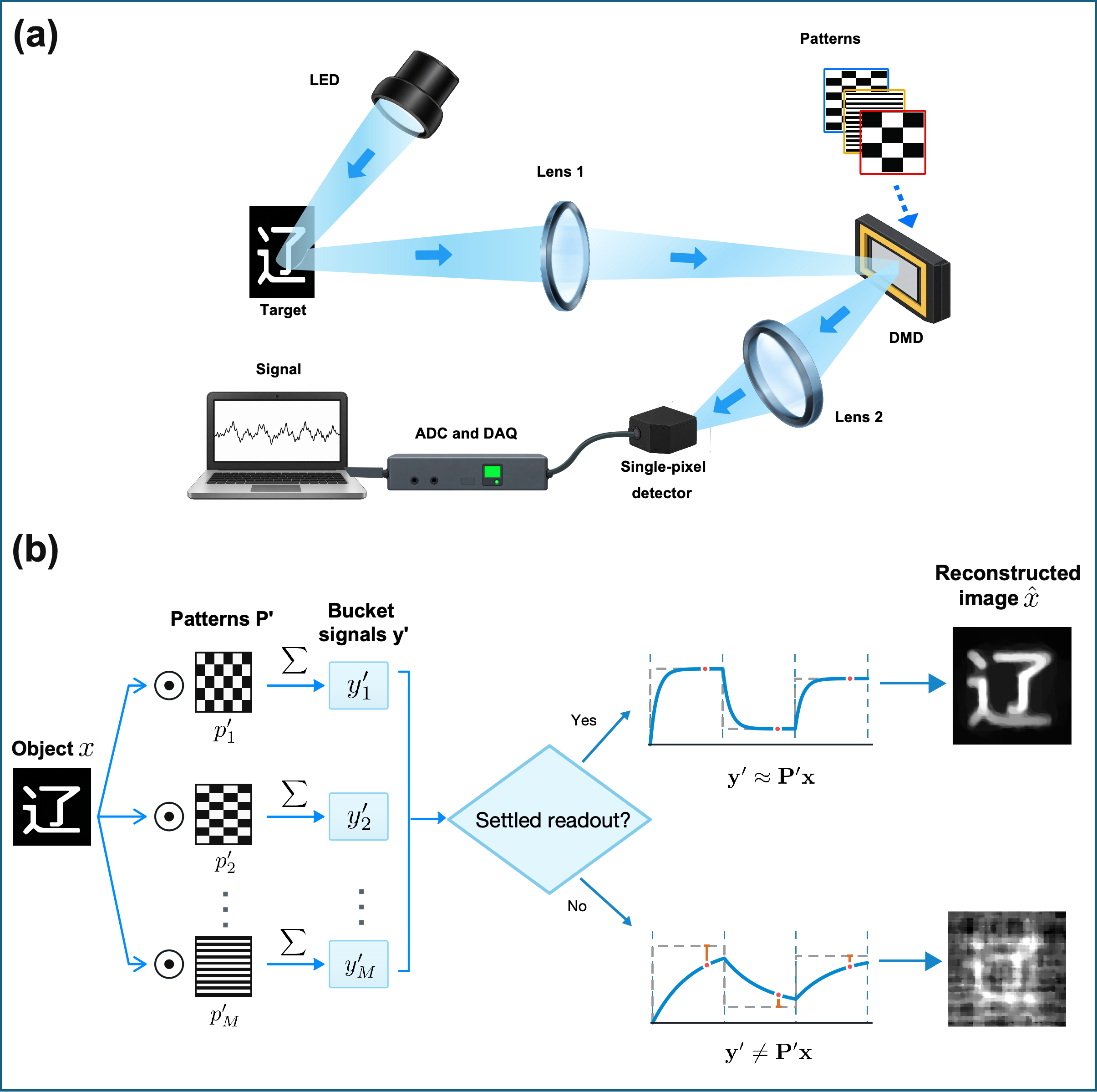}
   \caption{\textbf{Schematic layout and operational principle of the DMD-based SPI system.}
(a) Experimental architecture of the DMD-based SPI configuration, consisting of an LED illumination source, a target object, imaging lenses, a DMD deployed for spatial pattern modulation, a single-pixel detector, and a data acquisition (DAQ) module integrated with an analog-to-digital converter (ADC).
(b) Schematic workflow of the SPI principle. The target vector $\mathbf{x}$ and the structured-pattern matrix $\mathbf{P}'$ produce the bucket-measurement vector $\mathbf{y}'$ through optical inner-product modulation. Assessment of readout settling is followed by reconstruction of the estimated target $\widehat{\mathbf{x}}$.}
    \label{fig:1}
\end{figure}
Figure~\ref{fig:1} illustrates the passive SPI framework, which comprises an optical setup and a corresponding signal-processing model. As depicted in Fig.~\ref{fig:1}(a), a reflective target scene $\mathbf{x} \in \mathbb{R}^N$, where $N$ denotes the total dimension corresponding to the full-sampling pattern set, is illuminated by an LED and optically imaged onto a DMD loaded with a pattern vector sequence $\{\mathbf{p}_l'\}_{l=1}^M$ with $M \le N$, and the spatially modulated light is collected by a single-pixel detector. According to the standard measurement principle, the target scene $\mathbf{x}$ is probed by a spatial pattern vector $\mathbf{p}_l'$ at each time step $l$, yielding a scalar bucket signal $y_l'$ via optical inner-product modulation, with $y_l' = \mathbf{p}_l'^{\mathsf T}\mathbf{x}$ as illustrated in the left-hand panel of Fig.~\ref{fig:1}(b). By stacking these measurements over $M$ time steps, the forward model can be expressed in a compact matrix form as $\mathbf{y}' = \mathbf{P}'\mathbf{x}$~\cite{ref7}. At the central decision node, depending on whether the readout has sufficiently settled, the operational flow splits into two scenarios in the right panel of Fig.~\ref{fig:1}(b). In the upper ``Yes'' branch, these measurements are subsequently fed into standard optimization-based reconstruction to recover the target image. Nevertheless, the validity of this standard forward model rests on the assumption that each acquired bucket signal maps accurately to an isolated optical projection of the corresponding pattern. In high-speed or dynamic imaging, short pattern dwell times leave the readout response incompletely settled, causing recorded readings to retain contributions from preceding patterns~\cite{ref13,ref28}. In the lower ``No'' branch, factors such as finite system bandwidth and sluggish impulse tails induce strong temporal mixing among successive patterns, breaking the one-to-one pattern-to-sample correspondence and causing severe reconstruction failure if naively interpreted using the memoryless pattern matrix $\mathbf{P}'$.

To capture these dynamics without enforcing steady-state settling, we formulate a continuous-time measurement framework that embeds electro-optical memory directly into the spatial encoding process. To illustrate this framework, consider a static scene. Let $\mathbf{p}(t)\in\mathbb{R}^{N}$ be the spatial pattern vector at time $t$, and $\mathbf{x}^{\mathrm{opt}}\in\mathbb{R}^{N}$ be the static target vector in optical units, where the superscript $\mathrm{opt}$ identifies quantities before readout gain scaling. The instantaneous optical projection $s^{\mathrm{opt}}(t)$ is
\begin{equation}
    s^{\mathrm{opt}}(t) = \mathbf{p}(t)^{\mathsf T}\mathbf{x}^{\mathrm{opt}}.
    \label{Eq1}
\end{equation}
Assuming that the readout chain from the instantaneous optical projection to the recorded signal is causal, linear, and approximately time-invariant over the acquisition interval, let $y_i$ denote the recorded scalar measurement at sampling time $t_i$, where $i=1,\ldots,L$ indexes the $L$ retained samples. The recorded measurement is expressed as~\cite{ref34}
\begin{equation}
    y_i =
    \int_{-\infty}^{t_i}
    \kappa_i(u)s^{\mathrm{opt}}(u)\,\mathrm{d}u
    + b_i + \nu_i,
    \qquad i=1,\ldots,L,
    \label{Eq2}
\end{equation}
where $u$ is the integration time variable, $b_i$ is the baseline offset, $\nu_i$ is additive measurement noise, and $\kappa_i(u)$ is the effective temporal kernel associated with the $i$th sample. We write $\kappa_i(u)=G\,h(t_i-u)$, where $G$ is the overall readout gain and $h(\tau)$ is the normalized temporal response of the complete readout chain over time lag $\tau$, satisfying $h(\tau)=0$ for $\tau<0$ and $\int_{0}^{\infty}h(\tau)\,\mathrm{d}\tau=1$. This formulation captures the complete input--output mapping across detection, amplification, sampling aperture, and acquisition-side filtering without requiring separate identification of constituent device responses.
\begin{figure}[thb]
    \centering
    \includegraphics[width=\textwidth]{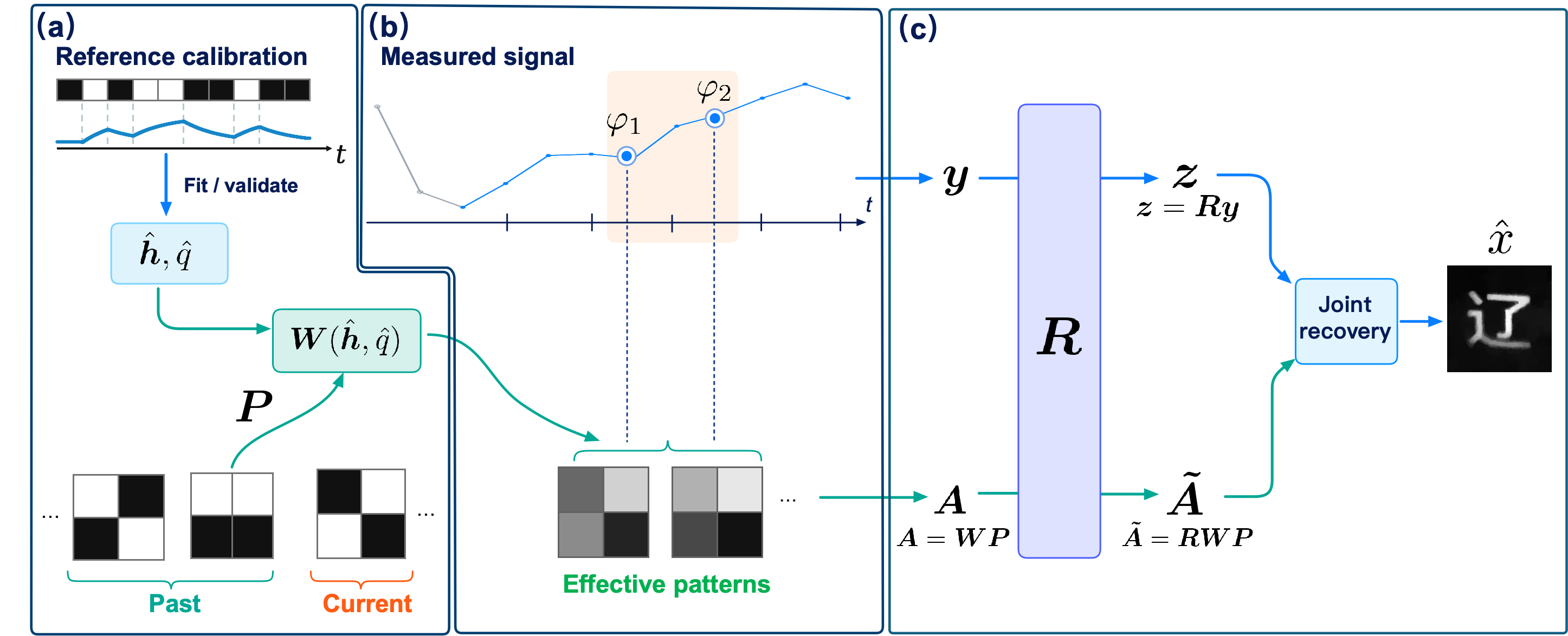}
    \caption{\textbf{Reference calibration, history-weighted effective patterns, and matched joint reconstruction.}
    (a) Known binary reference sequences and their recorded responses are used to estimate and validate the normalized response vector $\widehat{\mathbf{h}}$ and timing correspondence $\widehat{q}$. The resulting history-weight matrix $\mathbf{W}\equiv\mathbf{W}(\widehat{\mathbf{h}},\widehat{q})$ combines the current and preceding physical patterns in $\mathbf{P}$.
    (b) Measurements at different phases within the same pattern slot correspond to different history-weighted effective patterns. The dashed guides associate the highlighted sampling phases $\varphi_1$ and $\varphi_2$ with their corresponding effective patterns. Stacking the transposed effective-pattern vectors forms $\mathbf{A}=\mathbf{W}\mathbf{P}$. The shaded region highlights the selected phases rather than delimiting the pattern slot. The reference-response curve and pattern tiles are schematic.
    (c) The same linear readout operator $\mathbf{R}$ acts on the observations and effective measurement matrix, yielding $\mathbf{z}=\mathbf{R}\mathbf{y}$ and $\widetilde{\mathbf{A}}=\mathbf{R}\mathbf{W}\mathbf{P}$. Joint reconstruction from these matched observations and the effective operator yields the target estimate $\widehat{\mathbf{x}}$.}
    \label{fig:2}
\end{figure}

To identify $h(\tau)$ across diverse hardware configurations without component-level specifications, we establish a broadly applicable, data-driven calibration framework using spatially uniform all-black and all-white binary reference sequences alternating between states $0$ and $1$, as summarized in Fig.~\ref{fig:2}(a). Known reference inputs and their recorded waveforms are used to jointly identify and validate the temporal response and timing correspondence~\cite{ref35}. Because a spatially uniform all-white reference state reduces Eq.~\eqref{Eq1} to a scalar proportional to the total target reflectance $\mathbf{1}^{\mathsf T}\mathbf{x}^{\mathrm{opt}}$, its unknown optical amplitude and the readout gain $G$ contribute only a common scale factor that is automatically removed upon normalization, enabling calibration without a pixelwise model of the reference target. The reference block incorporates known prehistory, timing markers, training sequences, holdout validation, and tail checks, where fast switching, long holds, and multiscale pulses constrain recent-history responses, plateau levels, and response support, respectively, as detailed in Supplementary Section~S1.1.

The finite-lag response representation uses $H$ lag intervals and a coefficient vector $\mathbf{h}=(h_1,\dots,h_H)^{\mathsf T}\in\mathbb{R}^{H}$, where $h_j$ is the integrated response weight assigned to the $j$th lag interval. Let $q$ denote the continuous timing correspondence that maps ADC sample timestamps to physical pattern-slot coordinates, accommodating relative ADC--DMD clock drift without requiring hardware synchronization. For $L_{\mathrm{cal}}$ calibration samples, let $\mathbf{y}_{\mathrm{cal}}\in\mathbb{R}^{L_{\mathrm{cal}}}$ be the recorded reference-measurement vector and $\mathbf{X}_{\mathrm{cal}}(q)\in\mathbb{R}^{L_{\mathrm{cal}}\times H}$ be the design matrix constructed from the known reference histories under $q$. The corresponding calibration model is
\begin{equation}
    \mathbf{y}_{\mathrm{cal}} = \mathbf{X}_{\mathrm{cal}}(q)\boldsymbol{\alpha} + b_{\mathrm{cal}}\mathbf{1}_{L_{\mathrm{cal}}} + \boldsymbol{\eta}_{\mathrm{cal}},
    \label{Eq3}
\end{equation}
where $\boldsymbol{\alpha} \in \mathbb{R}^H$ contains the scaled response coefficients representing the weighted contribution of each historical lag, $b_{\mathrm{cal}}$ is the baseline offset, $\mathbf{1}_{L_{\mathrm{cal}}}$ is the all-one vector, and $\boldsymbol{\eta}_{\mathrm{cal}} \in \mathbb{R}^{L_{\mathrm{cal}}}$ collects measurement noise and disturbances.

Following the fitting and validation workflow in Fig.~\ref{fig:2}(a), for each candidate timing correspondence $q$, the response coefficients $\boldsymbol{\alpha}$ and calibration baseline $b_{\mathrm{cal}}$ are fitted over the $L_{\mathrm{tr}}$ training observations $\mathbf{y}_{\mathrm{tr}}$ and corresponding design submatrix $\mathbf{X}_{\mathrm{tr}}(q)$ by minimizing the regularized loss $\|\mathbf{y}_{\mathrm{tr}}-\mathbf{X}_{\mathrm{tr}}(q)\boldsymbol{\alpha}-b_{\mathrm{cal}}\mathbf{1}_{L_{\mathrm{tr}}}\|_2^2+\lambda_{\mathrm{cal}}\mathcal{R}(\boldsymbol{\alpha})$~\cite{ref36}, where $\mathcal{R}(\boldsymbol{\alpha})$ denotes the response regularization functional weighted by $\lambda_{\mathrm{cal}}\ge 0$, subject to a linear response-centroid constraint that resolves the coordinate ambiguity between timing onset and response lag. Evaluating these refitted models across candidates selects the timing correspondence $\widehat{q}$ yielding the smallest squared prediction residual on the same training observations~\cite{ref37}. Normalizing the coefficients fitted at $\widehat{q}$ then yields the reusable response vector $\widehat{\mathbf{h}}=(\widehat{h}_1,\dots,\widehat{h}_H)^{\mathsf T}$. Under the adopted piecewise-constant representation, these weights define the calibrated temporal response $\widehat{h}(\tau)$ governing how historical patterns are weighted and superposed into each measurement, as detailed in Supplementary Section~S1.2.

To translate this calibrated memory into a forward imaging operator, we absorb the overall readout gain into the target vector as $\mathbf{x}=G\mathbf{x}^{\mathrm{opt}}$ and let $\mathbf{P} \in \mathbb{R}^{K \times N}$ collect the physical pattern sequence across $K$ time slots. Directly integrating $\widehat{h}(\tau)$ over each registered pattern display interval $[T_k, T_{k+1})$ yields the history-weight matrix $\mathbf{W} \in \mathbb{R}^{L \times K}$ with entries $W_{ik} = \int_{T_k}^{T_{k+1}} \widehat{h}(t_i - u)\,\mathrm{d}u$, defining the effective measurement matrix $\mathbf{A} = \mathbf{WP} \in \mathbb{R}^{L \times N}$. Its $i$th row, $\mathbf{a}_i^{\mathsf T}$, is the effective spatial pattern of the $i$th sample. As shown in Fig.~\ref{fig:2}(a) and (b), each effective pattern is a history-weighted superposition of the current and preceding physical patterns; consequently, two sampling phases $\varphi_1$ and $\varphi_2$ within the same pattern slot correspond to distinct effective patterns even though the displayed physical pattern remains unchanged.

As illustrated in Fig.~\ref{fig:2}(c), a linear readout operator $\mathbf{R}\in\mathbb{R}^{Q\times L}$ is then applied identically to the recorded observations and their matched effective patterns to form inter-slot and intra-slot combinations that retain complementary target-dependent variations while suppressing common-mode offsets and disturbances. This establishes the central analytical result of our framework, an end-to-end unified observation relation:
\begin{equation}
\mathbf{z}=\mathbf{R}\mathbf{y}=\widetilde{\mathbf{A}}\mathbf{x}+\boldsymbol{\eta},
\label{Eq4}
\end{equation}
where $\widetilde{\mathbf{A}}=\mathbf{RWP}\in\mathbb{R}^{Q\times N}$ serves as the matched effective operator, $\mathbf{z}\in\mathbb{R}^{Q}$ is the processed observation vector, and $\boldsymbol{\eta}\in\mathbb{R}^{Q}$ aggregates measurement noise and residual mismatches. Crucially, Eq.~\eqref{Eq4} embodies our direct common-target recovery principle: because all temporally mixed measurements originate from the same underlying scene $\mathbf{x}$, the composite operator $\widetilde{\mathbf{A}}=\mathbf{R}\mathbf{W}\mathbf{P}$ bypasses ill-posed intermediate projection estimation and enforces spatial consistency directly when fitting the mixed observations, as analyzed in Supplementary Section~S2.3. By binding spatial modulation $\mathbf{P}$, temporal memory $\mathbf{W}$, and composite readout $\mathbf{R}$ directly to the common target $\mathbf{x}$, temporal mixing is transformed from an intermediate waveform distortion into a joint spatial constraint.

Building upon Eq.~\eqref{Eq4}, the matched observations $\mathbf{z}$ and effective operator $\widetilde{\mathbf{A}}$ directly define the single-stage inverse problem for recovering the target image $\mathbf{x}\in\mathbb{R}^{N}$, which interfaces seamlessly with variational optimization~\cite{ref38}, deep algorithm unrolling~\cite{ref39,ref40}, and hybrid architectures~\cite{ref15,ref41}:
\begin{equation}
\widehat{\mathbf{x}}\in\arg\min_{\mathbf{x}\in\mathbb{R}^{N}}\left\{\frac{1}{2}\Vert\widetilde{\mathbf{A}}\mathbf{x}-\mathbf{z}\Vert_2^2+\lambda\mathcal{J}(\mathbf{x})\right\},
\label{Eq5}
\end{equation}
where $\mathcal{J}(\mathbf{x})$ is the image regularization functional and $\lambda\geq0$ controls its weighting strength. Beyond eliminating intermediate projection separation, Eq.~\eqref{Eq5} establishes the reconstruction quality of the final target $\widehat{\mathbf{x}}$, rather than the fidelity of isolated slot waveforms, as the unified evaluation criterion governing system co-design. Indeed, not only do intermediate waveforms suffer inevitable distortion at high speeds, but accurate, undistorted waveforms are not necessarily preferable and often yield inferior reconstructions: pursuing strict waveform fidelity leaves measurements vulnerable to noise and external disturbances, whereas controlled mixing or smoothing in $\mathbf{W}$ acts in concert with $\mathbf{P}$, $\mathbf{R}$, and $\mathcal{J}(\mathbf{x})$ to suppress interference and reshape subspace constraints for superior $\widehat{\mathbf{x}}$. Anchoring evaluation directly on $\widehat{\mathbf{x}}$ thus allows the pattern sequence $\mathbf{P}$, the response and sampling timing determining $\mathbf{W}$, and the readout $\mathbf{R}$ to be jointly tailored without intermediate metric conflicts. Complete formulations of calibration reuse, valid-observation selection, composite-readout construction, and cross-frame extensions to dynamic scenes are provided in Supplementary Sections~S1.3--S1.5.

\subsection{Spatial Information in Temporal Mixing and Its Joint Exploitation}
\label{sec2.2}

With the complete-history forward operator $\widetilde{\mathbf{A}}=\mathbf{R}\mathbf{W}\mathbf{P}$ established in Section~\ref{sec2.1}, temporal mixing is described as a history-weighted redistribution of physical-pattern contributions across time and sampling phases. The acquisition response therefore shapes the effective spatial pattern associated with each recorded observation. Under the static-scene and linear-response assumptions established above, different sampling instants can carry distinct history-weighted superpositions of current and preceding patterns even while the displayed physical mask remains static. Guided by the common-target recovery criterion in Eq.~\eqref{Eq5}, this temporal redistribution transforms measurement design from enforcing isolated steady-state projections into coordinating four interacting degrees of freedom: within-slot phase readout, complementary pattern ordering, temporal filtering, and sampling timing.

\begin{figure}[thb]
    \centering
    \includegraphics[width=\textwidth]{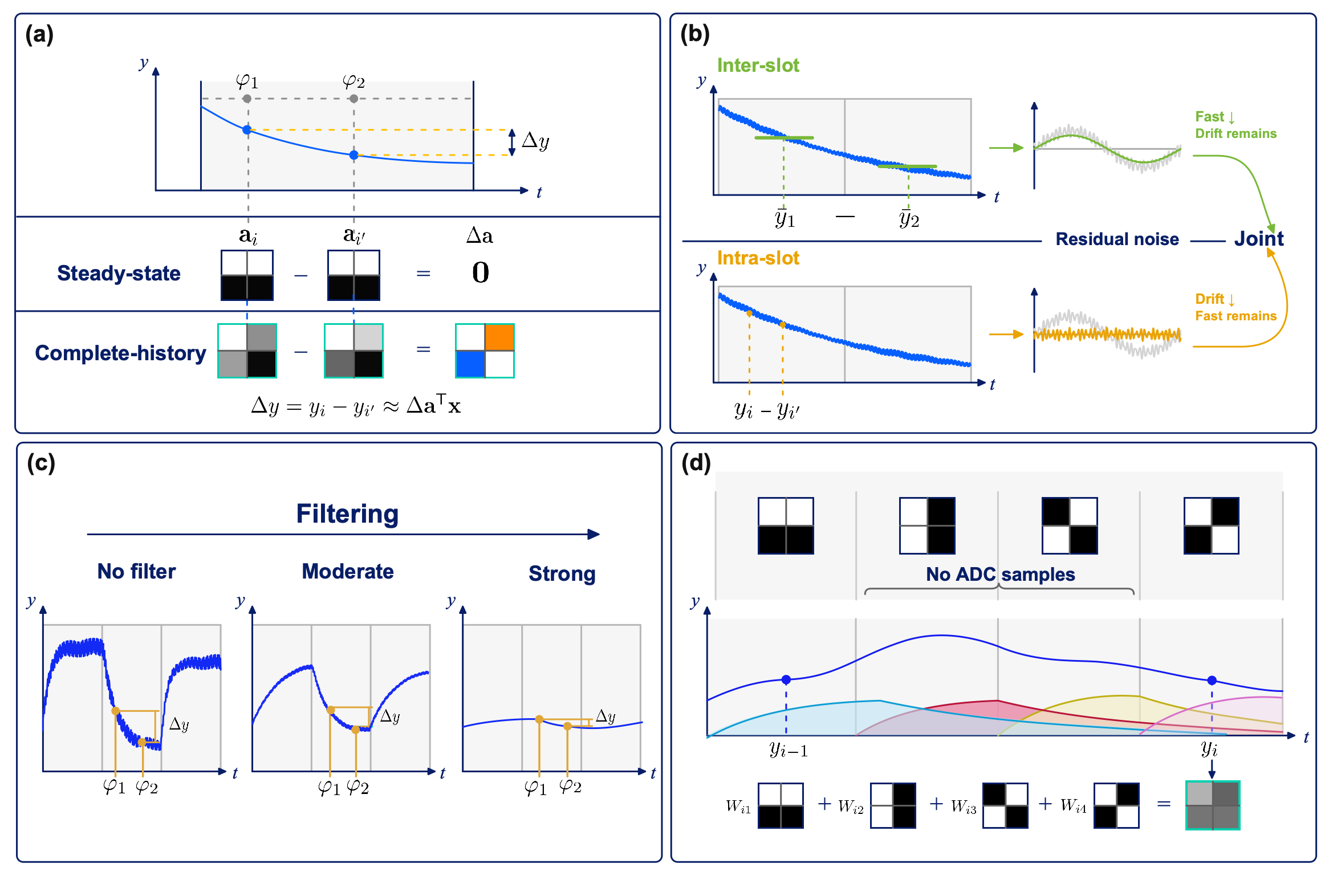}
    \caption{\textbf{Spatial information in temporal mixing and its joint exploitation.}
    (a) Intra-slot differential readout. In the steady-state regime, samples $i$ and $i'$ share the same effective pattern, whereas in the illustrated temporally mixed case, their effective-pattern difference $\Delta\mathbf{a}$ is nonzero and encodes the target through Eq.~\eqref{Eq6}, with orange and blue denoting negative and positive entries of $\Delta\mathbf{a}$, respectively.
    (b) Complementary inter- and intra-slot readouts. The former averages within complementary-pattern branches and can retain drift between them; the latter rejects common offsets and extracts within-slot transient variations while remaining sensitive to local temporal fluctuations.
    (c) Temporal filtering affects disturbance transmission and target-dependent transient responses, potentially shortening the steady-state window and reducing the intra-slot difference $\Delta y$.
    (d) When the DMD modulation rate exceeds the ADC sampling rate, unsampled patterns still contribute to later readings through the history weights $W_{ik}$.
    All curves and pattern tiles are schematic.}
    \label{fig:3}
\end{figure}

To explicitly harvest the spatial information redistributed within each modulation slot, intra-slot readout converts response variations across sampling phases into algebraic spatial constraints while achieving common-mode rejection~\cite{ref42}. In an ideal steady-state regime, the effective spatial pattern remains identical throughout a physical pattern slot, rendering within-slot phase differences zero and uninformative. In the presence of temporal mixing, however, the transient response continuously reweights current and historical pattern contributions across sampling phases, so that two samples $i$ and $i'$ acquired at phases $\varphi_1$ and $\varphi_2$ within the same slot correspond to distinct effective patterns $\mathbf{a}_i$ and $\mathbf{a}_{i'}$. Assuming an approximately constant baseline within the slot and neglecting noise, their differential readout satisfies
\begin{equation}
y_i-y_{i'}\approx\Delta\mathbf{a}^{\mathsf T}\mathbf{x},\qquad \Delta\mathbf{a}=\mathbf{a}_i-\mathbf{a}_{i'},
\label{Eq6}
\end{equation}
where $\Delta\mathbf{a}$ acts as a nonzero differential effective pattern that encodes the target even though the displayed physical pattern remains unchanged [Fig.~\ref{fig:3}(a)]. Summation by parts over the calibrated history weights reveals that $\Delta\mathbf{a}$ is governed by the continuing response to historical pattern transitions $\mathbf{p}_k-\mathbf{p}_{k-1}$ integrated between the two sampling phases. Extending this two-sample difference to all valid samples within a slot via a least-squares temporal slope readout guarantees exact zero-sum rejection of constant offsets while extracting a slope-based measure of the target-dependent historical evolution, as derived in Supplementary Section~S1.4.

Beyond within-slot evolution, the temporal ordering of complementary patterns satisfying $\mathbf{p}_{k_-}=\mathbf{1}-\mathbf{p}_{k_+}$ serves as a coupled design variable that governs how differential spatial encoding and disturbance sensitivity interact under temporal mixing, extending well beyond its traditional role in sub-Nyquist sampling~\cite{ref43}. Under low-pass temporal mixing, placing complementary patterns adjacently as in conventional drift-suppression schemes~\cite{ref44} causes their overlapping response histories to enter both readout branches with nearly equal weights, partially canceling their differential contributions along the coding direction, whereas increasing their temporal separation through grouped ordering reweights the positive and negative histories to preserve stronger differential contrast without altering the pattern count or slot duration. Changing the ordering also changes the temporal separation of complementary readout branches and their disturbance transmission, so the preferred arrangement depends on their combined contribution to target recovery. Inter-slot readout then extracts response differences between complementary pattern slots, whereas intra-slot readout extracts phase-dependent transient variations within each slot. These two readouts provide complementary spatial sensitivities and distinct disturbance-transmission characteristics [Fig.~\ref{fig:3}(b)]: inter-slot averaging attenuates rapid fluctuations but remains sensitive to baseline drift accumulated between complementary branches, while intra-slot slope readout eliminates constant offsets and captures local transient evolution. Rather than assigning mixed readings to the current pattern alone via a memoryless operator $\widetilde{\mathbf{A}}_0$, which introduces a history-dependent mismatch $(\widetilde{\mathbf{A}}-\widetilde{\mathbf{A}}_0)\mathbf{x}$, or attempting to separate individual projection amplitudes~\cite{ref13}, complete-history direct recovery allows inter-slot differences and intra-slot slopes from the same acquisition record to jointly constrain the common target in Eq.~\eqref{Eq5}, as analyzed in Supplementary Sections~S2.1--S2.3.

Co-designing temporal filtering with composite readout further transforms system bandwidth from a passive settling requirement into an active trade-off between disturbance suppression and historical spatial sensitivity. For low-pass responses, stronger filtering or lower receiver bandwidth attenuates broadband upstream fluctuations while prolonging response memory, shortening or even eliminating the steady-state plateau within each pattern slot [Fig.~\ref{fig:3}(c)]. Incorporating the calibrated composite response into $\widetilde{\mathbf{A}}=\mathbf{R}\mathbf{W}\mathbf{P}$ allows these unsettled waveforms to provide matched spatial constraints, where inter-slot and intra-slot readouts jointly exploit the retained target variations. Filter and bandwidth selection can therefore be guided by final target recoverability rather than the existence of an isolated steady-state window: even in an already bandwidth-limited detection chain, stronger low-pass filtering improves reconstruction whenever its noise-suppression benefit outweighs the attenuation of differential target sensitivity $\Delta\mathbf{a}$. Moreover, this formulation naturally distinguishes pre-ADC analog memory, which shapes the physical mixture of pattern projections entering each digital sample, from post-ADC causal digital filtering, which linearly reweights noise correlations and historical constraints within the sampled measurement subspace, as detailed in Supplementary Section~S2.4.

Finally, pre-sampling temporal memory unlocks a distinct historical encoding mechanism under limited digital sampling budgets, elevating sampling timing and modulation rate into active design variables. When the DMD modulation rate exceeds the ADC sampling rate, intervening patterns may complete their display intervals entirely between consecutive digital samples, yet their persisting responses remain present in subsequent readings through nonzero history weights $W_{ik}$ [Fig.~\ref{fig:3}(d)], analogous to multi-code integration in random demodulation~\cite{ref45} and snapshot compressive imaging~\cite{ref23,ref46}. Coupling these unsampled intervening patterns into later observations alters which target variations are distinguishable under a fixed digital observation budget, enabling memory-dependent measurements to distinguish target variations that remain indistinguishable under ideal memoryless instantaneous sampling, even in completely noiseless regimes. Because different within-slot sampling phases capture different historical pattern combinations, this mechanism supports two flexible acquisition adjustments governed by common-target recoverability: reducing the ADC sampling rate at a fixed modulation schedule to lower digital acquisition and transmission overhead, or accelerating the physical modulation rate at a fixed ADC sampling rate and pattern count to shorten optical encoding time, as established in Supplementary Section~S2.5.

\subsection{Generality, Cross-Platform Transferability, and Broader Optoelectronic Implications}
\label{sec2.3}

Beyond specific spatial modulation implementations in SPI, the unified calibration and composite-readout framework embodies a broader physical and methodological generality across high-speed optoelectronic detection domains. Because the operator factorization $\widetilde{\mathbf{A}}=\mathbf{RWP}$ structurally decouples the acquisition-chain impulse response $h(\tau)$ from the physical probing matrix $\mathbf{P}$ and digital readout $\mathbf{R}$, the underlying mathematical structure is agnostic to specific modulation mechanisms. Across different platforms and operating conditions, each receiving chain identifies its effective response from its own reference sequences, constructs history weights from the actual spatial coding sequence and sampling timestamps, and applies matched composite readout with direct common-target recovery. What transfers across operating regimes and hardware implementations is therefore this complete workflow and its underlying design principles, rather than a device-specific heuristic. Any optoelectronic system governed by temporal memory, finite bandwidth, or transient integration can conceptually adopt this modular paradigm, in applications ranging from high-speed single-element photodetection and time-stretch imaging to distributed optical fiber sensing, coherent laser radar, and ultrafast optical communications.

Crucially, this structural decoupling naturally redefines temporal mixing as a controllable physical encoding resource rather than an unmitigated distortion, in a spirit akin to faster-than-Nyquist signaling, where controlled intersymbol interference is deliberately introduced and jointly resolved~\cite{ref47}. The response dynamics of the electro-optical acquisition chain directly shape the continuous mapping from the physical scene to the recorded observations, dictating which target variation directions remain distinguishable under finite sampling budgets. Spatial modulation, temporal memory, and readout processing therefore seamlessly coalesce into an integrated spatiotemporal encoding process. Consequently, measurement design transcends traditional sequential projections onto static individual patterns, advancing to joint spatiotemporal operator engineering: device bandwidth, transient response shapes, pattern ordering, and sampling schedules become coupled degrees of freedom for recovery-oriented measurement design. By incorporating calibrated temporal memory into matched measurement design, this framework provides a transferable basis for coordinating spatial probing, receiving dynamics, sampling, and computational recovery, opening a route to systematic joint optimization and the exploration of recovery limits across optoelectronic platforms.

\section{Calibration and Exploitation of Temporal Mixing: Numerical and Experimental Validation}
\label{sec3}

Validating the paradigm established in Section~\ref{sec2} requires demonstrating not merely that temporal mixing can be compensated, but that calibrated electro-optical memory converts hardware inertia from a speed-limiting ceiling into a portable, configurable encoding resource. To establish this progression across numerical controls and physical experiments, Section~\ref{sec3.1} first verifies joint response-timing identification, synchronization-free drift maintenance, cross-condition response reuse, and full-workflow transfer to an independent receiving chain. Building on this calibrated forward relation, Section~\ref{sec3.2} demonstrates single-stage common-target recovery directly from mixed observations, showing how co-designing inter- and intra-slot readouts with complementary pattern ordering turns modulation well beyond the receiver bandwidth into constructive spatial constraints. Section~\ref{sec3.3} then co-designs temporal filtering and receiver bandwidth with composite readout under structured disturbances, enabling strongly filtered and narrowband configurations to surpass wider-bandwidth steady-state baselines. Finally, Section~\ref{sec3.4} generalizes this framework across diverse objects while unlocking two operating regimes inaccessible to memoryless protocols: sub-modulation digitization that harvests pre-sampling analog memory from unsampled intervening patterns, and synchronization-free real-time video imaging that elevates direct recovery from a static target to a cross-frame coupled spatiotemporal sequence.

\subsection{Calibration and Validation of the Temporal Mixing Model}
\label{sec3.1}

To validate the forward measurement model, we use the reflective DMD-based SPI architecture illustrated in Fig.~\ref{fig:1}(a), where an LED-illuminated target is imaged onto a DMD for spatial modulation and collected by a single-pixel receiving chain. At the 20~kHz DMD modulation rate adopted for our principal imaging experiments, conventional pattern-wise SPI requires the detector response to settle within each $50~\mu\mathrm{s}$ pattern slot; when the modulation rate approaches or exceeds the receiver bandwidth, unmitigated inter-slot persistence causes severe model mismatch. Rather than slowing the modulation to restore isolated steady-state readings, we examine three OPT101-based receiving configurations with estimated bandwidths of approximately 350~Hz, 1.8~kHz, and 18~kHz digitized at a default ADC rate of 240~kHz, alongside an independent receiving chain based on a DET36A2 photodetector and a PDA200C transimpedance amplifier with a 5~kHz nominal bandwidth. Following the calibration protocol in Section~\ref{sec2.1} and Fig.~\ref{fig:2}(a), spatially uniform binary reference sequences are recorded, incorporating known prehistory holds, timing markers, training sequences that combine rapid switching, long holds, and multiscale pulses, independent Block and Interleaved holdout sets, and tail checks. For each configuration, the normalized response $\widehat{\mathbf{h}}$, timing correspondence $\widehat{q}$, and calibration baseline $b_{\mathrm{cal}}$ are jointly fitted and selected using the training sequences and then held fixed for independent Block and Interleaved validation, as detailed in Supplementary Section~S3.1 and Supplementary Table~S1.

\begin{figure*}[t]
    \centering
    \includegraphics[width=\textwidth]{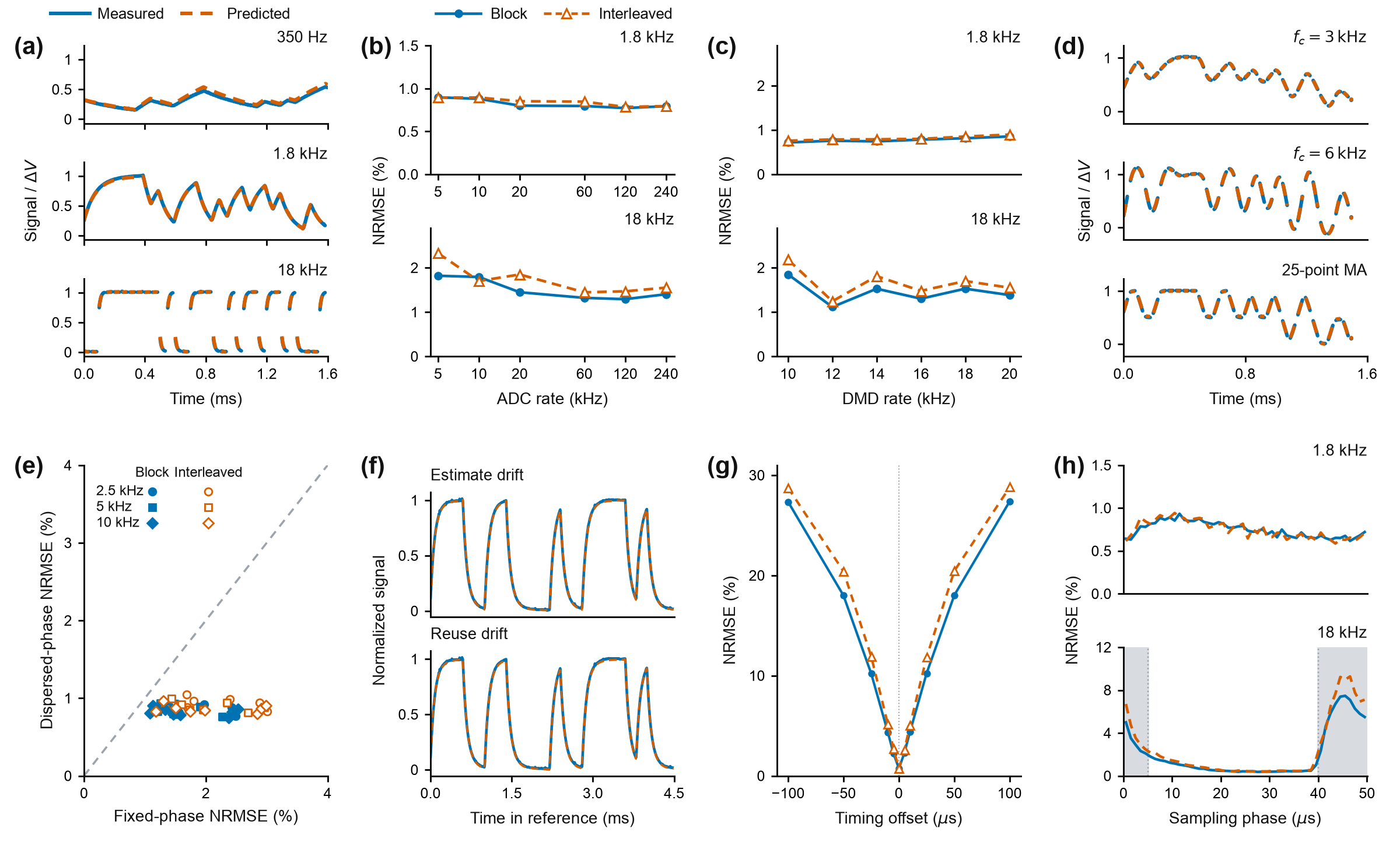}
    \caption{\textbf{Reference calibration and performance validation.}
(a) Waveform predictions across the 350~Hz, 1.8~kHz, and 18~kHz detector configurations.
(b, c) NRMSE vs.\ ADC sampling and DMD modulation rates for the 1.8~kHz and 18~kHz configurations.
(d) Waveform predictions under causal Butterworth low-pass and moving-average (MA) filters for the 18~kHz configuration.
(e) Comparison of fixed-phase and dispersed-phase calibration schemes.
(f) Timing-reference alignment in the Estimate drift and Reuse drift modes.
(g) Validation NRMSE under imposed timing offsets with $\widehat{\mathbf{h}}$ held fixed.
(h) Phase-resolved prediction NRMSE, where gray areas denote excluded switching intervals.
Panels (e, f, g) use the 1.8~kHz configuration, and waveforms for the 18~kHz configuration in (a, d) are plotted within the valid within-slot windows defined in (h).
In the waveform plots (a, d, f), solid blue and dashed orange curves denote measured and predicted signals, respectively; in the error plots (b, c, e, g, h), blue and orange curves and markers denote Block and Interleaved validation NRMSE, respectively, with all NRMSE values expressed as percentages. $\Delta V$ denotes the reference black--white plateau span.}
    \label{fig:4}
\end{figure*}

Figure~\ref{fig:4} summarizes the experimental validation of response identification and timing maintenance across operating conditions. Across the 350~Hz, 1.8~kHz, and 18~kHz receiving chains, the calibrated model closely predicts held-out reference waveforms despite substantial differences in response memory [Fig.~\ref{fig:4}(a)], with the 18~kHz waveforms displayed within their valid intra-slot observation windows. Independently scaling the ADC sampling rate from 5 to 240~kHz [Fig.~\ref{fig:4}(b)] and the DMD modulation rate from 10 to 20~kHz [Fig.~\ref{fig:4}(c)] yields consistently low prediction errors, with normalized root-mean-square error (NRMSE) remaining below $1.0\%$ for the 1.8~kHz chain and $2.4\%$ for the 18~kHz chain. Even when the ADC sampling rate falls below the DMD modulation rate, so that some physical patterns complete their display intervals between consecutive digital samples, the calibrated relation accurately predicts observations from the actual sampling timestamps and complete pattern histories. Furthermore, when causal digital filters, specifically second-order 3~kHz and fourth-order 6~kHz Butterworth low-pass filters as well as a 25-point moving average, are applied to continuous 18~kHz acquisition records, matched re-identification captures the composite acquisition-and-processing response with Block and Interleaved NRMSE values between $0.854\%$ and $1.306\%$ [Fig.~\ref{fig:4}(d)].

For the 1.8~kHz chain under limited training sampling budgets of 2.5, 5, and 10~kHz across 24 paired comparisons, within-slot phase distribution critically influences response identification [Fig.~\ref{fig:4}(e)]. Harnessing natural ADC--DMD clock drift for dispersed phase coverage outperforms fixed-phase sampling in all 24 comparisons, reducing Block NRMSE from $1.084\%\text{--}2.530\%$ to $0.749\%\text{--}0.963\%$ and Interleaved NRMSE from $1.178\%\text{--}3.006\%$ to $0.799\%\text{--}1.044\%$ under matched observation counts. For asynchronous continuous operation on the same 1.8~kHz chain, embedded reference sequences maintain slot coordinates without external hardware triggers while keeping the calibrated response $\widehat{\mathbf{h}}$ fixed [Fig.~\ref{fig:4}(f)]: both the Estimate drift mode, which updates initial phase and slot rate from two references, and the Reuse drift mode, which updates initial phase from one reference while retaining the calibrated slot rate, achieve a normalized correlation coefficient (NCC) of $0.9998$, with prediction NRMSE values of $0.784\%$ and $0.931\%$, respectively. Coordinated translations of the timing origin and response-centroid reference, with the response refitted at each offset, retain nearly unchanged holdout prediction accuracy within $0.739\%\text{--}0.765\%$ NRMSE across tested offsets up to $\pm100~\mu\mathrm{s}$, as shown in Supplementary Section~S3.3 and Supplementary Fig.~S2(a) and (b); in contrast, displacing the timing origin alone with $\widehat{\mathbf{h}}$ fixed already increases Block NRMSE from $0.752\%$ to $2.295\%\text{--}2.343\%$ and Interleaved NRMSE to $2.635\%\text{--}2.734\%$ at $\pm5~\mu\mathrm{s}$, and drives both errors steeply upward across offsets up to $\pm100~\mu\mathrm{s}$ [Fig.~\ref{fig:4}(g)], with the corresponding image degradation shown in Supplementary Fig.~S2(c). Finally, phase-resolved prediction diagnostics account for finite micromirror switching transients, characterized by a 4~$\mu$s crossover and 12.5~$\mu$s settling time, to define fixed valid observation windows prior to static imaging [Fig.~\ref{fig:4}(h)]: all within-slot samples are retained for the 350~Hz and 1.8~kHz chains, whereas the first 5~$\mu$s and final 10~$\mu$s of each slot are excluded for the 18~kHz chain while retaining the complete physical pattern history, as detailed in Supplementary Section~S3.2.

To examine whether an identified response is bound to a single calibration acquisition, we evaluate response reuse across three pairs of $15\times15$ cross-condition transfer matrices covering the 1.8~kHz, 18~kHz, and 350~Hz receiving chains, encompassing 675 source--receiver pairs and 1,350 Block and Interleaved NRMSE values across different ADC rates, DMD rates, and illumination levels, together with cross-date tests using archived calibrations acquired 28 or 38 days earlier. During transfer, the source-calibrated $\widehat{\mathbf{h}}$ remains fixed while history weights are reconstructed from the receiving record's timing correspondence and pattern sequence. For the 1.8~kHz chain, cross-condition transfer maintains Block and Interleaved NRMSE within $0.719\%\text{--}1.046\%$ and $0.749\%\text{--}1.205\%$, respectively, while cross-date reuse after 28 or 38 days yields $0.783\%\text{--}0.960\%$ across both timing-maintenance modes, whereas the 18~kHz and 350~Hz transfer matrices further reflect the disturbance levels of individual receiving records, as documented in Supplementary Section~S3.5 and Supplementary Figs.~S4 and S5.

Because reference-waveform NRMSE reflects both model fidelity and observation fluctuations, we further test whether useful responses can be identified from strongly fluctuating references. In the 350~Hz chain, pronounced low-frequency fluctuations inflate the receiving NRMSE on certain records such as record L20, where it reaches $16.42\%\text{--}19.05\%$, yet the response calibrated from L20 achieves $3.543\%\text{--}3.982\%$ NRMSE when transferred to cleaner receiving records. In a controlled disturbance experiment where a low-frequency fluctuation with an RMS amplitude of $7\%$ of the reference black--white plateau span is added to a filtered reference record, the response identified from the disturbed training data predicts the disturbed observations with $6.465\%\text{--}7.253\%$ NRMSE, but predicts the underlying pre-addition reference waveform with only $0.942\%\text{--}1.264\%$ NRMSE. Consistently, across 60 reconstructions of diverse targets using baseline and transferred responses, including L20, across the three receiving chains, the recovered images remain visually indistinguishable from those obtained with the reference calibration, confirming that direct image recovery depends on accurate operator identification rather than raw waveform error, as shown in Supplementary Sections~S3.6 and S3.7 and Supplementary Figs.~S6--S8.

Finally, we validate response adaptation under measurement-chain changes, both when digital filtering is incorporated into the 18~kHz receiving records and when the complete workflow is transferred to the independent DET36A2/PDA200C receiving chain operating at a $5~\mathrm{kHz}$ nominal bandwidth and a $250~\mathrm{kHz}$ ADC rate. Identifying the effective response of the filtered 18~kHz records preserves character recovery across all tested filters, while calibrating the independent DET36A2/PDA200C chain from its own reference sequences with $4.273\%\text{--}4.516\%$ validation NRMSE and constructing its matched history-weighted operator successfully recovers an independently acquired U-shaped target, as shown in Supplementary Section~S3.4 and Supplementary Fig.~S3. Together, these results establish a calibrated, maintainable, and transferable forward model for practical image recovery.

\subsection{Spatial Information Recovery and Joint Exploitation in Temporal Mixing}
\label{sec3.2}

Building on the calibrated forward model established in Section~\ref{sec3.1}, we next evaluate direct common-target recovery from temporally mixed observations and examine how complete-history modeling, composite readouts, and pattern ordering govern the exploitation of target-dependent spatial information. Following the variational framework in Eq.~\eqref{Eq5}, the combined-readout reconstruction is formulated as
\begin{equation}
\widehat{\mathbf{x}} \in \arg\min_{\mathbf{x}} \left\{ \frac{1}{2} \Vert\widetilde{\mathbf{A}}_{\mathrm{out}}\mathbf{x} - \mathbf{z}_{\mathrm{out}}\Vert_{2}^{2} + \frac{\beta}{2} \Vert\widetilde{\mathbf{A}}_{\mathrm{in}}\mathbf{x} - \mathbf{z}_{\mathrm{in}}\Vert_{2}^{2} + \lambda\,\mathrm{TV}(\mathbf{x}) \right\},
\label{Eq7}
\end{equation}
where $\mathbf{z}_{\mathrm{out}}$ and $\mathbf{z}_{\mathrm{in}}$ denote the inter-slot mean-difference and intra-slot temporal-slope readouts, respectively, and $\widetilde{\mathbf{A}}_{\mathrm{out}}$ and $\widetilde{\mathbf{A}}_{\mathrm{in}}$ are their matched history-dependent operators obtained by applying the same linear readout rows in $\mathbf{R}$. Both readouts constrain the same target image $\mathbf{x}$, where $\lambda \ge 0$ controls total variation (TV) regularization~\cite{ref48} and $\beta \ge 0$ determines the relative participation of the intra-slot branch, such that $\beta = 0$ reduces Eq.~\eqref{Eq7} to inter-slot-only recovery. To separate the effects of history modeling and composite readout, we compare three schemes: Base, which pairs inter-slot readouts with the memoryless current-pattern model $\widetilde{\mathbf{A}}_0$; Inter, which sets $\beta=0$ to pair the same inter-slot readouts with the complete-history operator $\widetilde{\mathbf{A}}_{\mathrm{out}}$; and Joint, which uses the combined-readout formulation in Eq.~\eqref{Eq7}. Across all static comparisons, each method selects its regularization and branch weights under a consistent evaluation protocol, as detailed in  Supplementary Section~S4.1.

\begin{figure*}[t]
    \centering
    \includegraphics[width=\textwidth]{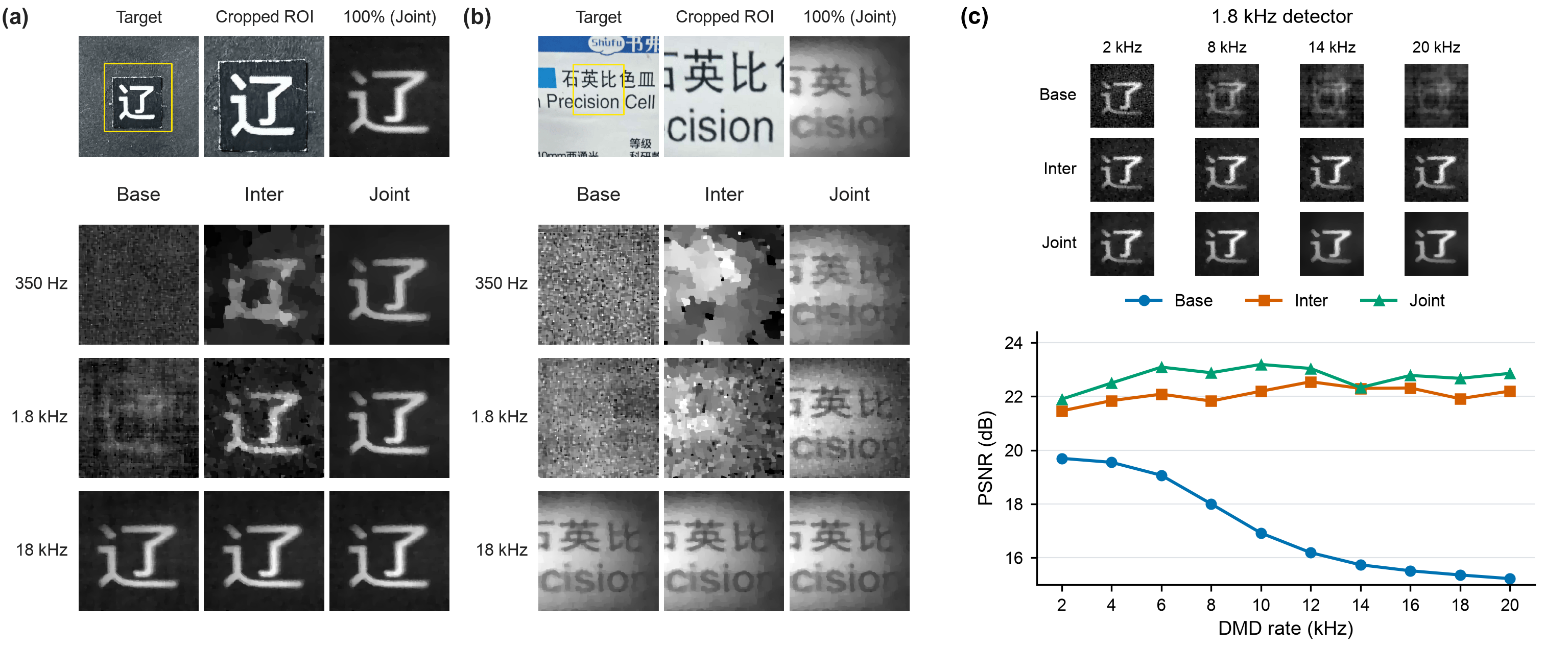}
    \caption{\textbf{Experimental image recovery under temporal mixing.}
    (a) and (b) Reconstructions of the character and printed targets at $12.5\%$ spatial sampling. The top row shows the target photograph with the imaging region outlined in yellow, the cropped region of interest, and the full-sampling Joint reference. Reconstruction rows correspond to the indicated detector bandwidths; columns compare current-pattern inter-slot reconstruction (Base), complete-history inter-slot reconstruction (Inter), and joint intra- and inter-slot reconstruction (Joint). Images of each target share the display range of its full-sampling reference.
  (c) Reconstructions at representative DMD modulation rates (top; 2, 8, 14, and 20~kHz) and peak signal-to-noise ratio (PSNR) curves across 2--20 kHz (bottom) for Base, Inter, and Joint with the 1.8~kHz detector, evaluated relative to the full-sampling reference in (a).}
    \label{fig:5}
\end{figure*}

Figure~\ref{fig:5}(a) and (b) present reconstructions of a character target and a printed target across the 350~Hz, 1.8~kHz, and 18~kHz detector configurations at $256 \times 256$ resolution, a $12.5\%$ spatial sampling ratio, and a 20~kHz DMD modulation rate. These experiments employ complementary binary Hadamard patterns~\cite{ref44} in Cake-cutting order~\cite{ref43} with a grouped C44 arrangement, in which four positive patterns from distinct encoding directions are followed by their four negative counterparts. At 18~kHz, all three methods yield comparable reconstructions. At 1.8~kHz, inter-pattern crosstalk causes severe structural distortion in Base; complete-history modeling in Inter restores the primary strokes, but noticeable background artifacts remain in the presence of the low-frequency fluctuations visible in the measured acquisition records. Joint substantially improves stroke continuity, background uniformity, and fine-feature visibility. At 350~Hz, where extended temporal memory and pronounced low-frequency fluctuations coexist, target structures are obscured in Base and severely degraded in Inter, whereas Joint recovers recognizable character strokes and printed text with a clean background, as contextualized by the measured imaging waveforms in Supplementary Section~S4.1 and Supplementary Fig.~S9.

Mechanistically, these comparisons reflect the complementary spatial sensitivities and disturbance responses of the two readout branches. While complete-history modeling in Inter resolves the deterministic mixing mismatch $(\widetilde{\mathbf{A}}_{\mathrm{out}}-\widetilde{\mathbf{A}}_0)\mathbf{x}$, inter-slot mean differences remain sensitive to slowly varying baseline variations accumulated between complementary branches [Fig.~\ref{fig:3}(b)]. The zero-sum intra-slot slope rejects constant offsets while retaining the local rate of change and the target-dependent historical evolution across sampling phases [Figs.~\ref{fig:2}(b) and \ref{fig:3}(a)]. Together with inter-slot differences, it provides complementary spatial sensitivities and disturbance responses for fitting the same target, as analyzed in Supplementary Section~S2.2.

Pattern ordering further interacts with temporal mixing and disturbance composition to shape these effective constraints. In controlled simulations with a fixed total post-response disturbance RMS of $0.3\%$ of the full-white response, varying the white-noise power fraction from $0.1\%$ to $20\%$ shifts the more favorable ordering from adjacent complementary pairing C11, which achieves $30.06~\mathrm{dB}$ compared with $29.70~\mathrm{dB}$ for C44 with Joint at $0.1\%$, to grouped pairing C44, which achieves $23.92~\mathrm{dB}$ compared with $21.50~\mathrm{dB}$ for C11 at $20\%$, corresponding to a $2.42~\mathrm{dB}$ gain. Moreover, measured C11 Joint reconstructions across 10--20~kHz exhibit corner-region extrema that compress apparent target contrast, while continuous-record power spectral density estimates reveal a prominent periodic component near half the DMD modulation frequency $f_{\mathrm{DMD}}/2$ in C11 that is substantially weaker in C44, consistent with their different periodic micromirror switching loads. Pattern ordering therefore modifies both retained historical spatial encoding and the periodic signal structure of physical acquisitions, acting as a system variable that must be co-designed with the receiving response and composite readout, as shown in Supplementary Section~S4.3 and Supplementary Figs.~S11 and S12.

Finally, sweeping the DMD modulation rate from 2 to 20~kHz with the 1.8~kHz detector demonstrates the shift from distortion mitigation to active mixing exploitation [Fig.~\ref{fig:5}(c)]. Whereas Base PSNR drops monotonically from $19.70$ to $15.23~\mathrm{dB}$ due to unmodeled history and fewer samples per slot, complete-history modeling in Inter preserves character structure across the sweep, and Joint consistently delivers the highest PSNR, stroke continuity, and background uniformity. Crucially, neither Inter nor Joint peaks at $2~\mathrm{kHz}$ near per-slot settling; instead, modulating well beyond the $1.8~\mathrm{kHz}$ receiver bandwidth yields superior recovery, as shown in Supplementary Section~S4.2 and Supplementary Fig.~S10(a). Under matched recovery, approaching steady-state settling weakens intra-slot differential sensitivity $\Delta\mathbf{a}$ and prolongs inter-branch exposure to low-frequency fluctuations. Conversely, faster modulation enriches multi-slot transient encoding and compresses drift intervals, with higher-rate gains bounded by reduced per-slot averaging and record-level noise variations. Once direct recovery removes the requirement for isolated projection settling, modulation rate, pattern ordering, and composite readout act as coupled co-design variables governed by final image recoverability and disturbance structure rather than waveform settling.

\subsection{Co-Design of Temporal Filtering and Joint Readout}
\label{sec3.3}

\begin{figure*}[t]
    \centering
    \includegraphics[width=\textwidth]{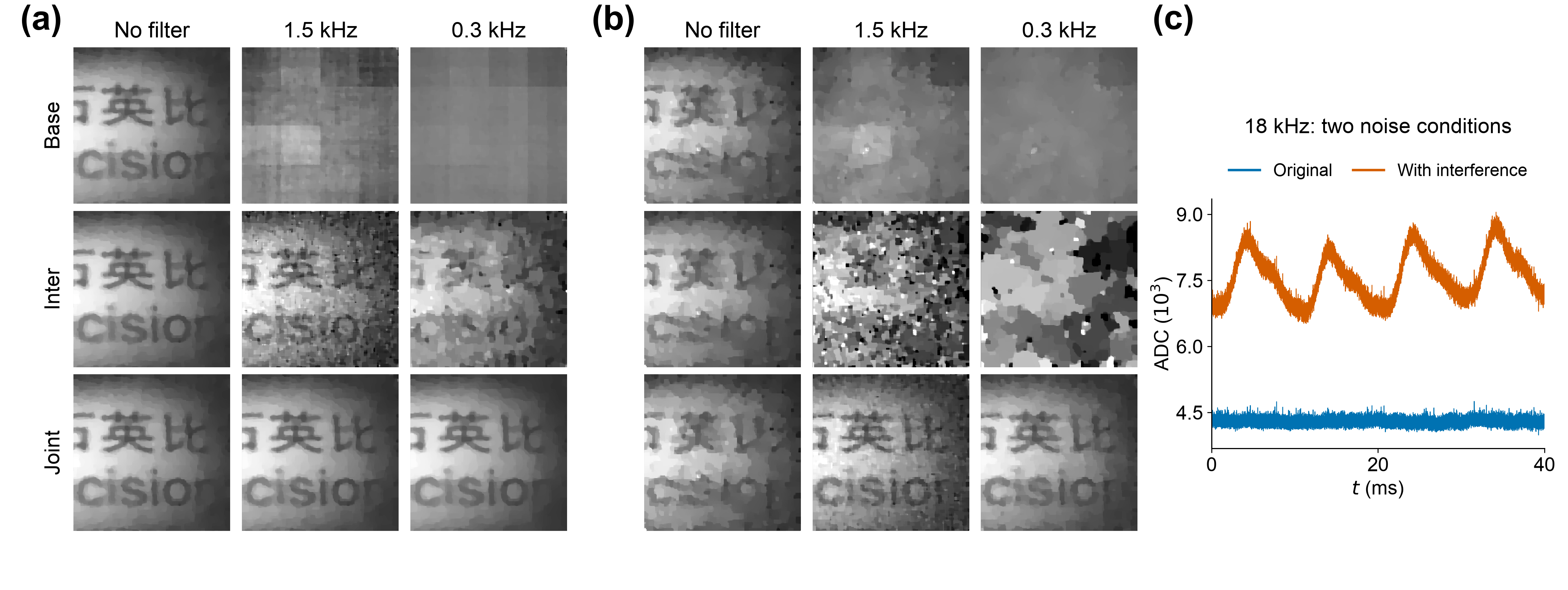}
    \caption{\textbf{Temporal filtering and joint readout with the 18~kHz detector.}
    (a) Reconstructions under the original illumination condition.
    (b) Reconstructions with additional optical interference.
    In (a) and (b), rows show Base, Inter, and Joint, respectively; columns show no additional filtering and first-order low-pass filtering with cutoff frequencies of 1.5~kHz and 0.3~kHz.
    All reconstructions share the grayscale range determined by the full-sampling reference in Fig.~\ref{fig:5}(b).
    (c) ADC waveform excerpts before additional filtering under the original illumination condition in blue and with additional optical interference in orange.}
    \label{fig:6}
\end{figure*}

Building on Section~\ref{sec3.2}, we now examine image recovery when temporal response memory is deliberately modified through additional low-pass filtering and receiver-bandwidth selection. Because temporal filtering alters both the historical superposition of target projections and the temporal correlation of disturbances, its effect cannot be separated from the subsequent composite readout, implementing the co-design mechanism introduced in Section~\ref{sec2.2} and Fig.~\ref{fig:3}(c). Using the printed target at $256\times256$ resolution and $12.5\%$ spatial sampling with the 18~kHz detector, Fig.~\ref{fig:6}(a) and (b) evaluate reconstructions under the original illumination condition and with additional optical interference, with corresponding unfiltered waveform excerpts shown in Fig.~\ref{fig:6}(c), comparing unfiltered records against causal first-order digital low-pass filtering at cutoff frequencies of $1.5$ and $0.3~\mathrm{kHz}$.

To quantify the relative participation of the two readout branches across filtering conditions, we define the scale-normalized intra-slot weight:
\begin{equation}
\rho_{\mathrm{in}} = \frac{\beta \Vert\widetilde{\mathbf{A}}_{\mathrm{in}}\Vert_F^2}{\Vert\widetilde{\mathbf{A}}_{\mathrm{out}}\Vert_F^2 + \beta \Vert\widetilde{\mathbf{A}}_{\mathrm{in}}\Vert_F^2},
\label{Eq8}
\end{equation}
where $\Vert\cdot\Vert_F$ denotes the Frobenius norm. Under the original illumination condition [Fig.~\ref{fig:6}(a)], lowering the filter cutoff frequency causes Base and Inter to progressively lose fine lettering, whereas Joint preserves the primary text structure as the selected intra-slot weight $\rho_{\mathrm{in}}$ increases from $0.08\%$ without filtering to $93.3\%$ at $1.5~\mathrm{kHz}$ and $99.8\%$ at $0.3~\mathrm{kHz}$, reflecting a systematic shift from inter-slot to intra-slot dominance. Under additional optical interference [Fig.~\ref{fig:6}(b) and (c)], unfiltered reconstructions exhibit pronounced granular artifacts across all methods. Applying causal low-pass filtering attenuates rapid fluctuations without reducing the in-band periodic interference, while the calibrated history model and intra-slot slope readout reject this slow variation and retain target-dependent transient constraints within the strongly mixed waveforms. Even at a cutoff frequency of $0.3~\mathrm{kHz}$, far below the $20~\mathrm{kHz}$ DMD modulation rate where steady-state plateaus vanish and heavy smoothing distorts isolated waveforms, direct Joint recovery avoids ill-conditioned waveform inversion and embeds this smoothing into the matched operator, substantially suppressing granular artifacts and structural fragmentation to outperform the unfiltered $18~\mathrm{kHz}$ reconstructions. Extending this causal digital filtering sweep on measured 18~kHz character and printed-target records across eight cutoff frequencies from $10,000$ down to $20~\mathrm{Hz}$ confirms that while Inter degrades under severe smoothing, Joint retains recognizable character strokes and principal printed features even at $20~\mathrm{Hz}$, as shown in Supplementary Section~S4.2 and Supplementary Fig.~S10(b) and (c).

To further examine pre-ADC bandwidth selection when disturbances enter before the receiving response, we compare matched Joint recovery against conventional pattern-wise single-pattern readout in controlled simulations with first-order receiving responses across six bandwidths of $100$, $50$, $18$, $10$, $1.8$, and $0.35~\mathrm{kHz}$ and four conventional within-slot sampling windows that retain the last 1, 3, 5, or all 10 permitted samples per slot. Under a $10\%$ RMS low-frequency upstream disturbance and $0.1\%$ RMS white noise relative to the full-white response, the $350~\mathrm{Hz}$ Joint reconstruction achieves $23.86~\mathrm{dB}$ for Cameraman and $28.47~\mathrm{dB}$ for Peppers, not only substantially outperforming conventional readouts at the same $350~\mathrm{Hz}$ bandwidth, but also surpassing the best conventional configurations selected across all six bandwidths and four sampling windows, which are achieved at $100~\mathrm{kHz}$, yielding net gains of $2.71$ and $4.44~\mathrm{dB}$. When the upstream white-noise RMS is increased to $1\%$, the $350~\mathrm{Hz}$ Joint reconstructions remain far superior to same-bandwidth conventional readouts and closely approach, though remain slightly below, the best conventional results across all tested bandwidths, which are achieved at $50~\mathrm{kHz}$. These comparisons demonstrate that temporal response and composite readout act as coupled design variables governed by the balance between retained historical target sensitivity and upstream disturbance transmission, as detailed in Supplementary Section~S4.4 and Supplementary Fig.~S13.

\subsection{Diverse Targets, Low Sampling Rates, and Real-Time Video Imaging}
\label{sec3.4}
\begin{figure*}[t]
    \centering
    \includegraphics[width=0.7\textwidth]{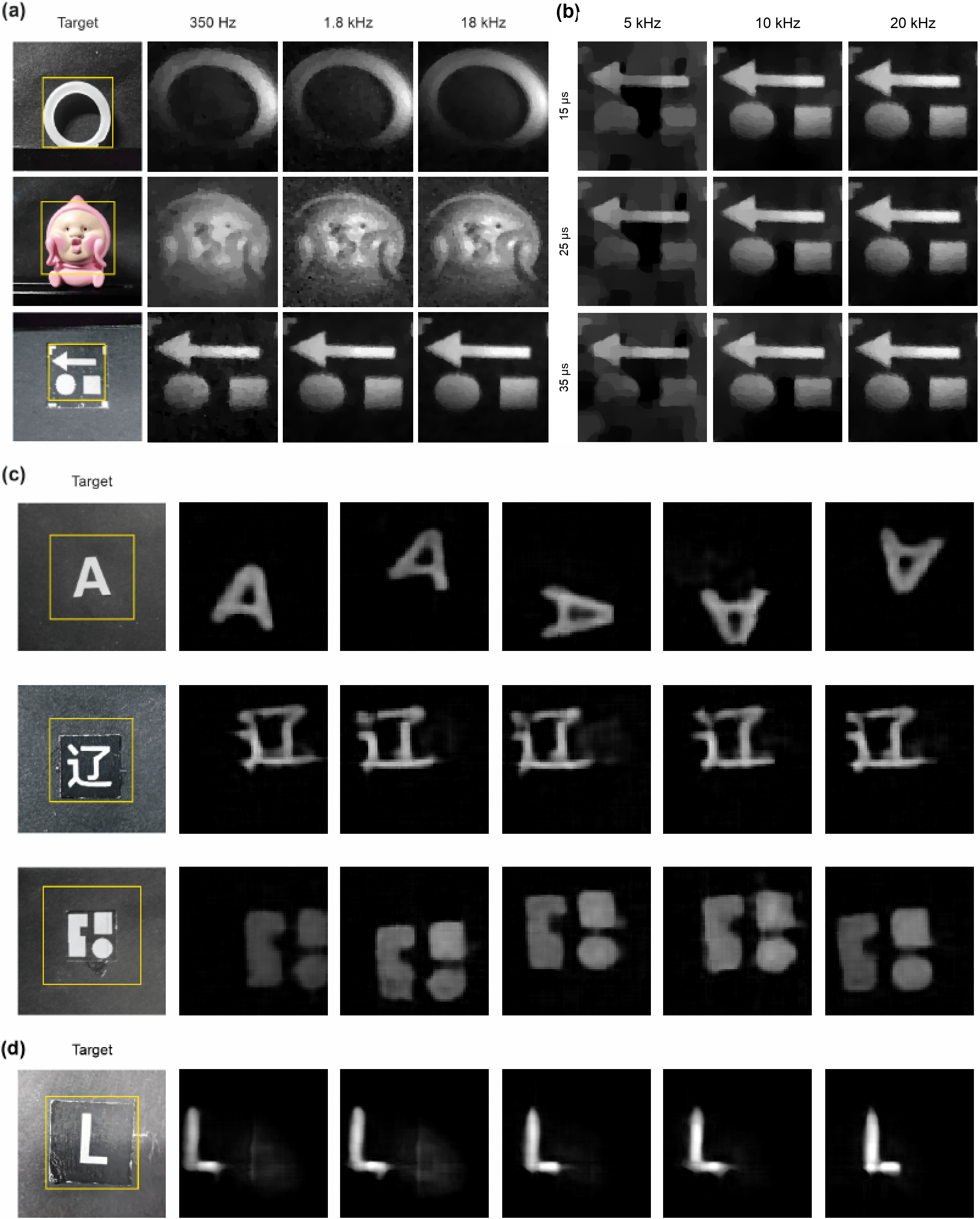}
    \caption{\textbf{Diverse targets, low sampling rates, and real-time video imaging.}
(a) Photographs and Joint reconstructions of the ring, figurine, and composite geometric targets under the 350~Hz, 1.8~kHz, and 18~kHz detector configurations.
(b) Complete-history Inter reconstructions of the composite geometric target across sparse sampling rates and within-slot phases extracted from a common 240~kHz record under the 1.8~kHz detector configuration. Columns correspond to effective sampling rates of 5, 10, and 20~kHz, and rows to target within-slot phases of 15, 25, and $35~\mu\mathrm{s}$, displayed using the fixed grayscale range adopted for this target in (a).
(c) Representative frames from $64\times64$ real-time imaging of the letter, character, and geometric targets using U-Net-based online reconstruction at approximately 25~frames/s, capturing continuous translation, in-plane rotation, and illumination-induced brightness variations as the primary dynamic-imaging demonstration.
(d) A higher-resolution demonstration of the current implementation, showing a moving letter L at $256\times256$ and approximately 25~frames/s.
In (c, d), each row shows a target photograph followed by five selected frames in chronological order from left to right. Yellow boxes indicate the imaged regions.}
    \label{fig:7}
\end{figure*}

Building on the established calibration and composite-readout framework, we finally demonstrate imaging across diverse target morphologies, sparse temporal sampling, and real-time dynamic scenes. Unless otherwise specified, static experiments use a $256 \times 256$ resolution, a $12.5\%$ spatial sampling ratio, and the parameter-selection protocol of Section~\ref{sec3.2}. Figure~\ref{fig:7}(a) presents Joint reconstructions of a ring, a figurine, and a composite geometric object across the 350~Hz, 1.8~kHz, and 18~kHz receiving chains. Across all three bandwidths, the reconstructions consistently resolve continuous surface shading, complex local reflectance, and sharp geometric boundaries, complementing the full-sampling $64\times64$ benchmarks for diverse targets in Supplementary Section~S5.1 and Supplementary Fig.~S15.

Figure~\ref{fig:7}(b) evaluates direct complete-history recovery across sparse temporal sampling rates and within-slot sampling phases. For the composite geometric target, sparse observations are extracted from a common $240~\mathrm{kHz}$ record under the $1.8~\mathrm{kHz}$ detector configuration and $20~\mathrm{kHz}$ DMD modulation at effective sampling rates of $5$, $10$, and $20~\mathrm{kHz}$ (one observation every four, two, or one physical pattern slots) and target within-slot phases of $15$, $25$, and $35~\mu\mathrm{s}$. Because each selected slot provides a single sample, all nine cases use complete-history Inter under identical calibrated forward parameters and regularization settings.

Under the common grayscale range used for this target in Fig.~\ref{fig:7}(a), target geometries are consistently recovered across all nine rate-and-phase combinations without requiring end-of-slot settling, with contour sharpness and background uniformity improving steadily from $5$ to $20~\mathrm{kHz}$. Notably, even at $5~\mathrm{kHz}$, where three of every four physical patterns carry no direct sample, their persisting analog responses remain captured in subsequent observations through the complete-history model [Fig.~\ref{fig:3}(d)]. At the same time, holding the observation budget fixed within each column reveals distinct phase-dependent variations in edge definition and low-rate background residuals across $15$, $25$, and $35~\mu\mathrm{s}$, confirming that within-slot sampling timing directly shapes final image quality; because shifting the selected timestamps alters both the historical weight conditioning and the sampled noise realizations, these variations reflect the combined effect of operator structure and experimental noise. Complementary experiments on directly acquired low-rate character records further confirm that single-stage complete-history recovery retains clearer strokes and a more coherent background than two-stage minimum-norm projection inversion, as shown alongside an extended $5\text{--}240~\mathrm{kHz}$ ADC-rate sweep in Supplementary Section~S4.5 and Supplementary Fig.~S14(a) and (b).

To isolate the spatial-encoding gain of temporal memory from the noise variations present in the experimental phase comparison, we compare ideal memoryless sensing against known first-order responses with $18$, $10$, and $1.8~\mathrm{kHz}$ bandwidths in noiseless simulations under matched C44 pattern sequences and identical digital observation budgets corresponding to $5$ and $10~\mathrm{kHz}$ sampling. When the sampled slots and within-slot phases are held fixed identically across configurations, direct recovery with temporal memory substantially improves image quality over ideal memoryless sampling, even though the latter provides exact isolated projections at every sampled instant. Representative fixed-schedule comparisons yield PSNR gains of $2.20\text{--}4.38~\mathrm{dB}$ with the $1.8~\mathrm{kHz}$ response. When each sensing configuration is instead allowed to select its best sampling schedule from the prescribed set of slots and phases, the best mixed configurations still retain net gains of $0.68$ and $0.79~\mathrm{dB}$ over the best memoryless results for Peppers at $5$ and $10~\mathrm{kHz}$, while achieving nearly identical peak PSNR for Cameraman. Because these controls contain no noise, the improvements arise directly from historical spatial encoding and its joint selection with sampling timing, as analyzed in Supplementary Section~S4.5 and Supplementary Fig.~S14(c) and (d).

Finally, Fig.~\ref{fig:7}(c) and (d) extend the framework to continuous real-time dynamic imaging with Joint readout at $20~\mathrm{kHz}$ DMD modulation on the $1.8~\mathrm{kHz}$ receiving chain, elevating direct recovery from a single static image to the spatiotemporal video sequence itself. Removing per-slot steady-state settling decouples modulation speed from the narrow receiver bandwidth, compressing $784$ C44 pattern slots into a $39.2~\mathrm{ms}$ nominal frame interval ($9.57\%$ spatial sampling at $64\times64$ resolution and $0.60\%$ at $256\times256$ resolution) to curb intra-frame motion blur while composite Joint readout suppresses motion- and illumination-induced flux drift. Brief timing references periodically embedded in the modulation stream maintain alignment online via the frozen calibrated response $\widehat{\mathbf{h}}$, retaining the preceding valid timing model through causal holdover whenever a reference is obscured during target entry, exit, or saturation. Rather than introducing an intermediate waveform-deconvolution stage prior to frame reconstruction, a causal two-frame joint historical operator models cross-frame persistence across the $3.20~\mathrm{ms}$ response support as physical coupling constraints linking consecutive scene states. Coupling this operator with a physics-guided hybrid neural solver that alternates data-consistency gradient updates and learned spatial residual corrections from a shared-weight U-Net regularizer frees the network from learning hardware-specific temporal deconvolution; acting purely as a spatial prior, the network is trained offline exclusively on synthetic two-frame sequences rendered through the calibrated forward model without requiring experimental ground-truth videos, as detailed in Supplementary Section~S5.2.

Executed on an Apple M4 platform, the mean online processing time per frame is $19.48\text{--}20.91~\mathrm{ms}$ at $64\times64$ resolution and $30.38~\mathrm{ms}$ at $256\times256$ resolution, both comfortably below the $39.2~\mathrm{ms}$ nominal frame interval to support continuous synchronization-free operation at approximately $25~\mathrm{frames/s}$. Across unconstrained translation, large-angle in-plane rotation, and illumination-induced frame-to-frame brightness variations in the $64\times64$ sequences [Fig.~\ref{fig:7}(c)], together with the $256\times256$ moving-letter sequence [Fig.~\ref{fig:7}(d)], target structures and motion trajectories remain clearly resolved. Moreover, held-out observation evaluations across the measured $64\times64$ dynamic streams confirm that incorporating intra-slot transient slopes reduces the median held-out inter-slot prediction RMSE by 15.80\%--24.74\% relative to inter-slot-only recovery, as documented alongside cross-frame coupling controls in Supplementary Section S5.3. Runtime and extended video evaluations are reported in Supplementary Section S5.4, Supplementary Table S2, and Supplementary Videos 1--5.

\section{Discussion}\label{sec4}

Multiplexed optical sensing and computational imaging rest on an implicit premise of temporal separability, requiring the acquisition chain to settle within each modulation slot as a memoryless transducer~\cite{ref10,ref13,ref26,ref28}. Under this convention, finite photodetector bandwidths and relaxation tails impose a rigid speed ceiling, forcing an unfavorable trade-off between wideband receivers that sacrifice transimpedance sensitivity~\cite{ref26,ref29} and post-acquisition deconvolution that amplifies noise under ill-conditioned overlap~\cite{ref13,ref28,ref31}. Embedding calibrated complete-history memory into the matched forward operator $\widetilde{\mathbf{A}}=\mathbf{R}\mathbf{W}\mathbf{P}$ overturns per-slot settling as an acquisition prerequisite and recasts hardware inertia into a configurable spatiotemporal encoding dimension~\cite{ref16,ref30}. Crucially, our noiseless sub-modulation controls prove that this gain is geometric rather than merely statistical: pre-sampling analog memory couples unsampled optical states into finite digital readings, strengthening algebraic distinguishability within a fixed observation budget. Together with our disturbance evaluations, where strongly bandwidth-limited configurations surpass wider-bandwidth steady-state setups under pronounced low-frequency fluctuations within limits set by excessive smoothing, coding cancellation, broadband noise, and calibration residuals, these results overturn the longstanding assumption that wider receiver bandwidth is intrinsically superior for high-speed optical acquisition.

Methodologically, this formulation breaks with the paradigm of intermediate waveform restoration. Hardware equalization~\cite{ref26} and two-stage numerical inversion~\cite{ref13} both treat step-settled projection waveforms as an obligatory intermediate target, leaving common-scene consistency unenforced during temporal inversion, propagating stage-one residuals, and even restoring upstream optical fluctuations that a low-pass receiver had physically attenuated. Our high-rate, strongly filtered, sub-modulation, and dynamic experiments establish a sharper principle: intermediate waveform fidelity is neither necessary nor optimal for recovery, as controlled mixing and smoothing can actively strengthen target constraints while suppressing disturbances. Dispensing with projection separation advances slow-detector exploitation from sparse waveform estimation~\cite{ref49} to direct variational image recovery~\cite{ref48} and synchronization-free real-time video reconstruction via two-frame historical coupling and physics-guided neural regularization. In dynamic scenes, calibrated persistence supplies physical coupling constraints between successive scene states, extending direct recovery from a single common image to an evolving scene. More fundamentally, images and videos are specific instances of a universal task-direct recovery paradigm. Across single-detector modalities where steady-state waveforms are conventionally acquired or deconvolved only to be processed a second time into a downstream physical quantity or semantic decision, coupling the calibrated complete-history model directly with the task mapping bypasses steady-state waveform restoration altogether. Evaluating acquisition directly by terminal task recoverability turns rigid hardware prerequisites into synergistic degrees of freedom across data-driven reference identification~\cite{ref36}, embedded timing maintenance, pattern ordering, pre- and post-ADC filtering, and composite inter- and intra-slot readouts~\cite{ref42,ref44}.

Structurally, isolating receiver memory $\mathbf{W}$ from spatial probing $\mathbf{P}$ and readout $\mathbf{R}$ in Section~\ref{sec2.3}, as validated on independent hardware in Section~\ref{sec3.1}, means that the calibrated kernel $\widehat{\mathbf{h}}$ characterizes the detector chain itself rather than the modulator, extracted directly from reference waveforms without internal circuit parameters~\cite{ref35}. This directly unlocks high-sensitivity, inertia-limited single-element detectors where focal-plane arrays remain prohibitive, spanning capacitance-limited near- and short-wave infrared photoreceivers~\cite{ref12,ref29}, thermally governed mid- and far-infrared pyroelectric sensors~\cite{ref50}, and terahertz or X-ray detectors~\cite{ref6,ref13,ref14} exhibiting carrier-recombination persistence or scintillator afterglow. Crucially, whereas fast programmable encoders such as DMDs, LED arrays~\cite{ref33}, and metasurfaces~\cite{ref6,ref11} can play multiscale binary references directly, the receiving response $\widehat{\mathbf{h}}$ need not be identified using the target modulator itself: pre-calibrating the detector chain on a DMD allows its frozen $\widehat{\mathbf{h}}$ to be transferred to modulators that lack crisp high-speed switching, such as liquid-crystal spatial light modulators, or cannot execute custom sequences in situ, such as spinning coding disks~\cite{ref25}, once the modulator's actual spatial patterns and acquisition timing are characterized on the target setup to form matched history weights. Paired with task-direct recovery, this portability extends immediately to single-pixel and dispersive Fourier spectroscopy~\cite{ref6,ref27}, compressive hyperspectral sensing~\cite{ref24}, fluorescence lifetime estimation, serial time-stretch~\cite{ref1,ref26}, computational ghost imaging~\cite{ref4,ref8}, and three-dimensional laser ranging~\cite{ref9,ref19,ref20}, as well as image-free tracking and classification~\cite{ref3,ref30}, recovering terminal parameters or decisions directly from mixed streams without restoring steady-state waveforms.

Viewing hardware inertia as physical encoding also clarifies the structural resilience and future trajectory of computational optics. First, freezing the reference-calibrated kernel $\widehat{\mathbf{h}}$ during target acquisition ensures that anomalous imaging data, such as specular glints or localized clipping, do not enter response updating or bias $\widehat{\mathbf{h}}$, avoiding a failure mode of blind deconvolution and online adaptive equalization. Nevertheless, anomalies or nonlinearities that distort recorded observations, embedded timing references, or front-end recovery tails, alongside sustained detector saturation, thermal drift, and intensity-dependent dynamics~\cite{ref32,ref35}, still perturb timing registration or forward consistency and motivate combining saturation screening with nonlinear history models. Second, analogous to faster-than-Nyquist signaling~\cite{ref47}, treating memory as an encoder defines an information-theoretic agenda to establish rigorous recoverability bounds where severe smoothing compresses target variations toward quantization, noise, and calibration-uncertainty floors. Third, scaling real-time direct recovery beyond $256\times256$ video under rapid intra-frame non-rigid motion motivates uniting continuous motion-and-memory operators with matrix-free transforms and parallelized deep unrolling~\cite{ref39,ref40}. Ultimately, embedding the differentiable operator $\widetilde{\mathbf{A}}=\mathbf{R}\mathbf{W}\mathbf{P}$ into end-to-end optimization enables optical coding, front-end electronics, sampling schedules, and reconstruction priors to be co-designed directly around terminal imaging, video, or task objectives, uniting optoelectronic dynamics and algorithmic inference into a single sensing instrument.

\section{Conclusion}\label{sec:conclusion}

In summary, this work establishes a unified calibration and composite-readout framework that recasts temporal mixing in high-speed single-pixel imaging from a speed-limiting artifact into an active spatiotemporal encoding resource. By integrating data-driven response identification and reference-based timing maintenance into the matched complete-history operator $\widetilde{\mathbf{A}} = \mathbf{R}\mathbf{W}\mathbf{P}$, our single-stage formulation bypasses steady-state waveform restoration, removing per-slot settling constraints while eliminating the conflict between intermediate waveform fidelity and terminal recovery quality. Anchored directly on target recoverability rather than isolated slot settling, jointly coordinating inter- and intra-slot readouts, complementary pattern ordering, temporal filtering, and sampling timing unlocks operating regimes inaccessible to memoryless protocols. Specifically, pre-sampling analog memory enables sub-modulation digitization well below the pattern switching rate and strengthens algebraic distinguishability even in noiseless controls, while bandwidth-limited or strongly low-pass-filtered receivers surpass wider-bandwidth steady-state setups under pronounced low-frequency fluctuations and optical interference. Extending direct recovery from static images to evolving video sequences, causal two-frame historical coupling turns cross-frame persistence into inter-state constraints and pairs with physics-guided neural regularization to achieve synchronization-free, real-time imaging up to $256 \times 256$ pixels at approximately $25~\mathrm{frames/s}$ on a cost-effective $1.8~\mathrm{kHz}$ detector. Validated across diverse operating conditions and independent receiving hardware, this decoupling of detector-chain memory from spatial modulation and readout establishes a portable foundation across optoelectronic platforms, charting a task-direct co-design paradigm that recovers images, videos, and downstream physical parameters directly from mixed observations without steady-state waveform restoration.

\section*{Code availability}
The custom response-calibration, matched-operator assembly, and static and dynamic image reconstruction codes used in this study are available from the corresponding author upon reasonable request.

\section*{Data availability}
Supplementary Data~1 and Supplementary Videos~1--5 are available from Jing-Sen Li (\texttt{luzi23887@163.com}) upon reasonable request.

\section*{Acknowledgements}
This research received no specific grant from any funding agency in the public, commercial, or not-for-profit sectors.

\section*{Author contributions}
J.-S.L. conceived the idea, developed the theoretical framework and reconstruction algorithms, built the optical setup, performed the experiments, and wrote the manuscript. E.-R.Z., K.-X.J., S.-Y.M., Q.-Q.B., and Y.-Z.Z. contributed to experimental validation, data analysis, and manuscript revision. X.-H.C. supervised the project and revised the manuscript. All authors discussed the results and approved the final manuscript.

\section*{Conflict of interest}
The authors declare no conflict of interest.

\section*{Supplementary information}
Supplementary Information accompanies this manuscript.

\bibliography{references}
\end{document}


\thispagestyle{plain}

\begin{center}
    {\Large \textbf{Supplementary Information}}\\[0.6em]
    {\Large \textbf{Real-Time Single-Pixel Imaging via Exploiting Temporal Mixing: A Unified Calibration and Composite Readout Framework}}\\[0.8em]
  {\normalsize Jing-Sen Li, En-Rui Zhang, Kan-Xu Jia, Yi-Zhu Zhang,
Qian-Qian Bao, Shao-Ying Meng, and \mbox{Xi-Hao Chen$^{\ast}$}}\\[0.4em]
    {\small \textit{Key Laboratory of Optoelectronic Devices and Detection Technology, School of Physics, Liaoning University, 66 Chongshan Middle Road, Shenyang, 110036, Liaoning, China}}\\[0.25em]
   {\small $^{\ast}$Corresponding author.
E-mail: xi-haochen@163.com}\\[0.2em]
{\small Tel.: +86~18704088217}
\end{center}

\vspace{0.4em}
\hrule
\vspace{0.4em}
{\small
\setlength{\parskip}{1pt}
\setcounter{tocdepth}{2}
\tableofcontents
}
\vspace{0.4em}
\hrule
\clearpage

\input{section_s1.tex}

\clearpage
\input{section_s2.tex}

\clearpage
\setcounter{equation}{19}
\input{section_s3.tex}

\clearpage
\input{section_s4.tex}

\clearpage
\input{section_s5.tex}

\renewcommand{\refname}{Supplementary References}

%% file: section_s1.tex
\section{Response calibration and construction of the matched observation model}
\label{sec:supp_s1}

\subsection{Reference input design and calibration constraints}

The reference calibration illustrated in Fig.~2(a) of the main text and Fig.~S1(a) and (b) uses spatially uniform all-black and all-white DMD patterns to provide a known binary temporal input. Let $r(u;q)\in\{0,1\}$ denote the binary reference state at time $u$ under the timing correspondence $q$, with $0$ and $1$ representing the all-black and all-white states, respectively. For a static reference target $\mathbf{x}^{\mathrm{opt}}\in\mathbb{R}^{N}$, the instantaneous optical projection in Eq.~(1) of the main text reduces to $s^{\mathrm{opt}}(u)=r(u;q)\mathbf{1}^{\mathsf T}\mathbf{x}^{\mathrm{opt}}$, where $\mathbf{1}\in\mathbb{R}^{N}$ denotes the pixelwise all-one vector. When the total optical projection of the all-white state remains approximately constant during calibration, its unknown magnitude and the readout gain $G$ contribute only a common scale factor to the interval-integrated response coefficients, $\alpha_j=G\left(\mathbf{1}^{\mathsf T}\mathbf{x}^{\mathrm{opt}}\right)h_j$ for $j=1,\ldots,H$. Normalization in Section~S1.2 removes this scale factor, allowing the temporal response to be identified without a pixelwise model of the reference target.

As shown in Fig.~S1(a), the reference sequence begins with long black--white--black holds and continues through repeated rounds containing timing markers, training sequences, Block and Interleaved holdout references, and black tails. The initial holds establish the steady-state black and white plateau levels ($V_B$ and $V_W$, defining the reference plateau span $\Delta V=V_W-V_B$) together with a known prehistory. Round-dependent timing markers locate individual rounds and initialize the correspondence between ADC samples and physical pattern slots. The training sequences combine rapid binary switching, long holds, and multiscale pulses to constrain short-lag transitions, plateau levels, and long-lag response support, respectively. Block and Interleaved holdout sequences with distinct pattern orderings provide independent validation references, while the black tails check the selected response support against the recorded decay. The complete reference history is retained when constructing the calibration design matrix.

\begin{figure}[!htb]
\centering
\includegraphics[width=\linewidth,height=0.57\textheight,keepaspectratio]{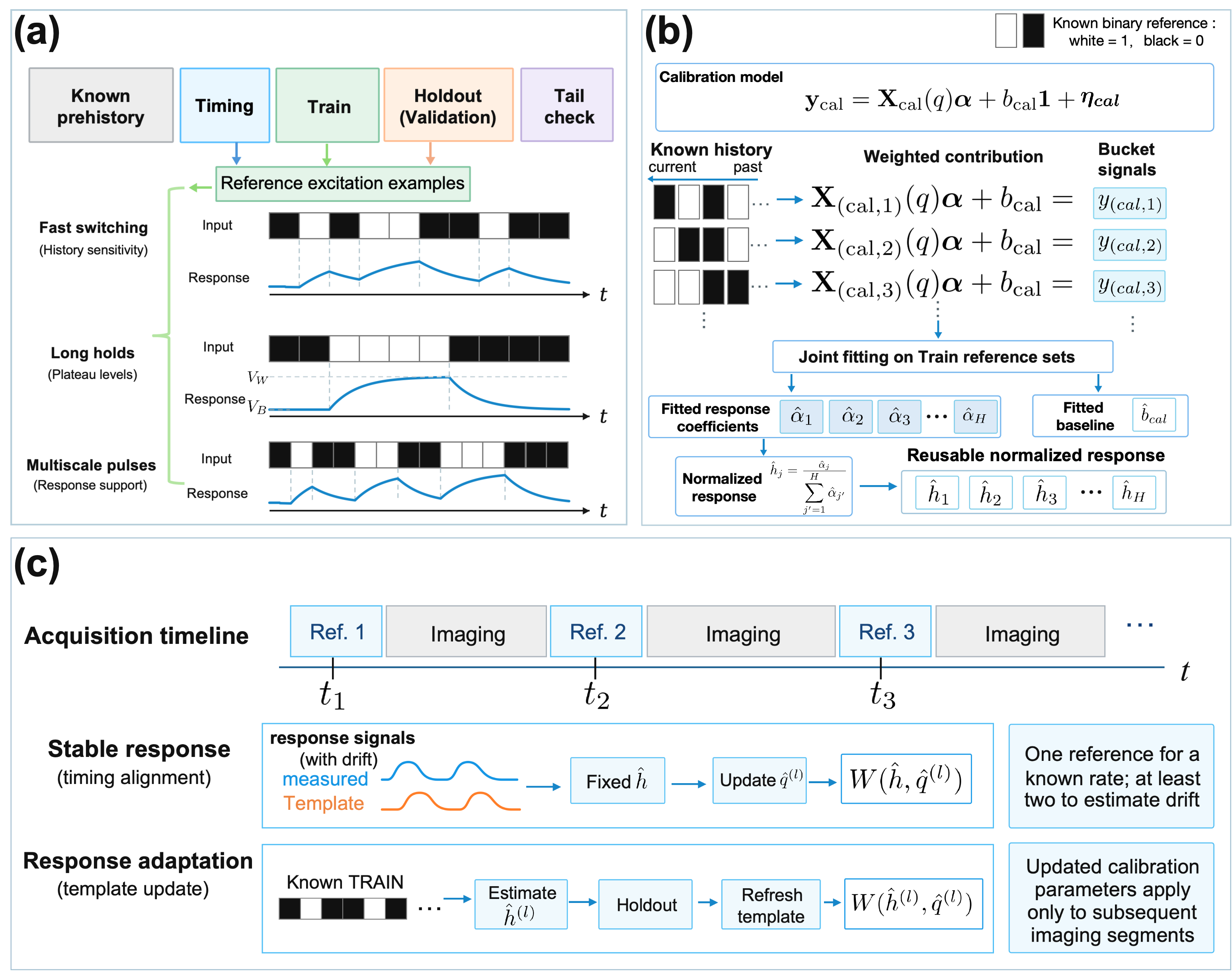}
\caption{Reference calibration and timing alignment.
(a) Organization of the reference block and representative binary excitation inputs, where $V_B$ and $V_W$ denote the steady-state black and white plateau levels, respectively. Reference patterns, response waveforms, and segment widths are schematic.
(b) The known reference history and a candidate timing correspondence $q$ determine $\mathbf{X}_{\mathrm{cal}}(q)$, whose $i$th row is denoted by $\mathbf{X}_{(\mathrm{cal},i)}(q)$ in the diagram. For each candidate timing, training (``Train'') observations jointly fit $\boldsymbol{\alpha}$ and $b_{\mathrm{cal}}$ under the response-centroid constraint. Normalization of the fitted coefficients $\widehat{\boldsymbol{\alpha}}$ yields the reusable response vector $\widehat{\mathbf{h}}$, while $\widehat{\boldsymbol{\alpha}}$, $\widehat{q}$, and $\widehat{b}_{\mathrm{cal}}$ are held fixed for independent holdout (``Holdout'') validation.
(c) Along a continuous acquisition timeline with embedded reference segments (``Ref.~1'' to ``Ref.~3'' at reference times $t_1,t_2,t_3$), timing alignment updates $\widehat{q}$ while holding $\widehat{\mathbf{h}}$ fixed: one reference updates the timing origin (initial slot phase) when the pattern-slot rate is retained, whereas two or more references also estimate relative clock drift. When the receiving-chain response changes, it is re-estimated on training (``TRAIN'') sequences and validated on holdout data before refreshing the reference template and history weights. Updated parameters apply causally to subsequent imaging segments, with superscripts $(l)$ indexing successive updates.}
\label{fig:supp_calibration}
\end{figure}

Partitioning the finite response support $[0,\tau_H)$ into $H$ contiguous lag intervals $[\tau_{j-1},\tau_j)$ ($j=1,\ldots,H$) with $0=\tau_0<\tau_1<\cdots<\tau_H$, the known reference input under a candidate timing correspondence $q$ is mapped onto the ADC time axis. Row $i$ ($i=1,\ldots,L_{\mathrm{cal}}$) of the calibration design matrix $\mathbf{X}_{\mathrm{cal}}(q)\in\mathbb{R}^{L_{\mathrm{cal}}\times H}$ corresponds to the ADC sample at $t_i$, and column $j$ corresponds to the response lag interval $[\tau_{j-1},\tau_j)$. Its $(i,j)$ entry in Eq.~(3) of the main text is
\begin{equation}
\left[\mathbf{X}_{\mathrm{cal}}(q)\right]_{ij}
=
\frac{1}{\tau_j-\tau_{j-1}}
\int_{t_i-\tau_j}^{t_i-\tau_{j-1}}
r(u;q)\,\mathrm{d}u.
\end{equation}

Each entry represents the fraction of the $j$th historical lag interval occupied by the all-white state: $0$ for an entirely black interval, $1$ for an entirely white interval, and an intermediate value when a state transition occurs within it. Distinct training histories and within-slot sampling phases therefore generate distinct rows of $\mathbf{X}_{\mathrm{cal}}(q)$ that jointly constrain the shared response coefficients $\boldsymbol{\alpha}$ and baseline offset $b_{\mathrm{cal}}$ in Eq.~(3) of the main text. This construction underlies the phase-coverage calibration comparison in Section~3.1 of the main text and Section~S3.2. Joint identification of the response and timing correspondence is detailed in Section~S1.2.

\subsection{Joint identification of response and timing}

The reference-measurement model in Eq.~(3) of the main text depends on both the acquisition-chain temporal response and the correspondence between ADC samples and physical pattern slots. Building on the reference-history construction in Section~S1.1, the procedure in Fig.~S1(b) jointly identifies the temporal response and timing correspondence under a common temporal reference, and then converts the calibrated response into history weights for imaging.

\paragraph{Timing coordinates and initial localization.}

With a reference time $t_0$ on the ADC clock, the continuous timing correspondence is parameterized as
\begin{equation}
q(t)=q_0+f_s(t-t_0),
\qquad
T_s=\frac{1}{f_s}.
\end{equation}
Here, $q(t)$ denotes the continuous pattern-slot coordinate, $q_0$ specifies the timing origin at $t_0$ (representing the initial slot phase), and $f_s$ and $T_s$ denote the effective pattern-slot rate and slot duration on the ADC time axis, respectively. Together, $(q_0,f_s)$ capture initial alignment and relative ADC--DMD clock drift over the acquisition record without requiring hardware clock synchronization, where $f_s$ absorbs the accumulated clock-rate offset relative to the nominal DMD modulation rate. Under $q(t)$, the $k$th physical pattern slot spans $[T_k,T_{k+1})$ with registered boundary $T_k=t_0+\frac{k-q_0}{f_s}$.

Initial estimates of $q_0$ and $f_s$ are obtained from the timing markers in Fig.~S1(a). Correlating templates synthesized from the known binary marker sequences and an approximate nominal response against the recorded waveform locates each marker on the ADC clock, pairing its matched ADC timestamp with its known physical pattern-slot coordinate. Fitting these coordinates against $t-t_0$ determines initial values of $q_0$ and $f_s$ as the intercept and slope in Eq.~(S2). The approximate response serves solely for coarse marker localization, while the final temporal response and refined timing are jointly identified from the training observations.

\paragraph{Response-weight estimation and centroid alignment.}

For each candidate timing correspondence $q$, Eq.~(S1) determines the calibration design matrix $\mathbf{X}_{\mathrm{cal}}(q)$, whose $L_{\mathrm{tr}}$ training rows form $\mathbf{X}_{\mathrm{tr}}(q)\in\mathbb{R}^{L_{\mathrm{tr}}\times H}$ paired with the recorded training vector $\mathbf{y}_{\mathrm{tr}}\in\mathbb{R}^{L_{\mathrm{tr}}}$. Following Eq.~(3) and Section~2.1 of the main text, the $H$ interval-integrated response coefficients $\boldsymbol{\alpha}$ and calibration baseline $b_{\mathrm{cal}}$ are fitted jointly by minimizing $\|\mathbf{y}_{\mathrm{tr}}-\mathbf{X}_{\mathrm{tr}}(q)\boldsymbol{\alpha}-b_{\mathrm{cal}}\mathbf{1}_{L_{\mathrm{tr}}}\|_2^2+\lambda_{\mathrm{cal}}\mathcal{R}(\boldsymbol{\alpha})$. Solving jointly for $\boldsymbol{\alpha}$ and $b_{\mathrm{cal}}$ across all $L_{\mathrm{tr}}$ training samples constrains a shared response and baseline across reference segments rather than separate parameters for individual segments. Under the piecewise-constant representation, the regularization functional is $\mathcal{R}(\boldsymbol{\alpha})=\sum_{j=1}^{H}\frac{\alpha_j^2}{\tau_j-\tau_{j-1}}$, which equals the integral of the squared piecewise-constant scaled response over $[0,\tau_H)$ and regularizes its amplitude when reference constraints are limited; its weight $\lambda_{\mathrm{cal}}\ge 0$ is selected via the training cross-validation protocol in Section~S3.1.

The timing origin $q_0$ and the position of the fitted response along the lag axis share a temporal-coordinate freedom. A displacement of the registered pattern boundaries can be accompanied by a corresponding translation of the response along the lag axis, producing nearly the same predicted waveform whenever the response remains within $[0,\tau_H)$. The measured waveform therefore constrains their relative alignment rather than independently specifying an absolute timing origin and response delay. To establish a common reference, let $\bar{\tau}_j=\frac{\tau_{j-1}+\tau_j}{2}$ denote the physical-time midpoint of lag interval $j$. The centroid of the interval-integrated response weights is $\frac{\sum_{j=1}^{H}\bar{\tau}_j\alpha_j}{\sum_{j=1}^{H}\alpha_j}$. Fixing this centroid at a preselected physical lag time $\tau_c\in(0,\tau_H)$ gives
\begin{equation}
\sum_{j=1}^{H}
\left(\bar{\tau}_j-\tau_c\right)\alpha_j=0.
\end{equation}
During the joint fit of $\boldsymbol{\alpha}$ and $b_{\mathrm{cal}}$ across candidate timing correspondences $q$, Eq.~(S3) acts as a homogeneous linear constraint on $\boldsymbol{\alpha}$ with the physical centroid $\tau_c$ held constant across candidates rather than rescaled with the candidate slot rate $f_s$. Response and timing parameters are thereby compared under a single lag-axis reference. Controlled coordinate translations that shift the timing origin together with $\tau_c$, as well as imposed timing mismatches with $\widehat{\mathbf{h}}$ fixed, are examined experimentally in Section~S3.3, Fig.~4(g) of the main text, and Fig.~S2; $\tau_c$ specifies the shared coordinate reference rather than an independently measured DMD switching delay.

\paragraph{Timing refinement and normalized response.}

Starting from the marker-based estimates, candidate values of $(q_0,f_s)$ are refined using training predictions. For each candidate $q$, the reference-slot boundaries and $\mathbf{X}_{\mathrm{tr}}(q)$ are reconstructed, and $\boldsymbol{\alpha}$ and $b_{\mathrm{cal}}$ are refitted subject to Eq.~(S3). Predictions are evaluated against the same training observations $\mathbf{y}_{\mathrm{tr}}$, and the candidate minimizing the squared prediction residual $\|\mathbf{y}_{\mathrm{tr}}-\mathbf{X}_{\mathrm{tr}}(q)\boldsymbol{\alpha}-b_{\mathrm{cal}}\mathbf{1}_{L_{\mathrm{tr}}}\|_2^2$ is selected. Each timing candidate is therefore assessed using the response refitted at that timing rather than one fitted at a different timing. The resulting $(\widehat{q}_0,\widehat{f}_s)$ with $\widehat{T}_s=1/\widehat{f}_s$ define the calibrated timing correspondence $\widehat{q}$, together with the fitted coefficients $\widehat{\boldsymbol{\alpha}}$ and baseline $\widehat{b}_{\mathrm{cal}}$.

The fitted response coefficients are normalized into interval response weights as
\begin{equation}
\widehat{h}_j
=
\frac{\widehat{\alpha}_j}
{\displaystyle\sum_{j'=1}^{H}\widehat{\alpha}_{j'}},
\qquad j=1,\ldots,H,
\end{equation}
which, together with Eq.~(S3), satisfies $\sum_{j=1}^{H}\widehat{h}_j=1$ and $\sum_{j=1}^{H}\bar{\tau}_j\widehat{h}_j=\tau_c$. The normalized vector $\widehat{\mathbf{h}}=(\widehat{h}_1,\ldots,\widehat{h}_H)^{\mathsf T}$ defines the piecewise-constant continuous-time response $\widehat{h}(\tau)=\widehat{h}_j/(\tau_j-\tau_{j-1})$ for $\tau\in[\tau_{j-1},\tau_j)$ ($j=1,\ldots,H$) and $\widehat{h}(\tau)=0$ outside $[0,\tau_H)$, satisfying $\int_0^{\infty}\widehat{h}(\tau)\,\mathrm{d}\tau=1$ independently of the reference-specific optical amplitude and readout gain. For independent Block and Interleaved holdout validation, $\widehat{q}$, $\widehat{\boldsymbol{\alpha}}$, and $\widehat{b}_{\mathrm{cal}}$ remain fixed after training selection, as illustrated in Fig.~S1(b) and evaluated in Section~3.1 of the main text. Subsequent acquisitions reuse $\widehat{\mathbf{h}}$ with their own timing correspondence according to Section~S1.3.

\paragraph{From the calibrated response to history weights.}

Under $\widehat{q}$, the registered boundaries of the $K$ physical pattern slots ($k=1,\ldots,K$) are $T_k=t_0+\frac{k-\widehat{q}_0}{\widehat{f}_s}$. Evaluating $\widehat{q}(t_i)$ at each ADC timestamp $t_i$ simultaneously locates its containing physical slot $k(i)=\lfloor\widehat{q}(t_i)\rfloor$ and its within-slot sampling phase $\varphi_i=\bigl(\widehat{q}(t_i)-k(i)\bigr)\widehat{T}_s=t_i-T_{k(i)}\in[0,\widehat{T}_s)$, corresponding to the within-slot phases (such as $\varphi_1$ and $\varphi_2$) in Figs.~2(b) and 3(a,c) of the main text. Substituting the piecewise-constant response $\widehat{h}(\tau)$ into the slot integral $W_{ik}=\int_{T_k}^{T_{k+1}}\widehat{h}(t_i-u)\,\mathrm{d}u$ in Section~2.1 of the main text yields
\begin{equation}
W_{ik}
=
\sum_{j=1}^{H}
\frac{\widehat{h}_j}{\tau_j-\tau_{j-1}}
\left|
[T_k,T_{k+1})
\cap
[t_i-\tau_j,t_i-\tau_{j-1})
\right|.
\end{equation}
Here, $|\cdot|$ denotes the duration of the interval intersection. The first interval $[T_k,T_{k+1})$ specifies when physical pattern $k$ is displayed; the second interval $[t_i-\tau_j,t_i-\tau_{j-1})$ selects the portion of input history associated with lag interval $j$ at ADC sample $i$. Multiplying their overlap duration by the calibrated response level $\widehat{h}_j/(\tau_j-\tau_{j-1})$ and summing over $j=1,\ldots,H$ yields $W_{ik}$, the total contribution weight of pattern $k$ to sample $i$. A single calibrated vector $\widehat{\mathbf{h}}$ thus generates the history weights for all ADC samples, completing the effective-pattern construction $\mathbf{A}=\mathbf{W}\mathbf{P}$ in Fig.~2(a) and (b) of the main text.

Changing ADC sampling times, within-slot phases $\varphi_i$, or physical pattern timing alters these interval overlaps, so the weights are reconstructed from the frozen $\widehat{\mathbf{h}}$ and the registered timing of each acquisition, as described in Section~S1.3. A physical pattern need not be directly sampled during its own slot $[T_k,T_{k+1})$ to contribute to a later ADC observation at $t_i>T_{k+1}$: its nonzero history weight $W_{ik}$ carries its spatial encoding into subsequent measurements while the calibrated response persists. This links the phase-dependent effective patterns in Fig.~3(a) of the main text and Section~S1.4 to the limited-ADC historical encoding developed in Section~S2.5 and examined in Section~S4.5.

\subsection{Calibration reuse and imaging-observation selection}

\paragraph{Response reuse and reference-based timing updates.}

The normalized response $\widehat{\mathbf{h}}$ identified in Section~S1.2 can be reused across acquisitions with a stable effective receiving response. For each acquisition, the timing correspondence $q$ is updated from its embedded references, and the history weights $W_{ik}$ are reconstructed using Eq.~(S5) and the complete physical pattern sequence. Figure~S1(c) illustrates this procedure, in which $\widehat{\mathbf{h}}$ remains fixed while the acquisition-specific timing is updated.

Let $r_k\in\{0,1\}$ denote the known black or white reference state in physical pattern slot $k$. For a candidate timing correspondence $q$, the reference-prediction template at sample $i$ is
\begin{equation}
g_i(q)
=
\sum_k W_{ik}(\widehat{\mathbf{h}},q)\,r_k,
\end{equation}
where $W_{ik}(\widehat{\mathbf{h}},q)$ follows Eq.~(S5) and retains the complete known reference-input history within $[0,\tau_H)$. Stacking $g_i(q)$ and the corresponding measured reference samples $y_{\mathrm{ref},i}$ over the valid reference sample set forms the template vector $\mathbf{g}(q)$ and reference-measurement vector $\mathbf{y}_{\mathrm{ref}}$. With the pattern-slot rate fixed, the timing origin (initial slot phase) is determined by maximizing the normalized correlation coefficient (NCC) between the recorded reference waveform and the prediction:
\begin{equation}
\widehat{q}_0
=
\underset{q_0}{\operatorname{arg\,max}}\,
\operatorname{NCC}
\left(
\mathbf{y}_{\mathrm{ref}},
\mathbf{g}(q)
\right),
\end{equation}
where NCC is defined as
\begin{equation}
\operatorname{NCC}
\left(
\mathbf{y}_{\mathrm{ref}},
\mathbf{g}(q)
\right)
=
\frac{
\displaystyle
\sum_i
\left(y_{\mathrm{ref},i}-\bar{y}_{\mathrm{ref}}\right)
\left(g_i(q)-\overline{g(q)}\right)
}{
\displaystyle
\sqrt{
\sum_i
\left(y_{\mathrm{ref},i}-\bar{y}_{\mathrm{ref}}\right)^2
}
\sqrt{
\sum_i
\left(g_i(q)-\overline{g(q)}\right)^2
}
},
\end{equation}
with $\bar{y}_{\mathrm{ref}}$ and $\overline{g(q)}$ denoting the arithmetic means of $\mathbf{y}_{\mathrm{ref}}$ and $\mathbf{g}(q)$, respectively, evaluated over the identical set of valid reference samples. Mean subtraction and $\ell_2$ normalization render the timing match insensitive to a constant baseline offset and a common amplitude scale.

The two timing-maintenance modes in Fig.~4(f) of the main text differ in the parameters updated while keeping $\widehat{\mathbf{h}}$ fixed. Reuse drift retains the calibrated pattern-slot rate $\widehat{f}_s$ and updates only the timing origin $\widehat{q}_0$ (initial phase) from one reference via Eqs.~(S6) and (S7). Estimate drift uses two or more temporally separated references with the frozen-$\widehat{\mathbf{h}}$ template in Eq.~(S6) to update both the timing origin $\widehat{q}_0$ and the pattern-slot rate $\widehat{f}_s$ in the linear timing model of Eq.~(S2), without refitting $\widehat{\mathbf{h}}$. Both modes then reconstruct the history weights using the updated timing, and their independent prediction performance is examined in Section~S3.3 and Fig.~4(f) of the main text.

\textit{Conditions for response reuse and recalibration.} Changes in pattern ordering, pattern hold duration, DMD modulation rate, and ADC sampling timing are accommodated by updating the timing correspondence $\widehat{q}$ and recomputing $W_{ik}$ from the retained response $\widehat{\mathbf{h}}$ and physical pattern history. Changes in illumination amplitude are accommodated through the overall gain and baseline fitted on receiving training references in the waveform-transfer evaluations of Section~S3.5 (or absorbed directly into $\mathbf{x}=G\mathbf{x}^{\mathrm{opt}}$ and zero-sum differential readouts during imaging). When the effective receiving response itself changes because of detector hardware, amplification, ADC sampling aperture, or incorporated signal processing, the corresponding response $\widehat{\mathbf{h}}$ is recalibrated and independently validated. Section~S3.4 demonstrates this adaptation under digital filtering and an independent receiving chain.

\paragraph{Reference protection and imaging-observation selection.}

Temporal memory allows reference patterns to contribute to ADC observations after the reference sequence has ended. Guard patterns (also referred to as buffer intervals in the acquisition records of Fig.~S9) separate the reference and imaging segments, and observation selection jointly accounts for the calibrated response support $\tau_H$ and the actual contributing pattern history.

\textit{Black-field guard.} With an all-black guard, the reference and guard observations are excluded, together with subsequent initial imaging samples whose lag support $[t_i-\tau_H,t_i]$ still overlaps reference or guard patterns. The all-black guard has a $0\%$ open fraction, whereas the binary Hadamard imaging patterns have a $50\%$ open fraction. Their transition alters the spatially uniform component and produces a step transient that would otherwise leak into differential observations; the prescribed exclusion removes these transition-affected samples before composite readout.

\textit{Half-black/half-white guard.} A half-black/half-white guard shares the $50\%$ open fraction of the imaging patterns, preserving their common uniform spatial component across the guard-to-imaging transition. Once the normalized response history no longer overlaps the preceding black/white reference, this uniform component remains constant across observations and cancels under zero-sum differential readouts, while the target-dependent spatial components of the guard patterns remain exactly represented in $\mathbf{W}\mathbf{P}$. Under this strategy, guard observations are excluded from the imaging readouts, whereas subsequent imaging samples are retained as soon as $[t_i-\tau_H,t_i]$ clears the preceding black/white reference, even when half-black/half-white guard patterns remain within that history.

\textit{Valid sampling and matched-operator construction.}

Using the within-slot phase $\varphi_i$ determined by $\widehat{q}$, imaging observations are selected according to the fixed valid phase windows established in Section~3.1 of the main text and evaluated in Sections~S3.1 and S3.2, combined with the history-clearance rule of the selected guard strategy. Retaining the $L$ valid ADC samples defines the observation vector $\mathbf{y}\in\mathbb{R}^{L}$ and selects the corresponding $L$ rows of $\mathbf{W}\in\mathbb{R}^{L\times K}$, while leaving all $K$ physical pattern slots in $\mathbf{P}\in\mathbb{R}^{K\times N}$ intact because patterns displayed near excluded samples still contribute to later retained observations through $W_{ik}$.

The composite readout operator $\mathbf{R}\in\mathbb{R}^{Q\times L}$ then applies identical linear combinations to the selected observations and their matched effective patterns $\mathbf{A}=\mathbf{W}\mathbf{P}\in\mathbb{R}^{L\times N}$, yielding $\mathbf{z}=\mathbf{R}\mathbf{y}\in\mathbb{R}^{Q}$ and $\widetilde{\mathbf{A}}=\mathbf{R}\mathbf{W}\mathbf{P}\in\mathbb{R}^{Q\times N}$ as defined in Eq.~(4) of the main text. A complementary inter-slot difference row is retained whenever both branches contain valid samples, while an intra-slot slope row uses the valid samples within its pattern slot. Detailed processing configurations are recorded in Supplementary Data~1.

\subsection{Construction of differential effective patterns and composite readouts}

Building on the calibrated history weights and $L$ valid observations from Sections~S1.2 and S1.3, this section specifies the inter-slot and intra-slot rows of $\mathbf{R}$ in Eq.~(4) of the main text. Stacking the inter-slot sub-operator $\mathbf{R}_{\mathrm{out}}$ and intra-slot sub-operator $\mathbf{R}_{\mathrm{in}}$ as $\mathbf{R}=[\mathbf{R}_{\mathrm{out}}^{\mathsf T},\mathbf{R}_{\mathrm{in}}^{\mathsf T}]^{\mathsf T}$ directly constructs the matched branch pairs $(\mathbf{z}_{\mathrm{out}}=\mathbf{R}_{\mathrm{out}}\mathbf{y},\,\widetilde{\mathbf{A}}_{\mathrm{out}}=\mathbf{R}_{\mathrm{out}}\mathbf{W}\mathbf{P})$ and $(\mathbf{z}_{\mathrm{in}}=\mathbf{R}_{\mathrm{in}}\mathbf{y},\,\widetilde{\mathbf{A}}_{\mathrm{in}}=\mathbf{R}_{\mathrm{in}}\mathbf{W}\mathbf{P})$ in Eq.~(7) of the main text.

\paragraph{Historical contributions to intra-slot differential patterns.}

Consider two retained ADC samples $i$ and $i'$ acquired at within-slot phases $\varphi_1$ and $\varphi_2$ of the same physical pattern slot, as in Fig.~3(a) of the main text. Their differential effective pattern $\Delta\mathbf{a}=\mathbf{a}_i-\mathbf{a}_{i'}$, introduced in Eq.~(6) of the main text, can be expressed through summation by parts over the complete history weights:
\begin{equation}
\begin{aligned}
\Delta\mathbf{a}
&=
\sum_k
\left(W_{ik}-W_{i'k}\right)\mathbf{p}_k
\\[4pt]
&=
\sum_k
\left[
\int_{t_{i'}-T_k}^{t_i-T_k}
\widehat{h}(\tau)\,\mathrm{d}\tau
\right]
\left(\mathbf{p}_k-\mathbf{p}_{k-1}\right).
\end{aligned}
\end{equation}
Here, $T_k$ is the registered pattern-slot boundary and $\widehat{h}(\tau)$ is the piecewise-constant calibrated response from Section~S1.2. Boundary terms vanish identically because causality sets $\widehat{h}(\tau)=0$ for future slots ($T_k>\max(t_i,t_{i'})$) and finite response support sets $\widehat{h}(\tau)=0$ beyond $[0,\tau_H)$ for distant past slots ($T_k<\min(t_i,t_{i'})-\tau_H$), with known prehistory retained across the active lag window. The first line of Eq.~(S9) describes the phase-dependent reweighting of historical patterns, whereas the second line reveals its spatial origin: each successive pattern transition $\mathbf{p}_k-\mathbf{p}_{k-1}$ contributes according to $\widehat{h}(\tau)$ integrated between the two sampling instants.

Even when the displayed DMD pattern remains unchanged within a slot, the transient response to preceding pattern transitions continues to evolve. Consequently, distinct intra-slot phases carry distinct target-dependent effective patterns, providing algebraic spatial constraints in addition to rejecting constant offsets. This historical-encoding mechanism, illustrated in Fig.~3(a) of the main text, is developed further through the complementary constraints in Section~S2.2 and the modulation-rate experiments in Section~S4.2.

\paragraph{Readout coefficients for an inter-slot difference.}

For the $m$th complementary pattern pair $(k_+,k_-)$ satisfying $\mathbf{p}_{k_-}=\mathbf{1}-\mathbf{p}_{k_+}$, let $\mathcal{I}_{m,+}$ and $\mathcal{I}_{m,-}$ denote the sets of valid retained ADC samples belonging to slots $k_+$ and $k_-$, respectively, with $|\mathcal{I}_{m,+}|\ge 1$ and $|\mathcal{I}_{m,-}|\ge 1$. The inter-slot readout in Section~3.2 of the main text averages the valid samples within each branch separately and subtracts the negative-branch mean from the positive-branch mean: in row $m$ of $\mathbf{R}_{\mathrm{out}}$, $R_{mi}=1/|\mathcal{I}_{m,+}|$ for $i\in\mathcal{I}_{m,+}$, $R_{mi}=-1/|\mathcal{I}_{m,-}|$ for $i\in\mathcal{I}_{m,-}$, and $R_{mi}=0$ otherwise, satisfying $\sum_{i=1}^{L}R_{mi}=0$.

Complementary patterns need not occupy adjacent slots: in the grouped C44 arrangement, other spatial patterns occur between $k_+$ and $k_-$. Acting identically on $\mathbf{y}$ and $\mathbf{A}=\mathbf{W}\mathbf{P}$, $\mathbf{R}_{\mathrm{out}}$ produces the matched pair $(\mathbf{z}_{\mathrm{out}},\widetilde{\mathbf{A}}_{\mathrm{out}})$ in Eq.~(7) of the main text. The influence of pattern ordering on these effective constraints is analyzed in Section~S2.1 and examined experimentally in Section~S4.3.

\paragraph{Readout coefficients for an intra-slot slope.}

The intra-slot readout extends the two-sample difference in Eq.~(6) of the main text to the set $\mathcal{I}_m$ of valid samples within an eligible physical pattern slot. Let $\bar{t}=\frac{1}{|\mathcal{I}_m|}\sum_{i'\in\mathcal{I}_m}t_{i'}$ denote the mean timestamp of those samples. A least-squares linear fit with an intercept gives the temporal-slope coefficients in row $m$ of $\mathbf{R}_{\mathrm{in}}$:
\begin{equation}
R_{mi}
=
\frac{t_i-\bar{t}}
{\displaystyle\sum_{i'\in\mathcal{I}_m}(t_{i'}-\bar{t})^2},
\qquad i\in\mathcal{I}_m,
\end{equation}
with $R_{mi}=0$ for $i\notin\mathcal{I}_m$. With two valid samples at distinct timestamps, Eq.~(S10) reduces to their signal difference divided by their time separation. For general $|\mathcal{I}_m|\ge 2$, the coefficients satisfy $\sum_{i=1}^{L}R_{mi}=0$ and $\sum_{i\in\mathcal{I}_m}R_{mi}(t_i-\bar{t})=1$, yielding exact zero response to a constant offset and unit response to a linear temporal slope.

The intra-slot slope therefore extracts the local rate of target-dependent historical evolution while suppressing slot-constant baselines; its disturbance transmission is examined in Section~S2.2. Because Eq.~(S10) requires $|\mathcal{I}_m|\ge 2$ at distinct timestamps, the sparse-sampling experiments with at most one sample per slot in Section~3.4 of the main text and Section~S4.5 use Inter ($\beta=0$). Applying $\mathbf{R}_{\mathrm{in}}$ identically to $\mathbf{y}$ and $\mathbf{A}=\mathbf{W}\mathbf{P}$ constructs $(\mathbf{z}_{\mathrm{in}},\widetilde{\mathbf{A}}_{\mathrm{in}})$ in Eq.~(7) of the main text. Since inter-slot mean differences (signal amplitude) and intra-slot slopes (signal amplitude per unit time) have different physical units and scales, $\beta\ge 0$ in Eq.~(7) of the main text controls their relative participation in Joint reconstruction, while $\rho_{\mathrm{in}}$ in Eq.~(8) of the main text provides the scale-normalized intra-slot weight.

\paragraph{Consistent filtering and matched readout construction.}

When causal digital filtering is applied to recorded ADC data, the reference records undergo the identical filtering prior to calibration so that the identified composite response $\widehat{\mathbf{h}}$ characterizes the combined acquisition-and-filtering chain. History weights, valid-observation selection, and composite readouts are then constructed consistently on the filtered imaging record, applying the same sample selection and linear combinations $\mathbf{R}$ to both measurements and effective patterns to preserve $\mathbf{z}=\mathbf{R}\mathbf{y}$ and $\widetilde{\mathbf{A}}=\mathbf{R}\mathbf{W}\mathbf{P}$ in Eq.~(4) of the main text. This procedure is validated in Section~S3.4, and the coupled effects of temporal filtering on target sensitivity and disturbance transmission are analyzed in Section~S2.4. Detailed processing and reconstruction settings are provided in Supplementary Data~1.

\subsection{Extension to dynamic targets}

\paragraph{Temporal response for dynamic optical projections.}

The continuous-time measurement relation in Eq.~(2) of the main text describes the causal linear response of the receiving chain to the instantaneous optical projection $s^{\mathrm{opt}}(u)$ and therefore remains valid when the scene evolves over time. Absorbing the overall readout gain $G$ into the time-varying target as $\mathbf{x}(u)=G\mathbf{x}^{\mathrm{opt}}(u)\in\mathbb{R}^{N}$, Eq.~(2) of the main text becomes $y_i=\int_{-\infty}^{t_i}h(t_i-u)\mathbf{p}(u)^{\mathsf T}\mathbf{x}(u)\,\mathrm{d}u+b_i+\nu_i$. Because $h(\tau)$ characterizes the electro-optical receiving chain rather than the static calibration target, the normalized response $\widehat{h}(\tau)$ calibrated on static binary references applies directly to dynamic optical projections whenever the linear time-invariant receiving behavior holds.

\paragraph{Complete-history measurements across scene frames.}

For frame-based dynamic reconstruction, consecutive physical pattern slots are partitioned into disjoint frame index sets $\{\mathcal{K}_n\}$, with the scene approximated as constant ($\mathbf{x}(u)\approx\mathbf{x}_n\in\mathbb{R}^{N}$) over the slots $k\in\mathcal{K}_n$ of frame $n$. Substituting $\widehat{h}(\tau)$ and the calibrated history weights $W_{ik}$ from Eq.~(S5) yields the discrete-frame dynamic observation model:
\begin{equation}
y_i
=
\sum_n
\sum_{k\in\mathcal{K}_n}
W_{ik}\mathbf{p}_k^{\mathsf T}\mathbf{x}_n
+b_i+\nu_i,
\end{equation}
where $\nu_i$ aggregates additive measurement noise together with minor calibration and within-frame motion residuals. Each historical pattern $k\in\mathcal{K}_n$ contributes according to $W_{ik}$ paired with the scene state $\mathbf{x}_n$ present during its display interval $[T_k,T_{k+1})$. Unlike static imaging, where all historical projections probe a single target $\mathbf{x}$, a single dynamic ADC sample $y_i$ near a frame boundary simultaneously encodes multiple scene states whenever $[t_i-\tau_H,t_i]$ spans more than one frame.

In particular, a pattern $k\in\mathcal{K}_{n-1}$ displayed near the end of frame $n-1$ contributes $W_{ik}\mathbf{p}_k^{\mathsf T}\mathbf{x}_{n-1}$ (rather than $W_{ik}\mathbf{p}_k^{\mathsf T}\mathbf{x}_n$) to early samples of frame $n$. Retaining the complete pattern history together with its true frame-state assignment avoids misattributing persistent cross-frame tails, with model accuracy governed by the scene motion within each frame interval.

\paragraph{Composite readout and cross-frame constraints.}

The valid-observation selection in Section~S1.3 and the inter-slot and intra-slot readout operators in Section~S1.4 apply directly to Eq.~(S11). Partitioning the columns of $\mathbf{W}$ and rows of $\mathbf{P}$ by $\mathcal{K}_n$, applying $\mathbf{R}$ via $\mathbf{z}=\mathbf{R}\mathbf{y}$ combines historical contributions into frame-partitioned composite operators $\widetilde{\mathbf{A}}_n=\mathbf{R}\mathbf{W}_{\mathcal{K}_n}\mathbf{P}_{\mathcal{K}_n}$ while retaining the full binary patterns (including their uniform spatial components), thereby coupling consecutive scene states whenever the response memory crosses frame boundaries and extending Eq.~(4) of the main text to dynamic sequences.

\paragraph{Two-frame implementation and dynamic reconstruction.}

When the response support $\tau_H$ is shorter than a single frame duration, cross-frame persistence at frame $n$ involves only the immediately preceding frame $n-1$ and the current frame $n$. The real-time dynamic implementation in Section~S5.2 accordingly instantiates Eq.~(S11) as a two-frame joint historical operator on $(\mathbf{x}_{n-1},\mathbf{x}_n)$ coupled with physics-guided hybrid neural recovery, supporting the video-rate experiments in Section~3.4 of the main text. Together, Sections~S1.2--S1.5 establish a unified forward foundation spanning static and dynamic regimes, with detailed dynamic solver and implementation settings provided in Section~S5 and Supplementary Data~1.

%% file: section_s2.tex
\section{Spatial information in temporal mixing and its joint exploitation}
\label{sec:supp_s2}

Conventional single-pixel imaging treats the electro-optical receiving chain as a scalar bottleneck whose impulse tail must vanish within each modulation slot so that spatial encoding resides solely in the static physical pattern matrix $\mathbf{P}$~\cite{ref1,ref2,ref3}. Under the calibrated complete-history formulation $\widetilde{\mathbf{A}}=\mathbf{R}\mathbf{W}\mathbf{P}$ established in Section~S1, spatial encoding is no longer fixed by $\mathbf{P}$ alone: instead, physical modulation $\mathbf{P}$, hardware temporal memory $\mathbf{W}$, and digital composite readout $\mathbf{R}$ jointly determine the effective measurement subspace and its disturbance statistics. Focusing on static scenes, this section establishes the theoretical basis for the co-design mechanisms in Section~2.2 of the main text. We first show how temporal ordering of a fixed pattern set reshapes differential spatial contrast alongside physical switching dynamics, and how inter-slot and intra-slot readouts provide complementary spatial sensitivities and disturbance responses for a common scene. Building on these properties, we examine why restoring disturbance-contaminated pre-response waveforms or unconstrained intermediate projections can conflict with final image quality, whereas single-stage direct recovery exploits mixed observations around a unified target-recovery objective without requiring unconstrained intermediate projection estimation. We then analyze the balance between disturbance suppression and target sensitivity under temporal filtering, and show how pre-ADC analog memory reshapes finite-sample spatial constraints under sub-modulation digitization so that target variations invisible to memoryless sampling can become identifiable even in noiseless regimes. Throughout, nonzero historical weights are evaluated by the effective target constraints they provide relative to disturbances, rather than assumed beneficial a priori.

\subsection{Complementary pattern ordering and retention of differential encoding}

In memoryless steady-state imaging of a fixed pattern set, permuting the display order merely permutes the rows of the forward system without altering the sensed spatial subspace, and leaves directional signal-to-noise ratios unchanged under independent, homoscedastic measurement noise. Under temporal mixing, or in the presence of temporally correlated drift, this permutation invariance breaks down. Consider a fixed set of $K$ complementary binary patterns with unchanged slot duration $T_s$ and spatial coding directions, varying only their temporal ordering. Under the pattern-independent linear response calibrated in Section~S1, the history-weight matrix $\mathbf{W}\in\mathbb{R}^{L\times K}$ is governed solely by the receiving response $\widehat{\mathbf{h}}$ and registered timing $\widehat{q}$. With hardware and sampling fixed, reordering the pattern sequence leaves $\mathbf{W}$ invariant while permuting the rows of $\mathbf{P}$, thereby assigning different combinations of historical lag weights to each spatial coding direction in $\mathbf{A}=\mathbf{W}\mathbf{P}$ and re-indexing the paired readout rows in $\mathbf{R}_{\mathrm{out}}$ and transition rows in $\mathbf{R}_{\mathrm{in}}$ accordingly. Temporal ordering thus serves as a hardware-free degree of freedom for shaping the directional sensitivities of $\widetilde{\mathbf{A}}=\mathbf{R}\mathbf{W}\mathbf{P}$.

Let $k_+$ and $k_-$ denote the physical slots occupied by a complementary binary pair~\cite{ref4,ref5} satisfying $\mathbf{p}_{k_+}\in\{0,1\}^{N}$ and $\mathbf{p}_{k_-}=\mathbf{1}-\mathbf{p}_{k_+}$, where $\mathbf{1}\in\mathbb{R}^{N}$ is the pixelwise all-one vector defined in Section~S1.1. Using the history weights $W_{ik}$ from Eq.~(S5), the combined contribution of this pair to the $i$th effective pattern $\mathbf{a}_i$ decomposes as
\begin{equation}
\begin{aligned}
&W_{ik_+}\mathbf{p}_{k_+}
+W_{ik_-}\mathbf{p}_{k_-}=
\frac{W_{ik_+}+W_{ik_-}}{2}\mathbf{1}
+
\frac{W_{ik_+}-W_{ik_-}}{2}
\left(\mathbf{p}_{k_+}-\mathbf{p}_{k_-}\right).
\end{aligned}
\end{equation}

Equation~(S12) separates the pair's historical contribution into a spatially uniform component along $\mathbf{1}$ and a zero-mean differential component along the signed coding direction $\mathbf{p}_{k_+}-\mathbf{p}_{k_-}\in\{-1,+1\}^{N}$. Although the physical masks achieve maximal binary separation, the directional contribution retained in sample $i$ scales with $W_{ik_+}-W_{ik_-}$: when overlapping impulse tails make $W_{ik_+}$ and $W_{ik_-}$ comparable, their signed spatial contributions partially cancel in the recorded sample, whereas unequal historical weights preserve directional contrast. Temporal placement controls the lag separation between $k_+$ and $k_-$ and thereby governs how strongly that pair's coding direction survives low-pass persistence.

Applying the inter-slot mean-difference operator $\mathbf{R}_{\mathrm{out}}$ from Section~S1.4 across the valid sample sets $\mathcal{I}_{m,+}$ and $\mathcal{I}_{m,-}$ of the $m$th complementary pair quantifies how self-slot retention competes with cross-branch tail leakage. Acting with row $m$ of $\mathbf{R}_{\mathrm{out}}$ on Eq.~(S12), the signed pair-specific directional coefficient multiplying $\mathbf{p}_{k_+}-\mathbf{p}_{k_-}$ in readout $m$ is
\begin{equation}
\begin{aligned}
&\frac{1}{2}\sum_{i=1}^{L} R_{mi}
\left(W_{ik_+}-W_{ik_-}\right)
=
\frac{1}{2}
\Big[
\left\langle W_{ik_+}\right\rangle_+
+
\left\langle W_{ik_-}\right\rangle_-
-
\left\langle W_{ik_-}\right\rangle_+
-
\left\langle W_{ik_+}\right\rangle_-
\Big],
\end{aligned}
\end{equation}
where $\langle u_i\rangle_+=\frac{1}{|\mathcal{I}_{m,+}|}\sum_{i\in\mathcal{I}_{m,+}}u_i$ and $\langle u_i\rangle_-=\frac{1}{|\mathcal{I}_{m,-}|}\sum_{i\in\mathcal{I}_{m,-}}u_i$ denote arithmetic means over the retained ADC samples in the positive and negative branches of pair $m$, respectively. The first two terms measure the direct responses of $k_+$ and $k_-$ within their assigned readout windows, whereas the last two capture cross-branch leakage between the complementary slots. By causality, the later-displayed pattern cannot influence samples in the earlier branch: for $k_+<k_-$, $\langle W_{ik_-}\rangle_+=0$ while the decaying tail of $k_+$ enters the negative branch through $\langle W_{ik_+}\rangle_-$ (and symmetrically when $k_-<k_+$). Together with any contributions from other historical patterns to readout $m$, Eq.~(S13) determines the pair-specific directional coefficient retained in row $m$ of $\widetilde{\mathbf{A}}_{\mathrm{out}}$.

Under low-pass responses with decaying persistence, adjacent complementary pairing C11 with $k_-=k_++1$ sends the persisting tail of $k_+$ directly into the immediately following slot $k_-$, enlarging $\langle W_{ik_+}\rangle_-$ and attenuating Eq.~(S13) through partial cancellation. Increasing the temporal separation via grouped ordering, such as C44 where four positive masks from distinct Hadamard directions precede their four negative counterparts, can reduce cross-branch tail overlap, retain a larger pair-specific coefficient along $\mathbf{p}_{k_+}-\mathbf{p}_{k_-}$, and redistribute historical cross-couplings across distinct coding directions in $\widetilde{\mathbf{A}}$.

At the same time, this spatial-encoding gain interacts with temporal drift sensitivity and physical switching dynamics. Adjacent pairing C11 minimizes the inter-branch time separation to suppress slow baseline drift, whereas grouped pairing C44 widens the temporal separation between branches in exchange for stronger differential signal retention. As analyzed in Section~S2.2 and validated in Section~S4.3, varying the relative strength of white noise and low-frequency drift shifts the more favorable ordering between C11 and C44 [Fig.~S11(a)]. Moreover, in physical DMD acquisitions, different pattern orderings impose different periodic micromirror switching loads, altering harmonic components in the recorded signal spectrum and reconstruction stability across modulation rates [Section~S4.3, Figs.~S11(b) and S12]. Pattern ordering therefore acts as a coupled design variable that must be coordinated with receiver dynamics and composite readout.

\subsection{Complementarity of inter-slot and intra-slot readouts}

Complementary binary masks achieve the largest squared Euclidean separation $\|\mathbf{p}_{k_+}-\mathbf{p}_{k_-}\|_2^2=N$ on $\{0,1\}^{N}$ with zero spatial mean at a $50\%$ open fraction, yet low-pass temporal mixing contracts inter-slot differences and couples successive coding directions. Rather than relying solely on slot-averaged levels, the composite readout $\mathbf{R}=[\mathbf{R}_{\mathrm{out}}^{\mathsf T},\mathbf{R}_{\mathrm{in}}^{\mathsf T}]^{\mathsf T}$ extracts two distinct physical signatures from the same recorded stream: $\widetilde{\mathbf{A}}_{\mathrm{out}}$ captures branch-averaged differences governed by Eq.~(S13), whereas $\widetilde{\mathbf{A}}_{\mathrm{in}}$ captures within-slot transient evolution governed by response-weighted historical transitions $\mathbf{p}_k-\mathbf{p}_{k-1}$ in Eq.~(S9). Whereas steady-state protocols wait for transients to decay so that $\Delta\mathbf{a}=\mathbf{0}$, temporal mixing turns within-slot phase evolution into informative spatial constraints.

\paragraph{Complementary spatial sensitivity.}

To characterize how the two branches jointly distinguish scene structures, consider a candidate target difference $\delta\mathbf{x}\in\mathbb{R}^{N}$. Under the Joint objective with weighting factor $\beta\ge 0$ in Eq.~(7) of the main text, the combined data-fidelity response to $\delta\mathbf{x}$ is
\begin{equation}
\left\|
\widetilde{\mathbf{A}}_{\mathrm{out}}
\delta\mathbf{x}
\right\|_2^2
+
\beta
\left\|
\widetilde{\mathbf{A}}_{\mathrm{in}}
\delta\mathbf{x}
\right\|_2^2.
\end{equation}

Expressing Eq.~(S14) as $\delta\mathbf{x}^{\mathsf T}\bigl(\widetilde{\mathbf{A}}_{\mathrm{out}}^{\mathsf T}\widetilde{\mathbf{A}}_{\mathrm{out}}+\beta\widetilde{\mathbf{A}}_{\mathrm{in}}^{\mathsf T}\widetilde{\mathbf{A}}_{\mathrm{in}}\bigr)\delta\mathbf{x}$ clarifies how the two branches supplement one another. Target variations $\delta\mathbf{x}$ that produce only weak responses in $\widetilde{\mathbf{A}}_{\mathrm{out}}$, such as under strong low-pass smoothing where slot means blur across multiple patterns, gain additional constraint from $\beta\widetilde{\mathbf{A}}_{\mathrm{in}}^{\mathsf T}\widetilde{\mathbf{A}}_{\mathrm{in}}$ whenever they elicit nonzero phase-dependent transition responses $\widetilde{\mathbf{A}}_{\mathrm{in}}\delta\mathbf{x}\neq\mathbf{0}$. Conversely, target variations weakly sensed by local within-slot slopes remain constrained whenever $\widetilde{\mathbf{A}}_{\mathrm{out}}\delta\mathbf{x}\neq\mathbf{0}$. Combining both readouts therefore strengthens distinguishability across directions with complementary branch sensitivities without requiring additional physical patterns or longer acquisition times.

\paragraph{Transmission of slowly varying disturbances.}

The same readout coefficients also govern the propagation of temporal disturbances. Approximating a slowly varying recorded-domain baseline $b_i$ locally around a reference time $t_{\mathrm{ref}}$ as $b_i\approx b_0+\dot{b}(t_i-t_{\mathrm{ref}})$ with local offset $b_0$ and drift rate $\dot{b}$, the zero-sum property $\sum_{i=1}^{L}R_{mi}=0$ and unit-slope normalization $\sum_{i\in\mathcal{I}_m}R_{mi}(t_i-\bar{t})=1$ in Section~S1.4 give
\begin{equation}
\sum_{i=1}^{L} R_{mi}b_i
\simeq
\begin{cases}
\dot{b}
\left(
\left\langle t_i\right\rangle_+
-
\left\langle t_i\right\rangle_-
\right),
&
\text{inter-slot mean difference},
\\[5pt]
\dot{b},
&
\text{intra-slot slope},
\end{cases}
\end{equation}
where $\langle t_i\rangle_+$ and $\langle t_i\rangle_-$ denote the mean sampling timestamps over $\mathcal{I}_{m,+}$ and $\mathcal{I}_{m,-}$, respectively.

Equation~(S15) shows that slow baseline leakage in an inter-slot mean difference scales directly with the inter-branch time separation $\langle t_i\rangle_+-\langle t_i\rangle_-$, linking drift sensitivity to pattern ordering and modulation rate. By contrast, an intra-slot slope operates within a single physical slot: its zero-sum weights eliminate slot-constant common-mode offsets $b_0$ while retaining the local rate of change $\dot{b}$ independently of complementary-pair separation. Because inter-slot differences have units of signal amplitude whereas intra-slot slopes have units of signal amplitude per unit time, these drift contributions must be evaluated alongside the corresponding target sensitivities and operator scales $\|\widetilde{\mathbf{A}}_{\mathrm{out}}\|_F$ and $\|\widetilde{\mathbf{A}}_{\mathrm{in}}\|_F$, consistent with Fig.~3(b) of the main text.

\paragraph{Transmission of correlated measurement noise.}

For the recorded-domain noise $\boldsymbol{\nu}=(\nu_1,\ldots,\nu_L)^{\mathsf T}$ in Eq.~(2) of the main text, the variance propagated into composite readout $m$ is
\begin{equation}
\operatorname{Var}
\left(
\sum_{i=1}^{L} R_{mi}\nu_i
\right)
=
\sum_{i=1}^{L}\sum_{i'=1}^{L}
R_{mi}R_{mi'}
\operatorname{Cov}(\nu_i,\nu_{i'}),
\end{equation}
where $\operatorname{Var}$ and $\operatorname{Cov}$ denote variance and covariance, respectively.

Equations~(S15) and (S16) capture the contrasting disturbance behavior of the two branches. Branch averaging in $\mathbf{R}_{\mathrm{out}}$ attenuates rapid uncorrelated noise in proportion to $1/|\mathcal{I}_{m,\pm}|$, whereas slope estimation in $\mathbf{R}_{\mathrm{in}}$ eliminates slot-constant offsets and bounds slow drift to $\dot{b}$, can amplify rapid noise when valid intra-slot timestamps span a narrow interval, and transmits general correlated fluctuations according to the full covariance quadratic form in Eq.~(S16). Moreover, because $\mathbf{R}_{\mathrm{out}}$ and $\mathbf{R}_{\mathrm{in}}$ act on shared ADC samples in $\mathbf{y}$, their output noise vectors exhibit cross-branch covariance $\operatorname{Cov}(\mathbf{R}_{\mathrm{out}}\boldsymbol{\nu},\mathbf{R}_{\mathrm{in}}\boldsymbol{\nu})=\mathbf{R}_{\mathrm{out}}\boldsymbol{\Sigma}_{\nu}\mathbf{R}_{\mathrm{in}}^{\mathsf T}$ with $\boldsymbol{\Sigma}_{\nu}=\operatorname{Cov}(\boldsymbol{\nu})$. Disturbances entering prior to the receiving response are analyzed separately in Section~S2.4.

\paragraph{Joint readout for common-target recovery.}

Equations~(S14)--(S16) show that the effective utility of each readout branch depends on its retained target constraints relative to the drift and noise transmitted through the same coefficients. Coupling both branches through $\beta\ge 0$ in Eq.~(7) of the main text, characterized across configurations by the scale-normalized weight $\rho_{\mathrm{in}}$ in Eq.~(8) of the main text, allows complementary spatial sensitivities and disturbance responses to participate jointly in reconstructing a common scene, as demonstrated in Sections~S4.2 and S4.3. Section~S2.3 next establishes why fitting these composite observations directly to the target image avoids the limitations of intermediate projection recovery.

\subsection{Direct recovery under a common-target constraint}

\paragraph{Spatial consistency of historical pattern projections.}

For a static target $\mathbf{x}\in\mathbb{R}^{N}$, let $s_k=\mathbf{p}_k^{\mathsf T}\mathbf{x}$ denote the noiseless spatial projection onto physical pattern $k$, collected as $\mathbf{s}=\mathbf{P}\mathbf{x}\in\mathbb{R}^{K}$. Although the $K$ masks are displayed at distinct times and temporally mixed by $\mathbf{W}$, the entries of $\mathbf{s}$ are not independent unknowns because every historical projection probes the same scene $\mathbf{x}$. For instance, every complementary binary pair satisfying $\mathbf{p}_{k_+}+\mathbf{p}_{k_-}=\mathbf{1}$ obeys
\begin{equation}
s_{k_+}+s_{k_-}
=
\left(\mathbf{p}_{k_+}+\mathbf{p}_{k_-}\right)^{\mathsf T}\mathbf{x}
=
\mathbf{1}^{\mathsf T}\mathbf{x},
\end{equation}
enforcing an identical projection sum regardless of temporal separation or mixing overlap. More generally, $\mathbf{s}$ must lie in the column space $\operatorname{range}(\mathbf{P})\subset\mathbb{R}^{K}$ and correspond to a spatially structured image $\mathbf{x}$ governed by $\mathcal{J}(\mathbf{x})$.

\paragraph{Waveform-restoration limitations across reconstruction and acquisition.}

Conventional approaches to temporally mixed observations $\mathbf{z}=\mathbf{R}\mathbf{y}$ encounter both algebraic and physical limitations. At the reconstruction level, pairing $\mathbf{z}$ with a memoryless current-pattern operator $\widetilde{\mathbf{A}}_0$ introduces a deterministic, scene-dependent mismatch $(\widetilde{\mathbf{A}}-\widetilde{\mathbf{A}}_0)\mathbf{x}$ whenever modulation rates approach or exceed the receiver bandwidth, as shown in Sections~2.2 and 3.2 of the main text. Projection-first two-stage methods instead model the mixed measurements through $\mathbf{R}\mathbf{W}$ and attempt to estimate intermediate slot amplitudes prior to spatial image reconstruction. Treating these amplitudes as unconstrained unknowns omits Eq.~(S17) and the column-space relation $\mathbf{s}=\mathbf{P}\mathbf{x}$ during temporal inversion: under strong low-pass smoothing or sub-modulation sampling where $Q<K$, unconstrained inversion of $\mathbf{R}\mathbf{W}\in\mathbb{R}^{Q\times K}$ is underdetermined or ill-conditioned, since any nonzero perturbation $\delta\mathbf{s}$ satisfying $\mathbf{R}\mathbf{W}\delta\mathbf{s}=\mathbf{0}$ produces no change in the predicted data while small singular values amplify recorded-domain drift residuals and measurement noise from Eqs.~(S15) and (S16). Although a two-stage pipeline could in principle stack composite readouts and propagate stage-one error covariances, practical unconstrained projection-first baselines that omit scene priors and full error-covariance propagation, such as the minimum-norm inversion evaluated in Section~S4.5, pass colored cross-slot estimation errors into image recovery and struggle when individual slot amplitudes are underdetermined.

More fundamentally, aiming to invert the receiver response and faithfully restore the pre-response optical waveform can conflict with final image quality under realistic disturbances. Unlike the ideal noiseless projection vector $\mathbf{s}=\mathbf{P}\mathbf{x}$, the physical pre-response input entering the receiving chain is already contaminated whenever upstream optical interference or illumination fluctuations are present, as modeled in Section~S2.4 and Eq.~(S18). Faithfully restoring this disturbance-contaminated pre-response waveform therefore recovers the upstream fluctuations that the low-pass receiving response had attenuated, while simultaneously amplifying recorded-domain noise during deconvolution. Conversely, regularizing or smoothing the intermediate waveform in stage one alters slot estimates without guidance from spatial image structure. At the acquisition-design level, enforcing isolated memoryless projections at sampled slots in sub-modulation regimes ($f_{\mathrm{ADC}}<f_{\mathrm{DMD}}$) also discards contributions from unsampled intervening masks that pre-ADC mixed waveforms retain, even in the absence of noise in Section~S2.5. Across both reconstruction and acquisition design, intermediate waveform fidelity is therefore an unreliable proxy for final image quality.

\paragraph{Waveform shaping for direct recovery and the target-centered co-design paradigm.}

Single-stage direct recovery resolves these limitations by dispensing with unconstrained intermediate waveform restoration and fitting the common target directly to the matched composite observations. Embedding $\mathbf{s}=\mathbf{P}\mathbf{x}$ into $\widetilde{\mathbf{A}}=\mathbf{R}\mathbf{W}\mathbf{P}$ in Eqs.~(4) and (5) of the main text restricts admissible projection variations to $\delta\mathbf{s}=\mathbf{P}\delta\mathbf{x}$, excluding projection ambiguities incompatible with a shared scene and applying the image prior $\mathcal{J}(\mathbf{x})$ directly in the measurement model. Even when $Q<K$ renders unconstrained recovery of a $K$-dimensional projection vector underdetermined, the target $\mathbf{x}$ can remain recoverable directly whenever $\widetilde{\mathbf{A}}$ retains sufficient distinguishability over the admissible target variations, as demonstrated against unconstrained minimum-norm two-stage recovery in Section~S4.5 and Fig.~S14(a).

Crucially, bypassing projection separation allows recorded waveforms to mix, smooth, and evolve in the manner best suited to assist target reconstruction rather than reproduce isolated step inputs. This establishes a target-recovery-centered co-design paradigm: every degree of freedom across spatial coding and ordering in $\mathbf{P}$, receiver bandwidth, filtering, modulation rate, and sampling timing in $\mathbf{W}$, and composite readouts in $\mathbf{R}$ is evaluated and coordinated around a single unified objective, namely the recoverability of the common target $\widehat{\mathbf{x}}$ in Eq.~(5) of the main text.

\subsection{Co-design of temporal filtering and composite readout}

\paragraph{Coupled transmission of upstream disturbances and temporal readout.}

Consider an additive upstream disturbance $\xi^{\mathrm{opt}}(u)$, such as optical interference or front-end flux fluctuation, entering the acquisition chain before readout gain scaling $G$. Because causality sets $\kappa_i(u)=G\,h(t_i-u)=0$ for $u>t_i$ in Eq.~(2) of the main text, extending the upper integration limit to $+\infty$ and applying row $m$ of $\mathbf{R}$ gives its contribution to composite readout $m$:
\begin{equation}
\begin{aligned}
&\sum_{i=1}^{L} R_{mi}
\int_{-\infty}^{\infty}
\kappa_i(u)\xi^{\mathrm{opt}}(u)\,\mathrm{d}u
=
\int_{-\infty}^{\infty}
\left[
\sum_{i=1}^{L} R_{mi}\kappa_i(u)
\right]
\xi^{\mathrm{opt}}(u)\,\mathrm{d}u.
\end{aligned}
\end{equation}

Equation~(S18) shows that upstream disturbances are filtered by the combined temporal kernel $\sum_{i=1}^{L}R_{mi}\kappa_i(u)$ formed jointly by the receiving response and composite readout weights, complementing Eqs.~(S15) and (S16) for recorded-domain drift and noise.

\paragraph{Balance between disturbance suppression and historical spatial sensitivity.}

Conventional pattern-wise SPI requires per-slot settling, demanding wide receiver bandwidths that admit broadband upstream fluctuations through Eq.~(S18) while trading away transimpedance gain. Incorporating the calibrated response into $\widetilde{\mathbf{A}}=\mathbf{R}\mathbf{W}\mathbf{P}$ turns receiver bandwidth and low-pass filtering into active co-design variables: stronger low-pass memory attenuates high-frequency upstream fluctuations while extending inter-slot overlap, narrowing or eliminating steady-state plateaus and redistributing historical target variations between $\widetilde{\mathbf{A}}_{\mathrm{out}}$ [Eq.~(S13)] and $\widetilde{\mathbf{A}}_{\mathrm{in}}$ [Eq.~(S9)].

Whether stronger filtering or lower receiver bandwidth improves reconstruction is governed by the balance between disturbance suppression across Eqs.~(S15), (S16), and (S18) and the retention of target-dependent sensitivity in Eq.~(S14), together with the directional noise structure and image prior $\mathcal{J}(\mathbf{x})$. As low-pass smoothing strengthens and contracts inter-slot contrast, Joint recovery can shift relative data participation $\rho_{\mathrm{in}}$ in Eq.~(8) of the main text  toward the intra-slot branch represented by $\widetilde{\mathbf{A}}_{\mathrm{in}}$. Consequently, under pronounced low-frequency fluctuations or upstream optical interference, strongly low-pass-filtered or bandwidth-limited Joint configurations can outperform wider-bandwidth steady-state readouts, while remaining bounded when broadband noise dominates, as demonstrated in Section~3.3 of the main text and Sections~S4.2, S4.4, Figs.~S10 and S13.

\paragraph{Pre-ADC analog memory versus post-ADC digital filtering.}

The stage at which temporal filtering occurs determines its role in spatial encoding. Pre-ADC analog memory integrates continuous-time optical projections $s^{\mathrm{opt}}(u)$ prior to digitization, attenuating analog fluctuations before sampling and coupling optical projections from intervals between ADC samples into recorded readings. Post-ADC causal digital filtering, by contrast, applies a linear transformation to an already digitized sample stream: while remaining within the measurement subspace of the acquired digital record, it reshapes noise correlations and reweights historical constraints across $\widetilde{\mathbf{A}}_{\mathrm{out}}$ and $\widetilde{\mathbf{A}}_{\mathrm{in}}$. Both forms of temporal processing are accommodated consistently through matched reference calibration, as established in Sections~S1.4 and S3.4.

\subsection{Historical encoding under limited sampling}

\paragraph{Historical contributions from unsampled intervening patterns.}

Under memoryless instantaneous sampling, each ADC sample probes only the physical pattern displayed at its sampling timestamp; when $f_{\mathrm{ADC}}<f_{\mathrm{DMD}}$, intervening patterns whose display intervals satisfy $t_i<T_k<T_{k+1}<t_{i+1}$ are absent from the instantaneous measurements. Under pre-sampling temporal memory with finite support $\tau_H$, causality sets $W_{ik}=0$ for $t_i<T_k$, whereas subsequent retained samples $t_{i'}>T_k$ for which pattern $k$ remains within the active lag support ($t_{i'}-T_{k+1}<\tau_H$) can carry nonzero history weights $W_{i'k}$ determined by the overlap integral in Eq.~(S5). Retaining all $K$ physical columns in $\mathbf{W}\in\mathbb{R}^{L\times K}$ in Section~S1.3 thus incorporates spatial contributions from unsampled intervening patterns without additional ADC conversions.

\paragraph{Historical encoding and noiseless target distinguishability.}

For a candidate target difference $\delta\mathbf{x}\in\mathbb{R}^{N}$, the resulting change in composite readout $m$ is
\begin{equation}
\left(
\widetilde{\mathbf{A}}\delta\mathbf{x}
\right)_m
=
\sum_{k=1}^{K}
\left(
\sum_{i=1}^{L} R_{mi}W_{ik}
\right)
\mathbf{p}_k^{\mathsf T}\delta\mathbf{x},
\end{equation}
where $c_{mk}=\sum_{i=1}^{L}R_{mi}W_{ik}$ denotes the net composite historical weight of physical pattern $k$ in readout $m$.

To separate spatial encoding from noise suppression, compare an ideal noiseless memoryless system against a noiseless memory-bearing system under the identical pattern sequence $\mathbf{P}$, sampling timestamps $\{t_i\}_{i=1}^{L}$, and digital readout count $Q$. Any nonzero target variation $\delta\mathbf{x}$ orthogonal to all directly sampled patterns ($\mathbf{p}_{k(i)}^{\mathsf T}\delta\mathbf{x}=0$ for every sampled slot $k(i)$) produces zero response in the memoryless measurements and is therefore invisible to instantaneous sampling. Under temporal memory, intervening unsampled patterns also enter Eq.~(S19) through $c_{mk}$: whenever their weighted contributions form a nonzero sum $\sum_{k=1}^{K}c_{mk}\mathbf{p}_k^{\mathsf T}\delta\mathbf{x}\neq 0$ in at least one composite readout $m$, we obtain $\widetilde{\mathbf{A}}\delta\mathbf{x}\neq\mathbf{0}$. Temporal memory therefore alters the spatial constraints captured within a fixed digital observation budget $Q$ and can render target variations invisible under memoryless sampling identifiable, even in the complete absence of measurement noise.

\paragraph{Sampling timing, modulation rate, and recovery-oriented co-design.}

Whereas shifting the within-slot sampling phase $\varphi_i$ leaves an instantaneous projection unchanged, under temporal memory distinct phases $\varphi_i$ yield distinct historical weightings in Eq.~(S5) and thus distinct effective spatial patterns, turning sampling phase into an active measurement-design variable. Guided by common-target recoverability, this supports two complementary adjustments: reducing the ADC sampling rate below the DMD modulation rate at a fixed modulation schedule to lower digitization, storage, and transmission overhead, or increasing the DMD modulation rate at a fixed ADC rate and pattern count $K$ to shorten optical encoding duration. When sparse sampling leaves fewer than two valid samples per slot so that intra-slot slopes in Eq.~(S10) cannot be formed, complete-history Inter with $\beta=0$ still exploits multi-slot historical encoding.

Section~S4.5 validates these mechanisms through both measured low-ADC reconstructions [Fig.~S14(a) and (b)] and noiseless numerical controls [Fig.~S14(c) and (d)], confirming that pre-sampling historical encoding improves target recovery under fixed sampling schedules and can retain net gains even when each configuration selects its own preferred sampling timing. Together, Sections~S2.1--S2.5 establish how pattern ordering, inter- and intra-slot readouts, temporal filtering, and sampling timing act as coupled dimensions of $\widetilde{\mathbf{A}}=\mathbf{R}\mathbf{W}\mathbf{P}$, turning calibrated hardware memory into a configurable imaging resource governed by final target recoverability.

%% file: section_s3.tex
\section{Experimental validation of response calibration, timing consistency, and transfer}
\label{sec:supp_s3}

Sections~S1 and S2 establish the calibrated complete-history measurement model $\widetilde{\mathbf{A}}=\mathbf{R}\mathbf{W}\mathbf{P}$ and the spatial constraints extracted through composite readouts. Complementing Sections~3.1--3.4 of the main text, this section experimentally validates temporal-response identification, phase coverage, reference-based timing maintenance, adaptation to modified acquisition chains, and cross-condition and cross-date transfer, culminating in imaging of diverse targets under transferred responses.

\subsection{Common calibration settings and validation protocol}
\label{sec:supp_s3_1}

Reference calibration and temporal-response identification follow Section~2.1 of the main text and Sections~S1.1--S1.3. Across the primary receiving chains, the piecewise-constant response vector $\widehat{\mathbf{h}}\in\mathbb{R}^{H}$ is discretized over $[0,\tau_H)$ using $(H,\tau_H)=(64, 3.20~\mathrm{ms})$ for the 350~Hz chain, $(384, 3.20~\mathrm{ms})$ for the 1.8~kHz chain, and $(96, 0.80~\mathrm{ms})$ for the 18~kHz chain, while the independent DET36A2/PDA200C chain ($5~\mathrm{kHz}$ nominal bandwidth) uses $(64, 0.80~\mathrm{ms})$, as summarized in Table~S1. For static imaging, all within-slot sampling phases are retained in the 350~Hz, 1.8~kHz, and DET36A2/PDA200C configurations, whereas the 18~kHz configuration excludes the first $5~\mu\mathrm{s}$ and final $10~\mu\mathrm{s}$ of each physical pattern slot. Valid-sample selection is applied identically to the recorded observations $\mathbf{y}$ and history-weight rows of $\mathbf{W}$ while retaining all $K$ physical pattern slots in $\mathbf{P}$.

\begin{table}[!htb]
\centering
\caption{Baseline piecewise-constant response representations for the receiving chains.}
\label{tab:supp_calibration_parameters}
\small
\setlength{\tabcolsep}{9pt}
\begin{tabular}{lcc}
\hline
Receiver configuration & $H$ & Response support $\tau_H$ (ms) \\
\hline
350~Hz & 64 & 3.20 \\
1.8~kHz & 384 & 3.20 \\
18~kHz & 96 & 0.80 \\
DET36A2/PDA200C (5~kHz) & 64 & 0.80 \\
\hline
\end{tabular}
\end{table}

Regularization selection for the calibration objective in Section~S1.2 uses four-fold cross-validation across the eight training reference rounds, which are paired into four groups: rounds 1 and 5, 2 and 6, 3 and 7, and 4 and 8. In each fold, $(\boldsymbol{\alpha},b_{\mathrm{cal}})$ are fitted on six rounds from three groups and evaluated on the two held-out rounds of the remaining group. Pooling the held-out prediction residuals across all eight training rounds defines the cross-validation root-mean-square error (RMSE), and the regularization weight $\lambda_{\mathrm{cal}}$ minimizing this pooled RMSE is selected. The final $(\widehat{\boldsymbol{\alpha}},\widehat{q},\widehat{b}_{\mathrm{cal}})$ are then fitted across all eight training rounds and held fixed to predict the independent Block and Interleaved holdout sets. Holdout prediction accuracy is reported as normalized RMSE, $\mathrm{NRMSE}=\mathrm{RMSE}/\Delta V\times 100\%$, normalized by the measured black--white reference plateau span $\Delta V$, or by the reference amplitude span fitted on training references in drift-tracking and cross-record transfer tests.

Our validation distinguishes response adaptation from response reuse. When detector hardware, amplification, or incorporated signal processing changes the effective temporal kernel, $\widehat{\mathbf{h}}$ is recalibrated from that configuration's own reference sequences and validated on independent holdout sets. Conversely, when a stable receiving chain operates under different ADC sampling rates, DMD modulation rates, illumination levels, or acquisition dates, the normalized source response $\widehat{\mathbf{h}}$ remains frozen while history weights $W_{ik}$ are recomputed from the receiving record's registered timing $\widehat{q}$ and complete pattern history; in waveform-transfer evaluations, only a scalar gain and baseline offset are fitted on receiving training references before predicting the Block and Interleaved holdout sets. In response-substitution imaging, the recorded imaging observations, valid-sample masks, composite readout operator $\mathbf{R}$, and reconstruction hyperparameters remain fixed while $\mathbf{W}$ is updated from each substituted $\widehat{\mathbf{h}}$ following the static reconstruction protocol in Section~S4.1. Detailed settings and numerical records are archived in \mbox{Supplementary Data~1}.

\subsection{Sampling-phase coverage and selection of valid observations}
\label{sec:supp_s3_2}

As established in Sections~S1.1 and S1.2, ADC samples acquired at distinct within-slot phases $\varphi_i$ generate distinct historical overlap rows in $\mathbf{X}_{\mathrm{tr}}(q)$ to constrain the shared response vector. Supporting Fig.~4(e) in Section~3.1 of the main text, we evaluate dispersed-phase versus fixed-phase calibration under matched low-rate budgets across eight measured records on the 1.8~kHz chain spanning DMD modulation rates of 10--20~kHz and three nominal training sampling rates of 2.5, 5, and 10~kHz, forming 24 paired comparisons.

Dispersed-phase subsets uniformly subsample the $240~\mathrm{kHz}$ training record to nominal rates of 2.5, 5, and 10~kHz, allowing natural asynchronous ADC--DMD clock drift to distribute samples across within-slot phases. Fixed-phase subsets select samples locked to candidate within-slot phase positions at 25\%, 40\%, 55\%, 70\%, or 85\% of the slot duration. Both schemes share identical training reference segments, matched physical pattern slots, equal observation counts, the timing registration of Section~S1.2, and a common discretization of $H=192$ over $\tau_H=3.20~\mathrm{ms}$, with $\lambda_{\mathrm{cal}}$ and the best fixed-phase position selected strictly by training cross-validation as in Section~S3.1 prior to evaluation on the same full-rate Block and Interleaved holdout sets.

Dispersed-phase calibration outperforms fixed-phase calibration across all 24 paired comparisons in Fig.~4(e) of the main text, lowering Block NRMSE from 1.084\%--2.530\% to 0.749\%--0.963\% and Interleaved NRMSE from 1.178\%--3.006\% to 0.799\%--1.044\%. At 10~kHz DMD modulation and a 2.5~kHz training budget, for instance, Block NRMSE drops from 2.488\% to 0.773\% and Interleaved NRMSE drops from 3.006\% to 0.832\%. Because observation counts, lag discretizations, and timing correspondences are matched, this gain stems directly from richer historical-lag coverage in $\mathbf{X}_{\mathrm{tr}}(q)$ across varied sampling phases rather than additional calibration samples, providing the calibration basis for the sub-modulation imaging experiments in Section~3.4 of the main text and Section~S4.5.

Within-slot phase $\varphi_i$ also governs the reliability of imaging observations near micromirror transitions. Grouping holdout residuals into $1~\mu\mathrm{s}$ within-slot phase bins in Fig.~4(h) of the main text shows uniform prediction accuracy across the slot for the 1.8~kHz chain, whereas the wider-bandwidth 18~kHz chain exhibits localized error increases near slot boundaries caused by finite DMD micromirror crossover and settling transients. Accordingly, all within-slot phases are retained for the 350~Hz and 1.8~kHz chains, whereas the first $5~\mu\mathrm{s}$ and final $10~\mu\mathrm{s}$ of each slot are excluded for the 18~kHz chain while preserving all physical patterns in $\mathbf{W}\mathbf{P}$ as formulated in Section~S1.3. These diagnostics underpin the phase-resolved effective patterns and composite readouts in Sections~S1.4 and S2.2, with complete numerical records provided in \mbox{Supplementary Data~1}.

\subsection{Timing updates and response--timing coordinate consistency}
\label{sec:supp_s3_3}

Accurate construction of the history weights $W_{ik}$ in Eq.~(S5) requires maintaining a consistent temporal coordinate reference between the frozen response $\widehat{\mathbf{h}}$ and each acquisition's physical pattern sequence. Complementing Fig.~4(f) and (g) in Section~3.1 of the main text, this subsection validates reference-based timing maintenance alongside coordinated versus mismatched timing shifts shown in Fig.~S2.

In the continuous asynchronous record of Fig.~4(f) on the 1.8~kHz chain, both timing-maintenance modes keep $\widehat{\mathbf{h}}$ fixed: Estimate drift uses the first two embedded 168-slot references to update both initial slot phase $\widehat{q}_0$ and effective slot rate $\widehat{f}_s$, whereas Reuse drift updates $\widehat{q}_0$ from the first reference alone while retaining the calibrated slot rate. Evaluated causally on the independent third embedded reference over identical valid samples, both modes achieve an NCC of 0.9998 with prediction NRMSE values of 0.784\% and 0.931\%, respectively, confirming synchronization-free timing registration without recalibrating $\widehat{\mathbf{h}}$ as formulated in Section~S1.3.

To verify the temporal coordinate reference formalized in Eq.~(S3) of Section~S1.2, Fig.~S2(a) and (b) apply 13 coordinated shifts ($0, \pm1, \pm5, \pm10, \pm25, \pm50$, and $\pm100~\mu\mathrm{s}$) simultaneously to the timing origin and the fixed physical-lag centroid $\tau_c$ on a $240~\mathrm{kHz}$ record from the 1.8~kHz chain while keeping the measured data, pattern sequence, and slot rate unchanged. Refitting the response with $H=384$ at each shifted reference translates the piecewise-constant curves along the lag axis in Fig.~S2(a), yet shifting them back by their imposed offsets collapses all profiles onto a common curve in Fig.~S2(b), with Block and Interleaved holdout NRMSE invariant within 0.739\%--0.765\% across all 13 shifts. Thus, $\tau_c$ acts purely as a shared temporal coordinate reference rather than an absolute hardware delay.

\begin{figure}[!htb]
\centering
\includegraphics[width=\linewidth,height=0.57\textheight,keepaspectratio]{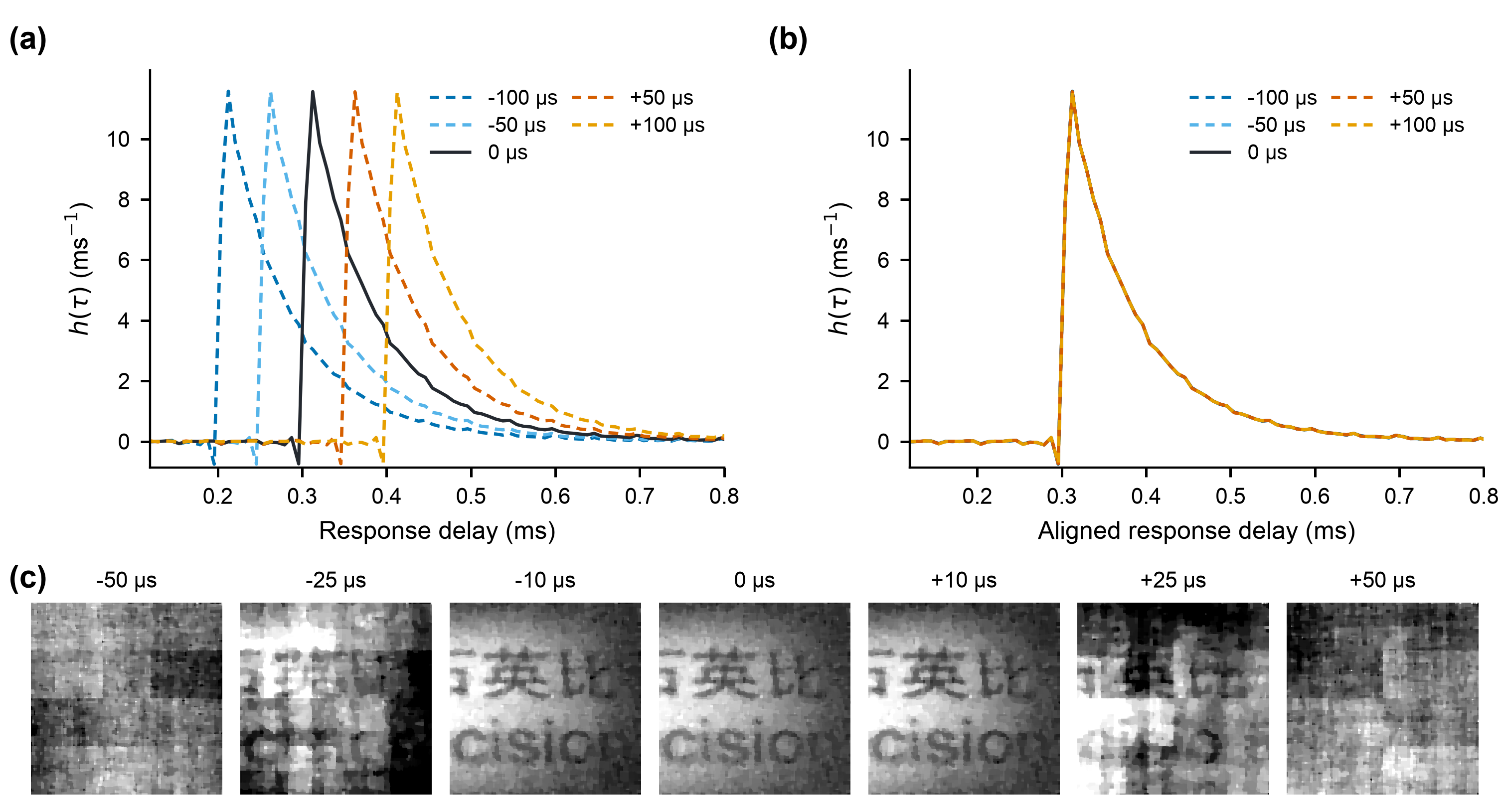}
\caption{Response--timing coordinate consistency and the imaging effects of timing mismatch on the 1.8~kHz chain. (a) Representative piecewise-constant responses $h(\tau)$ refitted after coordinated shifts of the timing origin and physical-lag centroid reference $\tau_c$. (b) The same responses after compensating for the imposed coordinate shifts. (c) Joint reconstructions of a printed target with measured observations, calibrated response $\widehat{\mathbf{h}}$, composite readouts, and reconstruction parameters held fixed while varying the timing offset applied to the matched operator.}
\label{fig:supp_timing}
\end{figure}

Conversely, displacing the timing origin $\widehat{q}_0$ while holding $\widehat{\mathbf{h}}$ fixed breaks this coordinate consistency. In Fig.~4(g) of the main text, holdout NRMSE rises sharply from approximately 0.752\% at zero offset to 2.295\%--2.343\% on Block and 2.635\%--2.734\% on Interleaved at $\pm5~\mu\mathrm{s}$, and continues climbing steeply up to $\pm100~\mu\mathrm{s}$. Propagating timing offsets of $-50, -25, -10, 0, +10, +25$, and $+50~\mu\mathrm{s}$ into Joint reconstruction of an independently measured printed target with observations, $\widehat{\mathbf{h}}$, readouts, and regularization parameters held fixed in Fig.~S2(c) preserves sharp text at zero offset but induces progressive structural blurring and background artifacts as mismatch grows, because misregistered slot boundaries corrupt $W_{ik}$ and perturb the matched operator $\widetilde{\mathbf{A}}=\mathbf{R}\mathbf{W}\mathbf{P}$ used for direct recovery in Section~S2.3. Together, these controls establish the timing-registration requirement underlying cross-condition and cross-date reuse in Section~S3.5, with full records provided in \mbox{Supplementary Data~1}.

\subsection{Composite-response identification under different measurement-chain configurations}
\label{sec:supp_s3_4}

As formulated in Section~2.1 of the main text and Section~S1.4, our forward model absorbs photodetection, amplification, digitization aperture, and causal signal processing into a single composite response $\widehat{\mathbf{h}}$. Extending the waveform validations in Fig.~4(d) of the main text, Fig.~S3 evaluates matched response recalibration and Joint image recovery across post-ADC causal digital filters and an independent hardware receiving chain.

\begin{figure}[!htb]
\centering
\includegraphics[width=\linewidth,height=0.57\textheight,keepaspectratio]{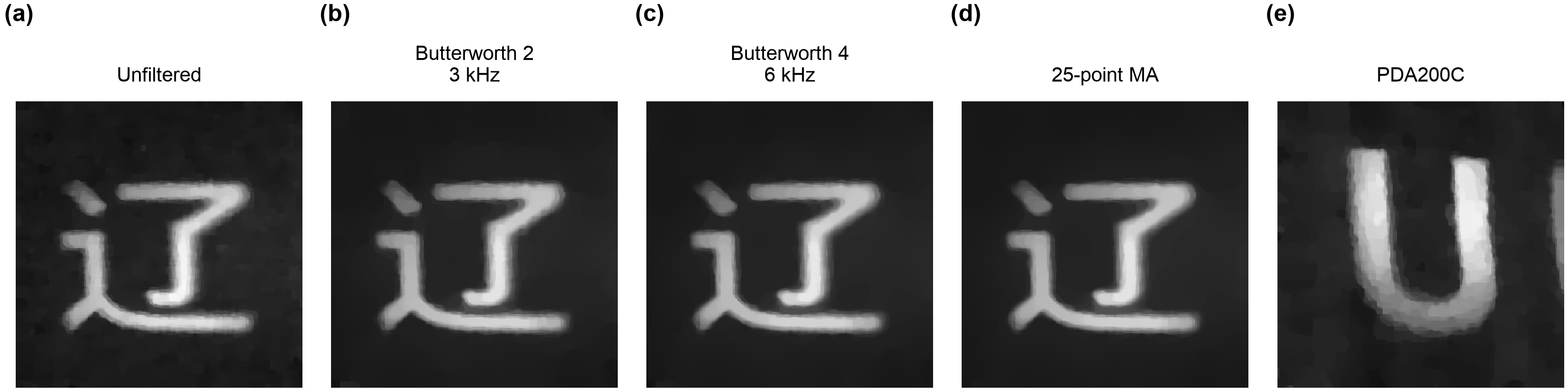}
\caption{Joint imaging with calibrated composite responses across digital filtering conditions and independent receiving hardware. (a) No additional digital filtering on the 18~kHz chain. (b) Causal second-order Butterworth low-pass filtering at 3~kHz. (c) Causal fourth-order Butterworth low-pass filtering at 6~kHz. (d) Causal 25-point moving-average filtering. Panels (a)--(d) reconstruct the same measured character record. (e) Joint reconstruction of an independently acquired U-shaped target on the independently calibrated DET36A2/PDA200C receiving chain.}
\label{fig:supp_composite_response}
\end{figure}

On the 18~kHz chain, measured reference and character-target records are processed under four conditions corresponding to Fig.~S3(a)--(d): no additional filtering, causal second-order Butterworth low-pass filtering at 3~kHz, causal fourth-order Butterworth low-pass filtering at 6~kHz, and a causal 25-point moving average. For the holdout reference-waveform prediction evaluations, all four conditions use a common response support of $\tau_H=1.20~\mathrm{ms}$. Each causal filter runs continuously across the raw ADC stream without state resets at slot or segment boundaries, and samples within the post-transition filter-settling interval immediately following segment boundaries are excluded alongside the valid-phase window of Section~S3.1. Applying identical processing to reference and imaging records yields holdout Block NRMSE of 0.783\%--1.127\% and Interleaved NRMSE of 0.963\%--1.306\%, while the resulting matched operators preserve crisp character strokes across all four conditions in Fig.~S3(a)--(d), directly supporting the filtering co-design in Section~3.3 of the main text and Section~S2.4.

Figure~S3(e) transfers the complete calibration-and-recovery workflow to an independent receiving chain comprising a Thorlabs DET36A2 photodetector, a PDA200C transimpedance amplifier ($10^7~\mathrm{V/A}$ gain, $5~\mathrm{kHz}$ nominal bandwidth), and a 16-bit DAQ sampling at $250~\mathrm{kHz}$. Calibrating $\widehat{\mathbf{h}}$ from this platform's own binary references yields Block and Interleaved holdout NRMSE values of 4.516\% and 4.273\%, and the resulting matched Joint operator cleanly reconstructs an independently acquired U-shaped target in Fig.~S3(e). Together, these experiments confirm that changes in either digital processing or physical detection hardware are accommodated seamlessly through matched recalibration under Eq.~(4) of the main text, with detailed records archived in \mbox{Supplementary Data~1}.

\subsection{Response transfer across acquisition conditions and time}
\label{sec:supp_s3_5}

Complementing the rate-variation validations in Fig.~4(b) and (c) of the main text, we evaluate whether a normalized response $\widehat{\mathbf{h}}$ calibrated under one operating condition transfers across diverse acquisition settings and multi-week intervals when paired with updated timing registration from Section~S3.3.

Figures~S4 and S5 present three pairs of $15\times15$ cross-condition transfer matrices for the 1.8~kHz, 18~kHz, and 350~Hz receiving chains, encompassing 15 reference records per chain across varied ADC rates, DMD modulation rates, and illumination levels (675 source--receiver pairs and 1,350 Block and Interleaved holdout NRMSE values in total, with rows indexing source responses and columns indexing receiving records). In every transfer pair, the source-calibrated $\widehat{\mathbf{h}}$ is held frozen, history weights $W_{ik}$ are recomputed from the receiving record's timing correspondence $\widehat{q}$ and pattern sequence, and only scalar gain and baseline offset are fitted on receiving training references prior to independent Block and Interleaved holdout evaluation.

\begin{figure}[!htb]
\centering
\includegraphics[width=\linewidth,height=0.57\textheight,keepaspectratio]{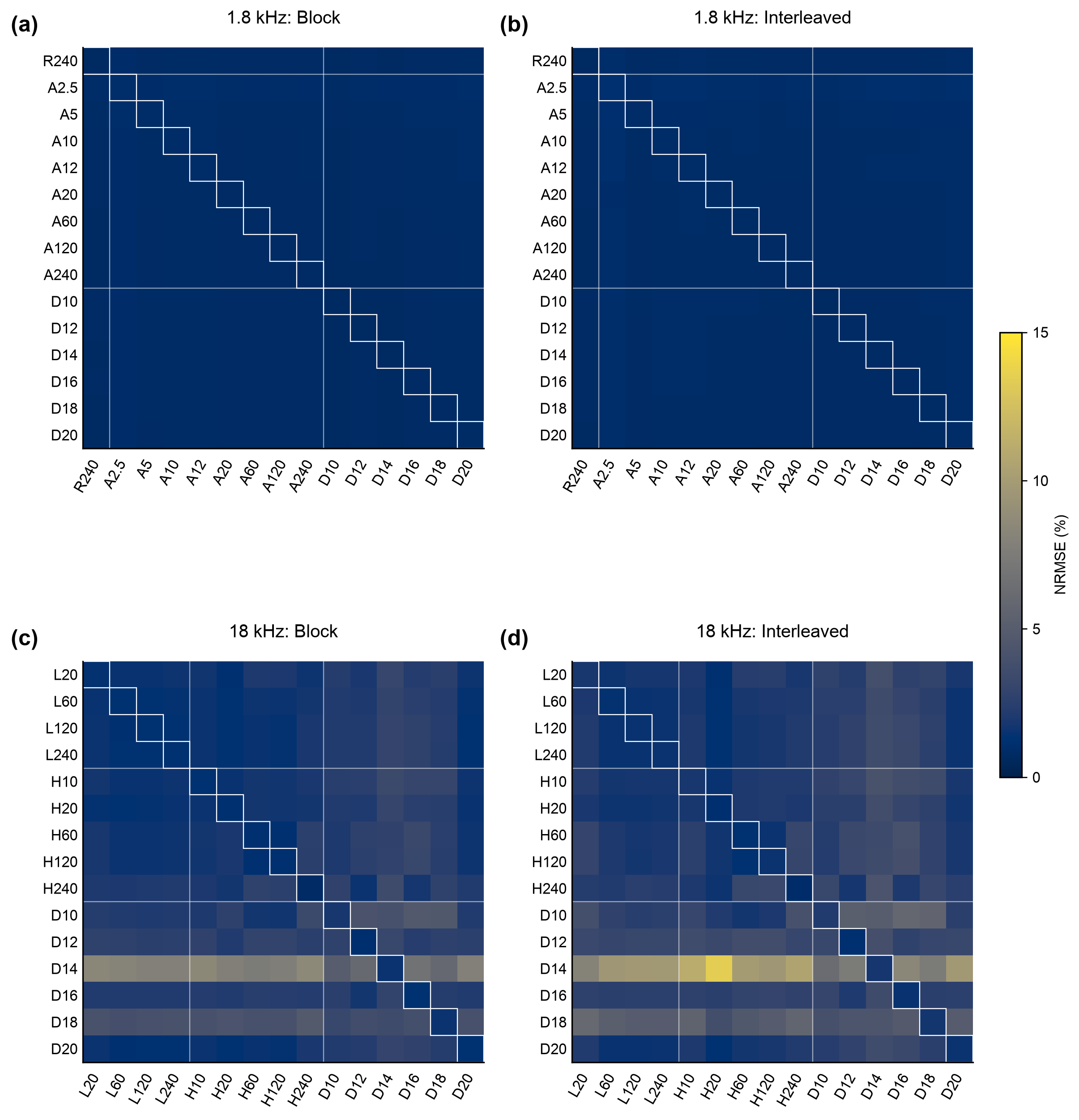}
\caption{Cross-condition response-transfer matrices for the 1.8~kHz and 18~kHz receiving chains. (a) Block and (b) Interleaved holdout NRMSE for the 1.8~kHz chain. (c) Block and (d) Interleaved holdout NRMSE for the 18~kHz chain. Matrix rows index frozen source responses $\widehat{\mathbf{h}}$ and columns index receiving records. R denotes the reference configuration, A and D denote ADC-rate and DMD-rate variations, and L and H denote lower- and higher-illumination records; numbers following R, A, L, and H specify ADC rates in kHz, whereas numbers following D specify DMD rates in kHz. The linear color scale spans 0\%--15\%.}
\label{fig:supp_transfer_main}
\end{figure}

\begin{figure}[!htb]
\centering
\includegraphics[width=\linewidth,height=0.57\textheight,keepaspectratio]{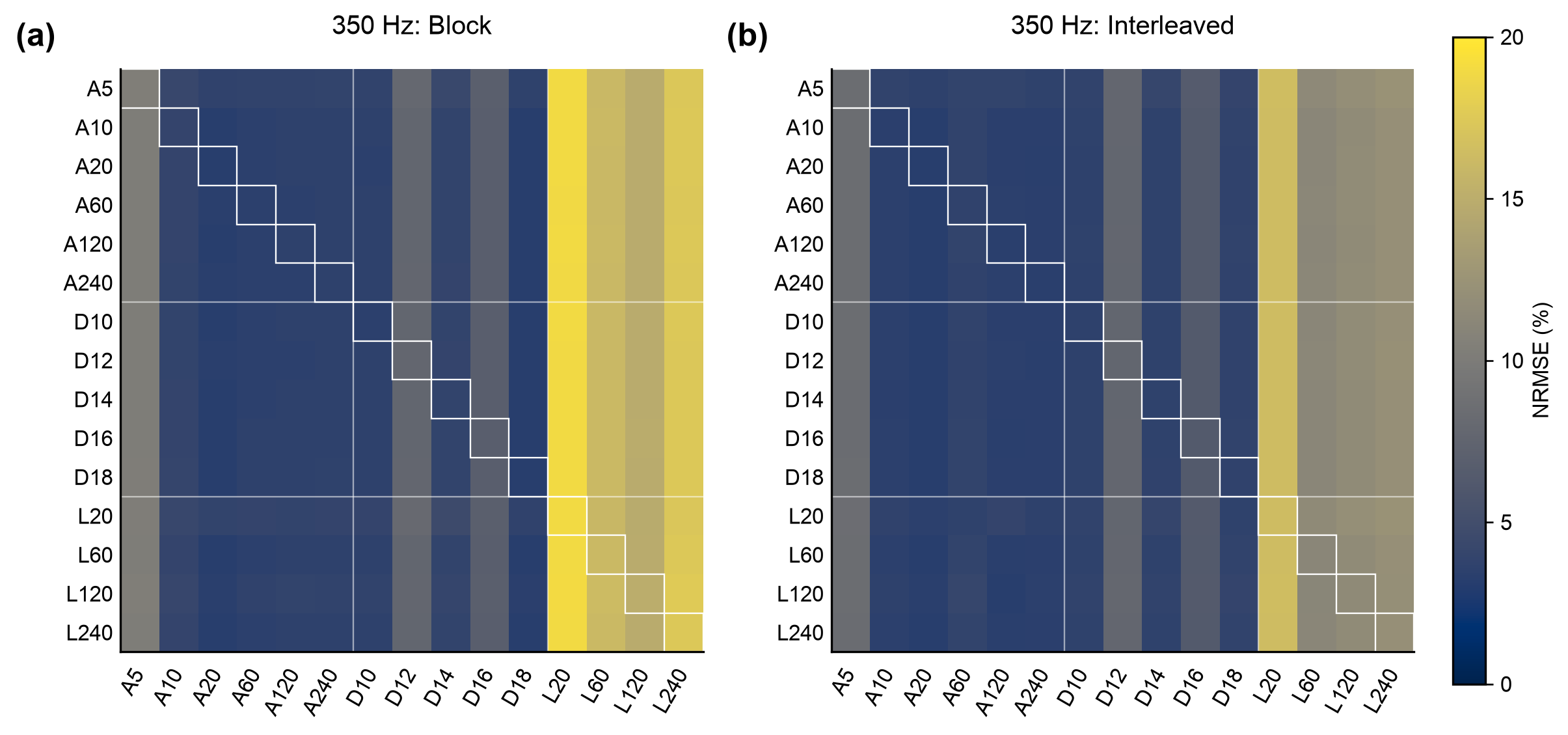}
\caption{Cross-condition response-transfer matrices for the 350~Hz receiving chain. (a) Block and (b) Interleaved holdout NRMSE. Matrix axes and record codes follow Fig.~S4, with the linear color scale spanning 0\%--20\%.}
\label{fig:supp_transfer_350}
\end{figure}

On the 1.8~kHz chain in Fig.~S4(a) and (b), Block and Interleaved NRMSE remain uniformly low within 0.719\%--1.046\% and 0.749\%--1.205\%, respectively, confirming accurate cross-rate reuse. On the 18~kHz chain in Fig.~S4(c) and (d), errors span 0.795\%--8.569\% on Block and 0.963\%--13.447\% on Interleaved, remaining low across most pairs while reflecting elevated noise sensitivity in lower-illumination or sparser calibration records.

The 350~Hz transfer matrices in Fig.~S5(a) and (b), spanning 3.208\%--19.047\% on Block and 3.218\%--16.541\% on Interleaved, reveal a striking row-versus-column asymmetry exemplified by record L20. Whenever L20 serves as the receiving column, prediction NRMSE stays high across all 15 source responses (18.88\%--19.05\% on Block and 16.42\%--16.54\% on Interleaved). Yet when the response $\widehat{\mathbf{h}}$ calibrated from L20 serves as the source row transferred to quieter receiving records A60, A240, D10, and D18, holdout NRMSE drops to 3.745\%--3.982\% on Block and 3.543\%--3.756\% on Interleaved, matching native calibrations on those records. High waveform residuals on L20 are therefore dominated by fluctuations in the receiving observations rather than a corrupted $\widehat{\mathbf{h}}$, motivating the direct fluctuation diagnostics in Section~S3.6.

Cross-date stability is further tested on the 1.8~kHz chain by transferring four archived calibrations acquired 28 or 38 days earlier to a current reference record. Re-estimating the receiving slot rate $\widehat{f}_s$ from receiving training references yields Block and Interleaved NRMSE of 0.786\%--0.918\% and 0.783\%--0.886\%, while reusing the archived source slot rate yields 0.795\%--0.949\% and 0.810\%--0.960\%. Both modes preserve sub-percent accuracy over multi-week intervals, confirming that $\widehat{\mathbf{h}}$ characterizes an intrinsic property of a stable receiving chain and establishing the foundation for imaging transfer in Section~S3.7. Detailed records are provided in \mbox{Supplementary Data~1}.

\Needspace{0.50\textheight}
\subsection{Effective response identification under substantial fluctuations}
\label{sec:supp_s3_6}

To resolve why fluctuating 350~Hz records such as L20 exhibit inflated receiving NRMSE yet yield accurate transferable responses in Section~S3.5, Figs.~S6 and S7 examine direct fixed-state recordings alongside a controlled disturbance-injection calibration test.

Figure~S6 records outputs from the 350~Hz, 1.8~kHz, and 18~kHz chains at $60~\mathrm{kHz}$ ADC sampling while holding the DMD in static all-black and all-white states. Both the full recordings in Fig.~S6(a), (c), and (e) and the zoomed white-plateau traces in Fig.~S6(b), (d), and (f) reveal substantial nonstationary fluctuations even under constant optical projection, most prominently in the high-gain 350~Hz configuration, directly explaining the elevated receiving-column residuals in Fig.~S5.

\begin{figure}[!htb]
\centering
\includegraphics[width=\linewidth,height=0.42\textheight,keepaspectratio]{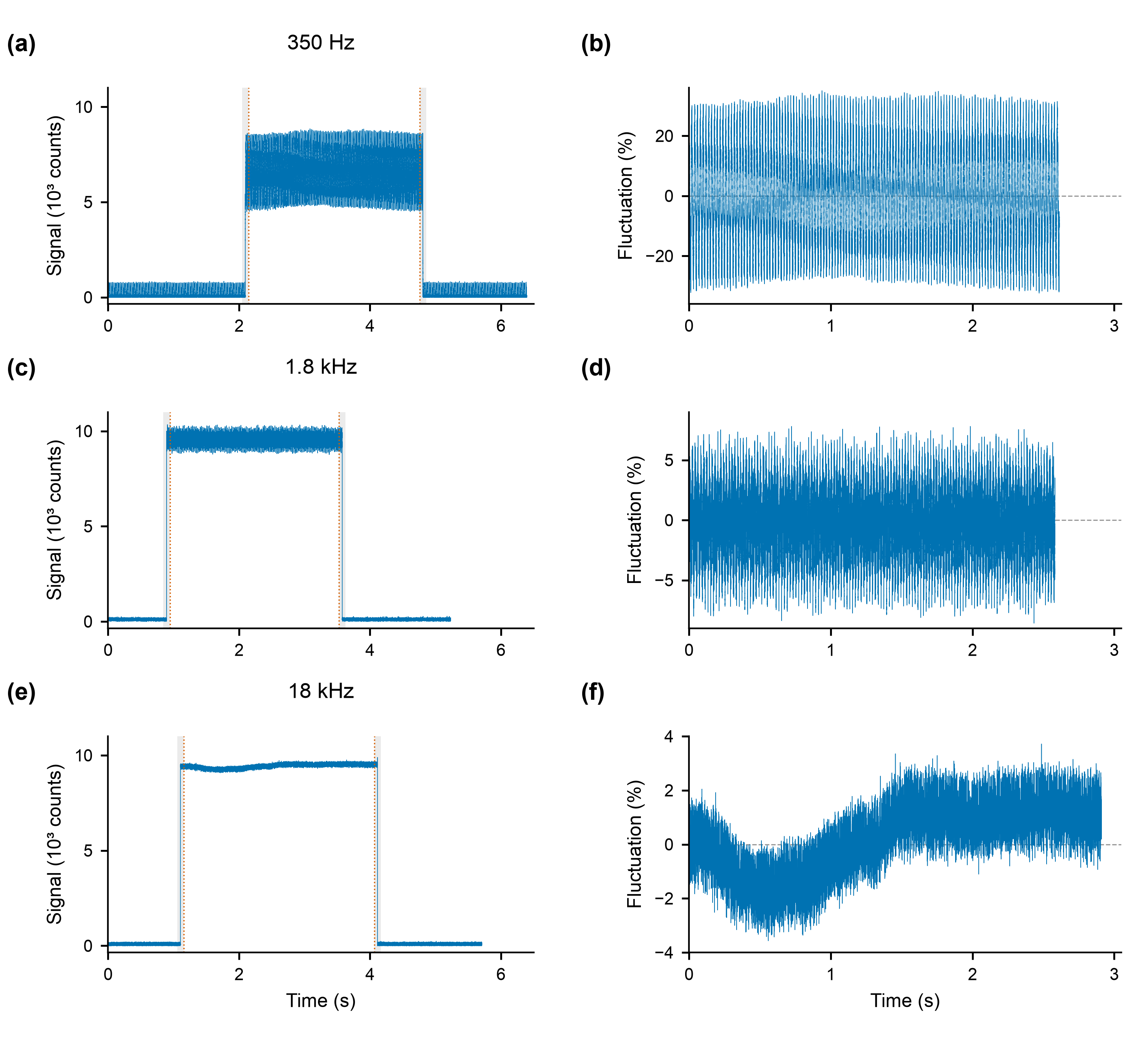}
\caption{Measured output fluctuations under static all-black and all-white DMD states at $60~\mathrm{kHz}$ ADC sampling. (a), (c), and (e) Complete recordings from the 350~Hz, 1.8~kHz, and 18~kHz receiving chains, respectively. (b), (d), and (f) Zoomed amplitude variations within the corresponding white-state plateaus.}
\label{fig:supp_fixed_state}
\end{figure}
\FloatBarrier

Despite these differences in background fluctuation, normalized responses $\widehat{\mathbf{h}}$ calibrated independently across diverse 350~Hz records exhibit closely aligned temporal profiles in Fig.~S7(a). To test this decoupling quantitatively against the pre-addition reference, Fig.~S7(b)--(d) pass a measured 18~kHz reference record through a causal first-order $350~\mathrm{Hz}$ low-pass filter to obtain a slow-response baseline before disturbance addition, and then add a low-frequency disturbance constructed from recorded fluctuation characteristics with an RMS amplitude of 7\% of the reference plateau span $\Delta V$. Calibrating separately before and after disturbance addition with $H=128$ over $\tau_H=5.35~\mathrm{ms}$ yields closely matching normalized responses in Fig.~S7(b).

\begin{figure}[!htb]
\centering
\includegraphics[width=0.85\linewidth,height=0.36\textheight,keepaspectratio]{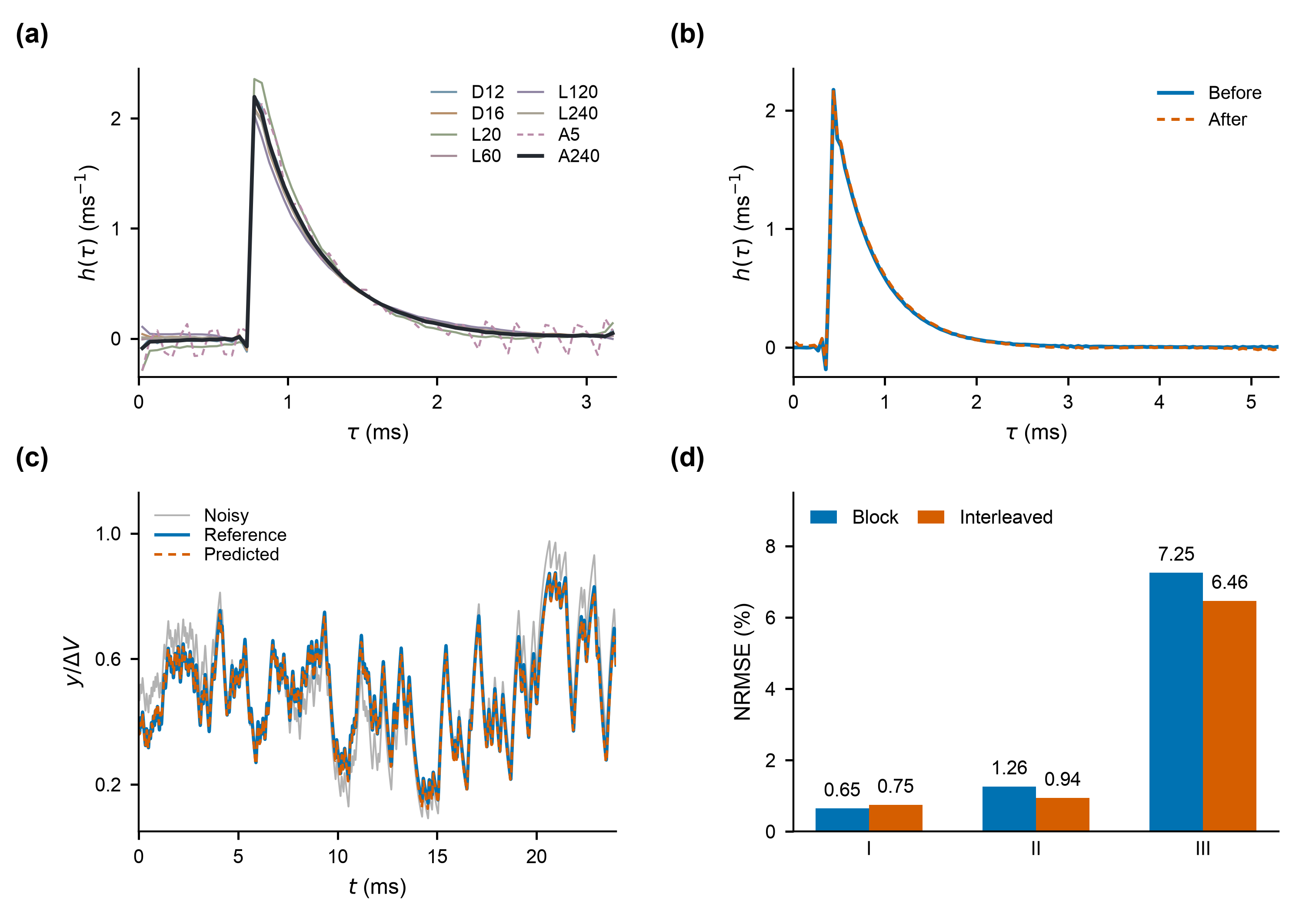}
\caption{Identification of the underlying temporal response under substantial low-frequency disturbances. (a) Normalized piecewise-constant responses $h(\tau)$ identified across distinct 350~Hz calibration records. (b) Responses calibrated before and after adding a 7\% RMS low-frequency disturbance to a $350~\mathrm{Hz}$-filtered reference record. (c) Disturbed observations, the pre-addition reference, and waveform prediction from the disturbed-training model over an independent Block holdout segment. (d) Block and Interleaved holdout NRMSE under condition I (pre-addition model versus pre-addition reference), condition II (disturbed-training model versus pre-addition reference), and condition III (disturbed-training model versus disturbed observations).}
\label{fig:supp_noisy_identification}
\end{figure}
\FloatBarrier

Evaluating predictions against both the pre-addition and disturbed holdout records in Fig.~S7(c) and (d) distinguishes calibration robustness from disturbance-dominated waveform residuals: in condition I, the pre-addition model predicts the pre-addition reference with Block and Interleaved NRMSE of 0.645\% and 0.747\%; in condition II, the model calibrated from disturbed training data predicts that same pre-addition reference with only 1.264\% and 0.942\% NRMSE; whereas in condition III, evaluating that identical disturbed-training model against the disturbed holdout observations yields 7.253\% and 6.465\% NRMSE, consistent with the scale of the injected 7\% RMS disturbance. These results show that multi-round reference fitting can retain an accurate effective response $\widehat{\mathbf{h}}$ under the tested low-frequency disturbance, even when raw holdout NRMSE is dominated by observation-level fluctuations that composite readouts subsequently suppress during image recovery in Sections~S2.2 and S2.4. Detailed records are provided in \mbox{Supplementary Data~1}.

\subsection{Imaging validation across diverse targets using transferred responses}
\label{sec:supp_s3_7}

Sections~S3.5 and S3.6 demonstrate that raw reference-waveform NRMSE reflects both response-model fidelity and observation-level fluctuations. Consistent with the target-centered paradigm established in Section~S2.3, the decisive test of a calibrated response $\widehat{\mathbf{h}}$ is whether its matched operator $\widetilde{\mathbf{A}}=\mathbf{R}\mathbf{W}\mathbf{P}$ supports high-fidelity reconstruction of independent scenes. Extending the baseline reconstructions in Figs.~5(a), 5(b), and 7(a) of the main text, Fig.~S8 evaluates 60 Joint reconstructions using baseline and transferred responses.

Five diverse targets, namely the character, printed target, ring, figurine, and composite geometric object, are imaged across the 350~Hz, 1.8~kHz, and 18~kHz receiving chains. For every target and chain, Joint recovery is performed using the baseline calibration and three transferred source responses: A60, D12, and L20 at 350~Hz; A240, A60, and D12 at 1.8~kHz; and L240, L60, and D12 at 18~kHz. Within each four-way comparison, the raw imaging record, valid-sample selection, readout operator $\mathbf{R}$, regularization weights $(\lambda,\beta)$, and grayscale display range remain strictly identical, updating only $\mathbf{W}$ from the substituted $\widehat{\mathbf{h}}$ and the imaging record's registered timing $\widehat{q}$ according to Sections~S1.3 and S1.4.

\begin{figure}[!htb]
\centering
\includegraphics[width=\linewidth,height=0.41\textheight,keepaspectratio]{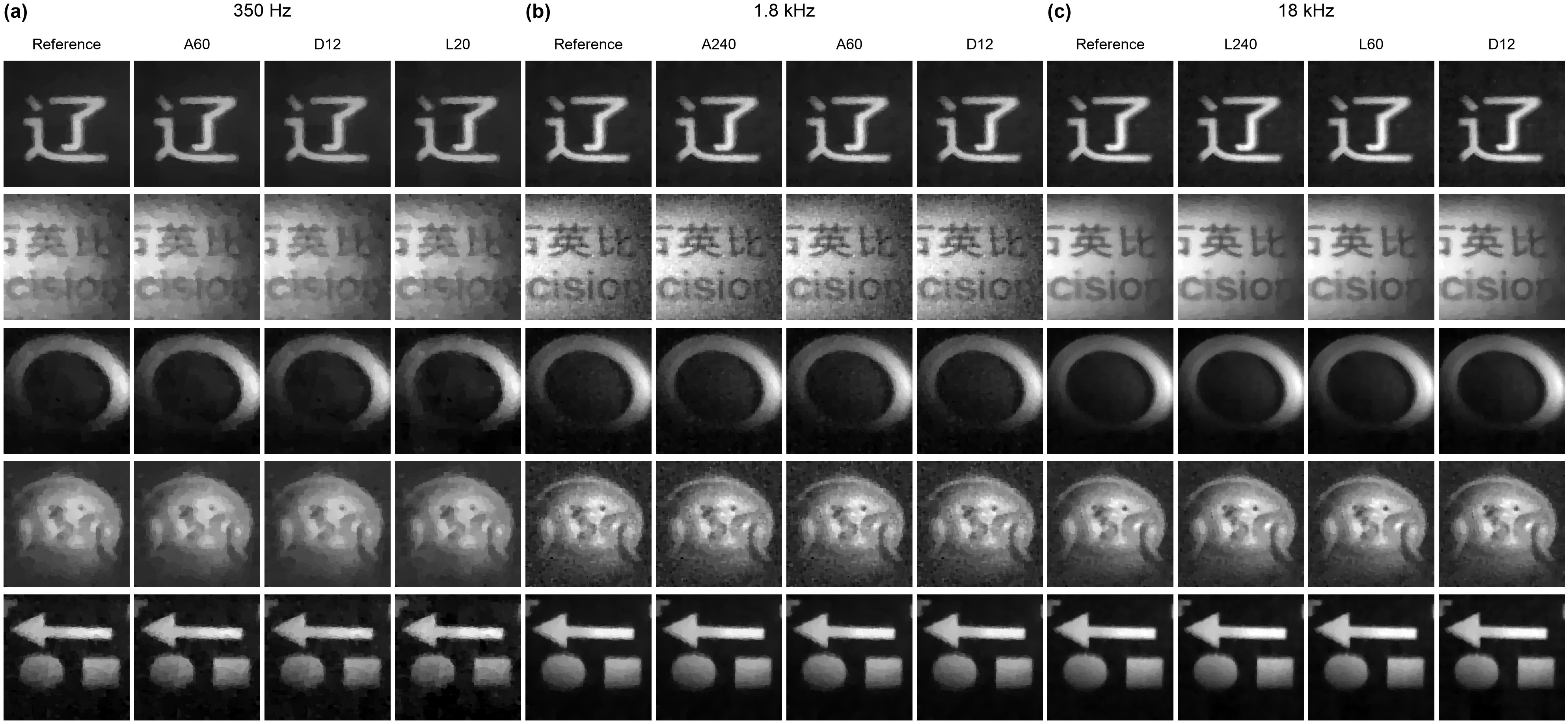}
\caption{Joint reconstructions of diverse targets using baseline and transferred calibration responses across 60 test cases. (a) 350~Hz receiving chain using the baseline reference, A60, D12, and L20 responses. (b) 1.8~kHz chain using the baseline reference, A240, A60, and D12 responses. (c) 18~kHz chain using the baseline reference, L240, L60, and D12 responses. Each panel presents the character, printed target, ring, figurine, and composite geometric target. Within each target and chain, imaging observations, composite readouts, regularization parameters, and display ranges are held fixed while updating only $\widehat{\mathbf{h}}$ and the resulting history-weight matrix $\mathbf{W}$. Record codes follow Fig.~S4.}
\label{fig:supp_imaging_transfer}
\end{figure}
\FloatBarrier

Across the 350~Hz substitutions in Fig.~S8(a), A60, D12, and crucially L20 all recover character strokes, fine printed text, ring contours, figurine shading, and sharp geometric facets that are visually indistinguishable from the baseline reconstructions, despite L20 exhibiting 16.42\%--19.05\% receiving waveform NRMSE in Fig.~S5. Likewise, substituting responses calibrated across varied ADC rates, DMD rates, and illumination levels on the 1.8~kHz and 18~kHz chains in Fig.~S8(b) and (c) yields virtually identical reconstructions across all five targets.

Overall, Section~S3 establishes an end-to-end validation chain from joint response--timing identification and phase-window diagnostics to hardware adaptation, multi-week reuse, and disturbance-resilient operator recovery. Confirming Section~S2.3 experimentally, practical image recovery is governed by accurate identification of the effective historical operator $\widetilde{\mathbf{A}}$ paired with disturbance-rejecting composite readouts rather than raw waveform error on fluctuating references, providing a solid forward-model foundation for the co-design studies in Section~S4. Detailed settings and numerical records are provided in \mbox{Supplementary Data~1}.

%% file: section_s4.tex
\section{Static image recovery and joint measurement design under temporal mixing}
\label{sec:supp_s4}

Sections~S1 and S2 established the calibrated complete-history measurement relation and its effective spatial constraints, while Section~S3 validated response identification, timing consistency, transfer, and reconstruction on independently measured targets. Building on that validation, this section examines temporal mixing as a controllable imaging resource. Measured reconstructions and controlled simulations investigate how modulation rate, temporal filtering, pattern ordering, and sampling timing interact with composite readout to shape recovery performance. The comparisons complement the common-target recovery mechanism in Section~2.2 of the main text and the imaging results in Sections~3.2--3.4.

\subsection{Reconstruction settings and measured-waveform context}
\label{sec:supp_s4_1}

This section summarizes the common reconstruction settings and parameter-selection protocol for static imaging in the main text and Supplementary Information. Spatial encoding, pattern ordering, response identification, timing registration, observation selection, and matched historical-operator construction follow the preceding sections. Base, Inter, and Joint use the reconstruction definitions and TV objectives in Section~3.2 of the main text; their matched inter- and intra-slot readouts are described in Section~S1.4, and their complementary spatial constraints in Sections~S2.2 and S2.3.

All static TV reconstructions~\cite{ref6} use predefined parameter libraries, with every candidate evaluated using a fixed budget of 5000 iterations. Simulations use 48 TV coefficients and, where applicable, 20 Joint weights; the parameters maximizing ground-truth PSNR are selected from the corresponding library using unclipped floating-point reconstructions. Measured reconstructions use 28 TV coefficients and 16 Joint weights, selected by target structure, detail preservation, and reconstruction artifacts. Methods without an intra-slot branch search only their applicable parameters. Response-substitution and timing-mismatch imaging comparisons keep the selected reconstruction parameters fixed while updating the matched measurement operator. As in Section~S3.1, the 350~Hz and 1.8~kHz measured receiving chains retain all intra-slot samples, whereas the 18~kHz chain excludes the first 5~$\mu$s and final 10~$\mu$s of each pattern slot.

Figure~S9 provides the measured-waveform context for the static imaging experiments. The upper row presents character-target ADC records from the 350~Hz, 1.8~kHz, and 18~kHz receiving chains, and the lower row presents the corresponding printed-target records. Both targets exhibit prominent, approximately periodic amplitude variations in the 350~Hz configuration, smaller variations in the 1.8~kHz configuration, and comparatively concentrated signal levels in the 18~kHz configuration. Unlike the fixed-state fluctuations examined in Section~S3.6 and Fig.~S6, these imaging records also contain the pattern-dependent target projections and their temporally mixed histories. They therefore provide the observed signal context for the subsequent comparisons of filtering, pattern ordering, and joint recovery. Complete numerical and processing settings for this subsection are given in Supplementary Data~1.

\begin{figure}[!htb]
\centering
\includegraphics[width=\linewidth,height=0.57\textheight,keepaspectratio]{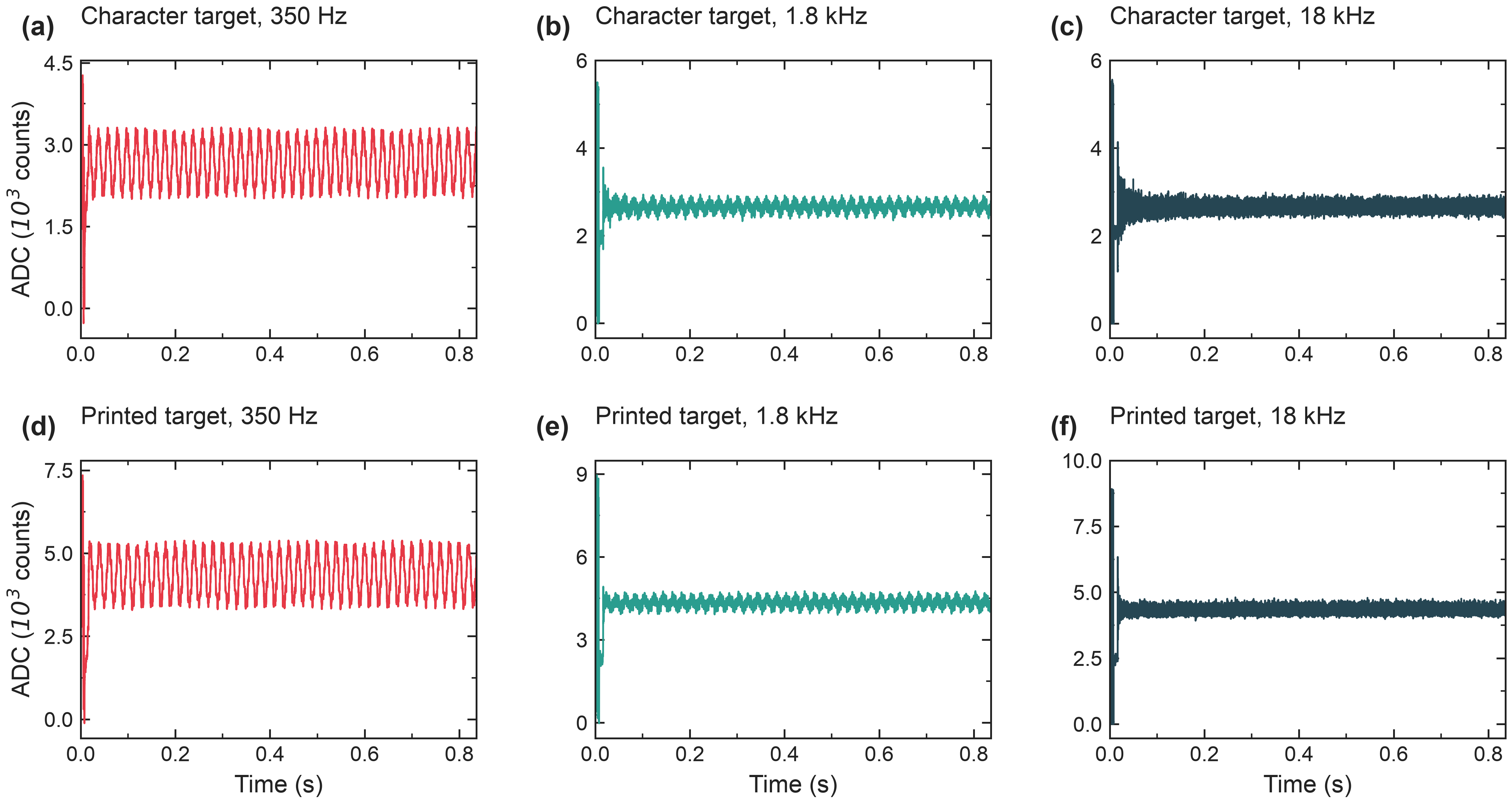}
\caption{\textbf{Measured ADC waveforms during spatially encoded imaging.}
(a--c) Character-target records from the 350~Hz, 1.8~kHz, and 18~kHz receiving chains, respectively.
(d--f) Corresponding printed-target records.
The signals include pattern-dependent target projections, temporal-response effects, and measurement disturbances.}
\label{fig:supp_imaging_waveforms}
\end{figure}

\subsection{Beneficial temporal mixing under faster modulation and stronger temporal filtering}
\label{sec:supp_s4_2}

The measurements in Section~3.2 of the main text show that complete-history recovery can remain effective when the pattern modulation rate exceeds the receiving bandwidth. Figure~S10(a) extends this comparison across DMD rates from 2 to 20~kHz in increments of 2~kHz using the 1.8~kHz receiving chain. The same C44 spatial encoding directions and ADC sampling configuration are used across the rate sweep; historical operators are constructed from the calibrated response and the registered pattern timing of each acquisition. As the DMD rate increases, Base progressively loses character strokes and introduces substantial background artifacts. Inter, which accounts for the complete pattern history, retains the main structure even under stronger mixing. Joint additionally combines intra-slot information with the inter-slot observations, producing more continuous strokes and a cleaner background over a range of modulation rates.

\begin{figure}[!htb]
\centering
\includegraphics[width=\linewidth,height=0.57\textheight,keepaspectratio]{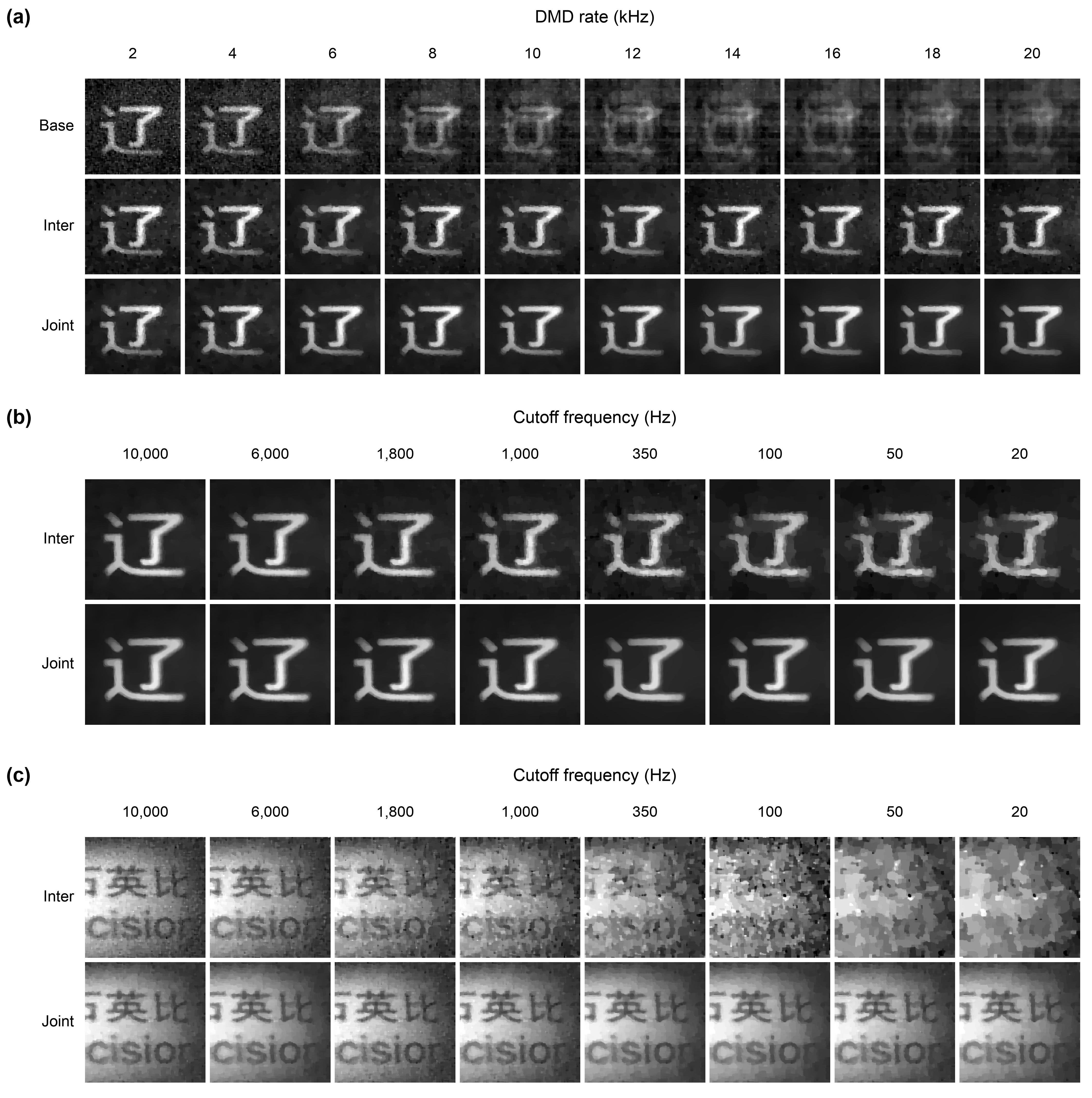}
\caption{Measured reconstruction under faster modulation and stronger temporal filtering. (a) Base, Inter, and Joint character reconstructions from the 1.8~kHz receiving chain as the DMD modulation rate increases from 2 to 20~kHz. (b) Character and (c) printed-target reconstructions using Inter and Joint after causal first-order digital low-pass filtering of measured 18~kHz receiving-chain records. Filter cutoff frequencies are 10,000, 6,000, 1,800, 1,000, 350, 100, 50, and 20~Hz.}
\label{fig:supp_mixing_filtering}
\end{figure}

An important feature of Fig.~S10(a) is that Joint does not necessarily give its best reconstruction at the slowest modulation rates. In several higher-rate conditions, where the response spans a greater number of physical pattern slots, its recovery is better than at lower rates. The phase-dependent effective patterns in Section~S1.4, particularly Eq.~(S9), explain how the calibrated history converts ongoing pattern transitions into different within-slot spatial sensitivities. Combining these sensitivities with the inter-slot observations can improve the common-target constraints described by Eq.~(S14) in Section~S2.2. Thus, increasing the degree of temporal mixing can improve reconstruction when the resulting effective patterns and composite readouts provide more useful constraints, rather than merely increasing distortion.

Figure~S10(b) and (c) examine a complementary way of strengthening the temporal memory in the recorded data. Character and printed-target records acquired using the 18~kHz receiving chain are processed using causal first-order digital low-pass filters at cutoff frequencies of 10,000, 6,000, 1,800, 1,000, 350, 100, 50, and 20~Hz. The filtering is applied to the continuous imaging records, with matched processing used to construct the corresponding observation operators. As the cutoff frequency decreases, Inter develops stronger background artifacts and loses fine structure. Joint maintains recognizable character strokes and principal printed features even at 20~Hz, demonstrating the contribution of intra-slot measurements when the filtered temporal history is modeled consistently.

The rate sweep and filtering comparison probe different controls over the same complete-history observation relation. Increasing the DMD rate changes the history covered by each physical slot, while digital filtering reshapes the target and disturbance components already present in the acquired record. The latter is an acquisition-side processing operation after ADC conversion, rather than a change in the physical pre-ADC receiving bandwidth. Together with the filtering analysis in Section~S2.4 and the composite-response calibration in Section~S3.4, these observations establish modulation rate, temporal response, and readout choice as coupled variables in target recovery. The complete experimental settings and reconstruction parameters are listed in Supplementary Data~1.

\subsection{Pattern ordering as a design variable for temporal encoding and switching dynamics}
\label{sec:supp_s4_3}

Section~2.2 of the main text and Section~S2.1 establish that temporal pattern ordering changes the historical spatial encoding retained by complementary readout, even when the physical pattern set and digital sampling budget are fixed. We compare C11, in which each positive pattern is immediately followed by its corresponding negative pattern, with C44, in which four positive patterns from distinct encoding directions are followed by their four corresponding negative patterns. Their different temporal separations alter the mixed effective patterns and the response of complementary differencing to disturbances.

Figure~S11(a) compares C11 and C44 using the Peppers target, a first-order 1.8~kHz receiving response, and the same spatial encoding directions and sampling budget. The added disturbance combines a low-frequency component obtained by Gaussian temporal smoothing with a 2~ms standard deviation and a white-noise component. Noise is added after the receiving response, and its total RMS is fixed at 0.3\% of the full-white response while the white-noise \emph{power fraction} (the white-noise variance divided by the sum of the white- and low-frequency noise variances) is varied from 0.1\% to 20\%. This isolates the influence of ordering and disturbance composition without changing the response-dependent transmission of the injected noise. At a 0.1\% white-noise power fraction, Joint reaches 30.06~dB for C11 and 29.70~dB for C44, with C11 slightly higher. At 20\%, the corresponding PSNR values are 21.50 and 23.92~dB, favoring C44. Both Inter and Joint display ordering-dependent trends, consistent with the different historical difference coefficients and slow-disturbance responses derived in Sections~S2.1 and S2.2.

\begin{figure}[!htb]
\centering
\includegraphics[width=\linewidth,height=0.57\textheight,keepaspectratio]{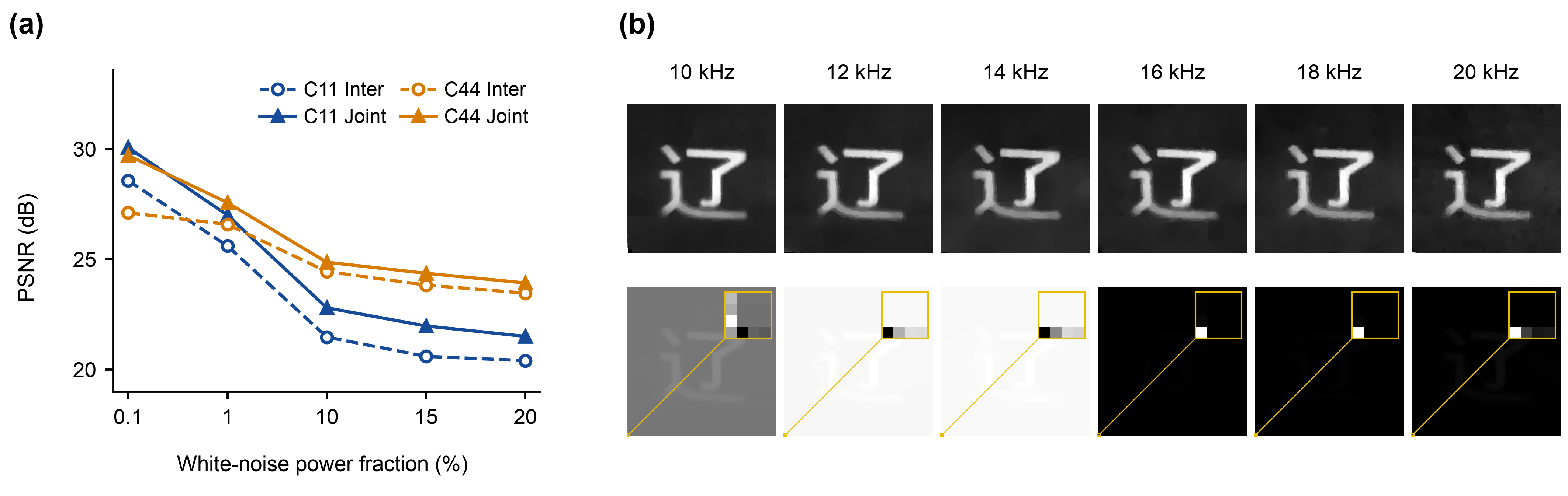}
\caption{Pattern ordering, disturbance composition, and C11 reconstruction anomalies. (a) PSNR of C11 and C44 with Inter and Joint as the white-noise power fraction varies; the total disturbance RMS after the receiving response is fixed at 0.3\% of the full-white response. (b) Measured C11 Joint reconstructions at DMD rates of 10, 12, 14, 16, 18, and 20~kHz. Upper images use a fixed reference grayscale, and lower images show the full range of the corresponding reconstruction arrays. Insets magnify the corner region in each lower image.}
\label{fig:supp_pattern_order}
\end{figure}

Measured records reveal an additional ordering-dependent effect. Figure~S11(b) shows C11 Joint reconstructions from the 1.8~kHz receiving chain at DMD rates of 10, 12, 14, 16, 18, and 20~kHz. The principal character remains recognizable, but the reconstruction arrays exhibit anomalous values in a corner region. The fixed reference grayscale in the upper row preserves visibility of the character, whereas displaying the full value range of the same arrays in the lower row reveals how the corner extrema compress the apparent target contrast. The magnified insets show that the anomalous values remain in the reconstruction, rather than being removed by the display setting.

The switching sequence provides a plausible source of ordering-dependent periodic components. In C11, a transition from a positive pattern to its complementary negative pattern reverses all encoded micromirrors, while the subsequent transition to the next positive direction typically reverses approximately half. The alternating switching load has a two-slot period. C44 instead predominantly places transitions between different directions or corresponding groups, giving a different switching sequence. As discussed in Section~3.1 of the main text and Section~S3.2, finite receiving bandwidth can spread switching-related deviations beyond the immediate pattern boundary and into otherwise retained measurements. Their relationship with the historical coding therefore depends on both ordering and receiving response.

To characterize the recorded periodic components, Fig.~S12 presents measured spectra for both arrangements at DMD rates of 10, 12, 14, 16, 18, and 20~kHz. The left and right columns correspond to C11 and C44, respectively. For each record, the power spectral density (PSD) is estimated from a continuous, equally spaced ADC segment using a mean-subtracted, Hann-windowed periodogram~\cite{ref7} and normalized by the squared measured black--white reference span; the result is expressed in dB/Hz. The dotted and dashed markers indicate half and full DMD modulation frequencies. C11 exhibits a prominent spectral component near half the DMD rate that shifts with modulation rate, whereas the corresponding component is substantially weaker in C44. This behavior is consistent with their different periodic switching loads. The spectra characterize the combined recorded signal and do not independently isolate one physical source of the corner-region anomaly.

\begin{figure}[!htb]
\centering
\includegraphics[width=\linewidth,height=0.57\textheight,keepaspectratio]{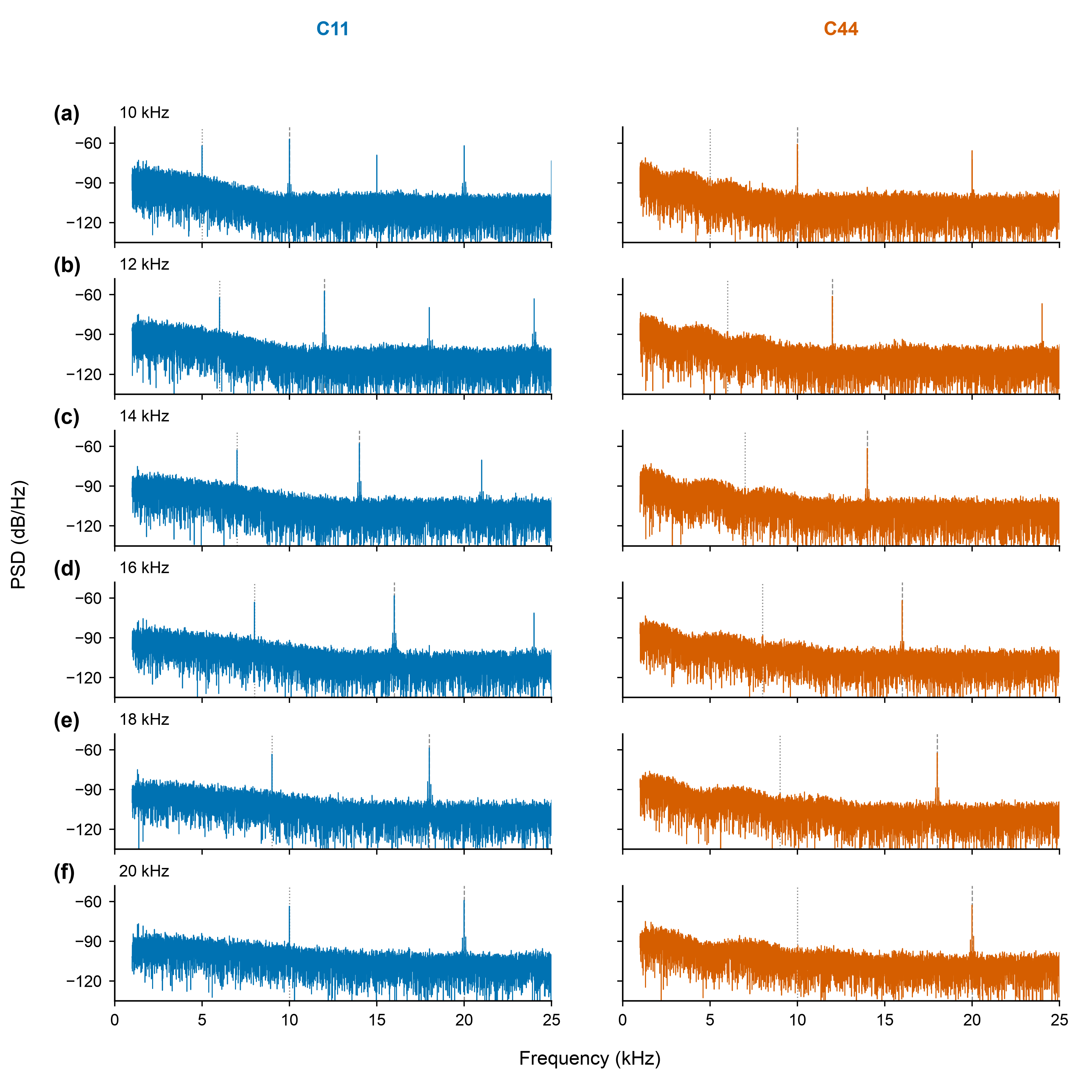}
\caption{Measured power spectral density for C11 (left column) and C44 (right column) at DMD modulation rates of (a) 10, (b) 12, (c) 14, (d) 16, (e) 18, and (f) 20~kHz. Spectra are estimated from continuous ADC records and normalized by the squared black--white reference span. Vertical markers indicate half and full modulation frequencies.}
\label{fig:supp_order_spectra}
\end{figure}

Together, Figs.~S11 and S12 show that pattern ordering changes both the retained historical spatial encoding and the periodic structure of the recorded signals. These results extend the ordering mechanism in Section~S2.1 to practical acquisition conditions and support the coordinated design of pattern ordering, receiving response, and composite readout. Simulation and imaging reconstruction settings for this subsection are provided in Supplementary Data~1.

\subsection{Temporal mixing as an imaging resource: comparison with conventional pattern-wise readout}
\label{sec:supp_s4_4}

The preceding measurements show that modulation and filtering can alter the usefulness of temporal mixing for joint recovery. We next compare matched complete-history reconstruction against conventional pattern-wise sampling under controlled upstream disturbances, extending the noise-transfer and effective-constraint analyses in Sections~S2.2 and S2.4 and the filtering experiments in Section~3.3 of the main text. Figure~S13 evaluates Cameraman and Peppers across first-order receiving responses with bandwidths of 100, 50, 18, 10, 1.8, and 0.35~kHz under an identical digital observation budget. To hold the physical pattern sequence strictly fixed as a controlled variable while preserving the standard adjacent-pair differential baseline for conventional pattern-wise readout, all conventional and Joint reconstructions use the same C11 encoding sequence.

\begin{figure}[!htb]
\centering
\includegraphics[width=\linewidth,height=0.57\textheight,keepaspectratio]{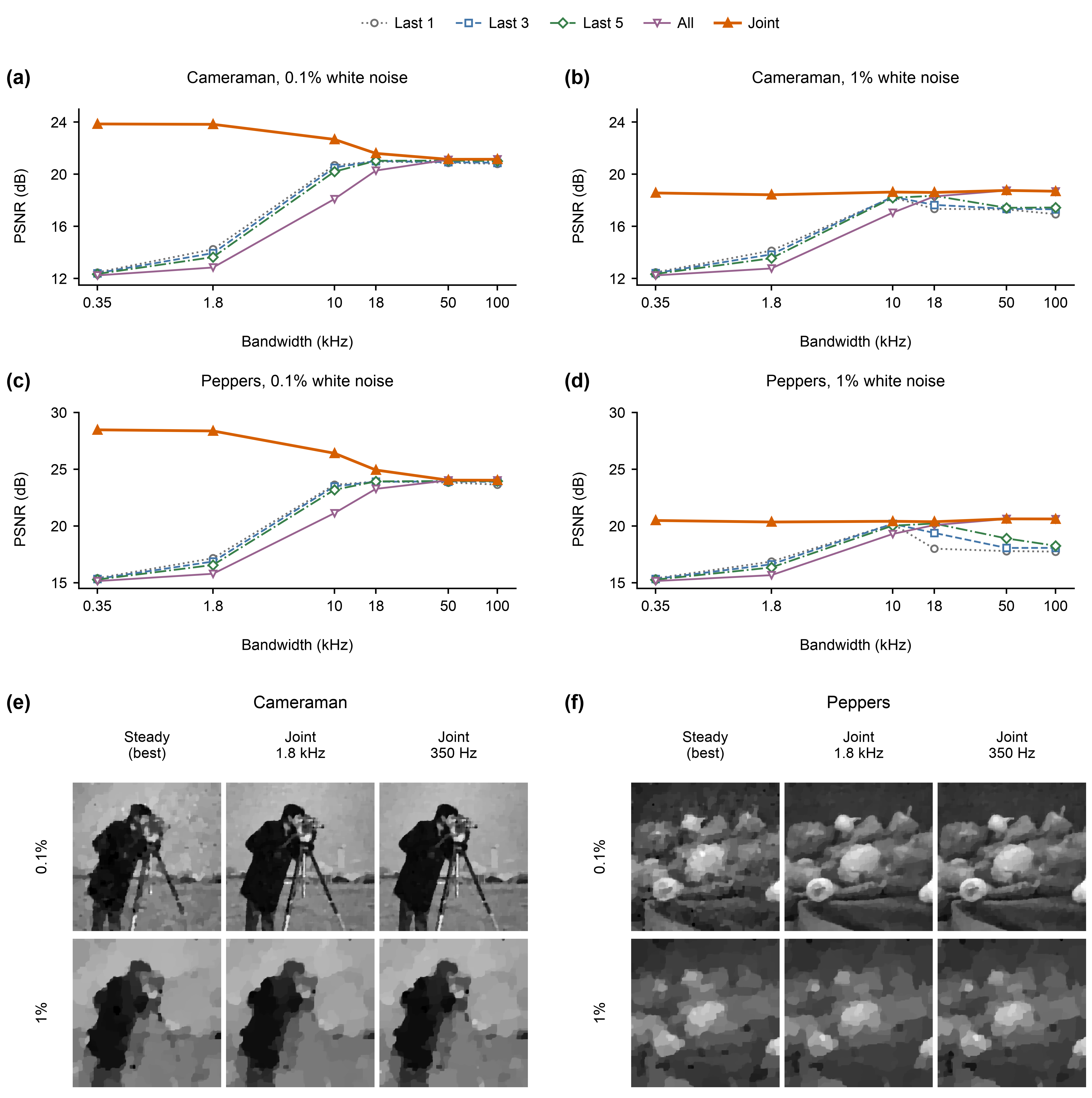}
\caption{Conventional pattern-wise readout and matched Joint reconstruction across receiving bandwidths under upstream disturbances. (a) Cameraman with 0.1\% white-noise RMS; (b) Cameraman with 1\%; (c) Peppers with 0.1\%; and (d) Peppers with 1\%. Last 1, Last 3, Last 5, and All use the indicated number of permitted samples near the end of each physical pattern slot. (e) Cameraman and (f) Peppers representative reconstructions compare the best conventional configuration with Joint at 1.8~kHz and 350~Hz; upper and lower rows correspond to 0.1\% and 1\% white-noise RMS. Each condition includes 10\% RMS low-frequency disturbance injected before the receiving response.}
\label{fig:supp_steady_comparison}
\end{figure}

Motivated by the pronounced low-frequency fluctuations measured in the 350~Hz receiving chain in Section~S3.6 and Fig.~S6, the simulations use a controlled disturbance environment dominated by a low-frequency component. Before the receiving response, a 10\% RMS low-frequency disturbance and white noise of either 0.1\% or 1\% RMS are added relative to the full-white response. The low-frequency component is formed by Gaussian temporal smoothing with a 2~ms standard deviation. Each bandwidth is tested with the same input disturbance realization, allowing the receiving response to modify both the historical target coding and the transmission of upstream noise. This differs from the post-response disturbance injection in Section~S4.3.

Four conventional pattern-wise readouts use the last one, last three, last five, or all ten permitted ADC samples from each physical pattern slot. For this simulation, samples in the first 10~$\mu$s of each slot are excluded. Each conventional readout forms a complementary difference and reconstructs using the ideal single-pattern spatial model, with its own TV parameter selection. Joint processes the same available record using the matched full-history operator and combined inter- and intra-slot readouts. As the receiver bandwidth decreases, the conventional pattern-wise assumption becomes progressively less consistent with the recorded mixed observations, whereas Joint uses the historical contributions within those samples to recover the common target.

As shown in Fig.~S13(a) and (c), at 0.1\% white-noise RMS, the 350~Hz Joint reconstruction reaches 23.86~dB for Cameraman and 28.47~dB for Peppers. The best conventional readouts at that same bandwidth reach 12.47 and 15.39~dB, respectively. More importantly, the 350~Hz Joint results also exceed the best conventional configurations selected across all six bandwidths and four sampling windows: these achieve 21.15 and 24.02~dB at 100~kHz. The corresponding gains are 2.71 and 4.44~dB. Figure~S13(e) and (f) show the associated improvements in structure, detail, and background quality for representative reconstructions.

At 1\% white-noise RMS, Fig.~S13(b) and (d) show that the 350~Hz Joint reconstructions achieve 18.56 and 20.48~dB, remaining substantially better than conventional readout at the same low bandwidth. Their performance is close to, but slightly below, the best conventional configurations across all tested bandwidths, which reach 18.75 and 20.62~dB at 50~kHz. Thus, disturbance composition affects the balance between noise attenuation and retention of target sensitivity. The low-bandwidth response reduces part of the upstream disturbance while increasing overlap between successive physical patterns; the calibrated complete-history model and complementary composite readout determine whether the resulting measurements remain useful for image recovery.

These results demonstrate that, under the tested upstream-noise conditions, strong calibrated mixing can outperform conventional pattern-wise readout even when the latter is allowed to use a more favorable receiver bandwidth. The useful regime is governed jointly by target sensitivity in the effective operator and disturbance transmission through the receiving response and readout, as discussed in Sections~S2.2 and S2.4. Complete noise-generation, sampling-window, and reconstruction settings are given in Supplementary Data 1.

\subsection{Historical encoding and sampling-timing design under limited ADC acquisition}
\label{sec:supp_s4_5}

The low-budget calibration controls in Section~S3.2 and the timing-registration results in Section~S3.3 establish the prerequisites for using calibrated temporal memory at ADC rates below the DMD modulation rate. The imaging experiments in Section~3.4 of the main text show that such acquisition can recover recognizable targets. Here, Fig.~S14 first examines direct common-target reconstruction from mixed measurements and then uses noiseless simulations to identify the contribution of historical encoding apart from disturbance suppression. This tests the spatial-history mechanism developed in Section~2.2 of the main text and Section~S2.5, particularly Eq.~(S19).

\begin{figure}[!htb]
\centering
\includegraphics[width=\linewidth,height=0.57\textheight,keepaspectratio]{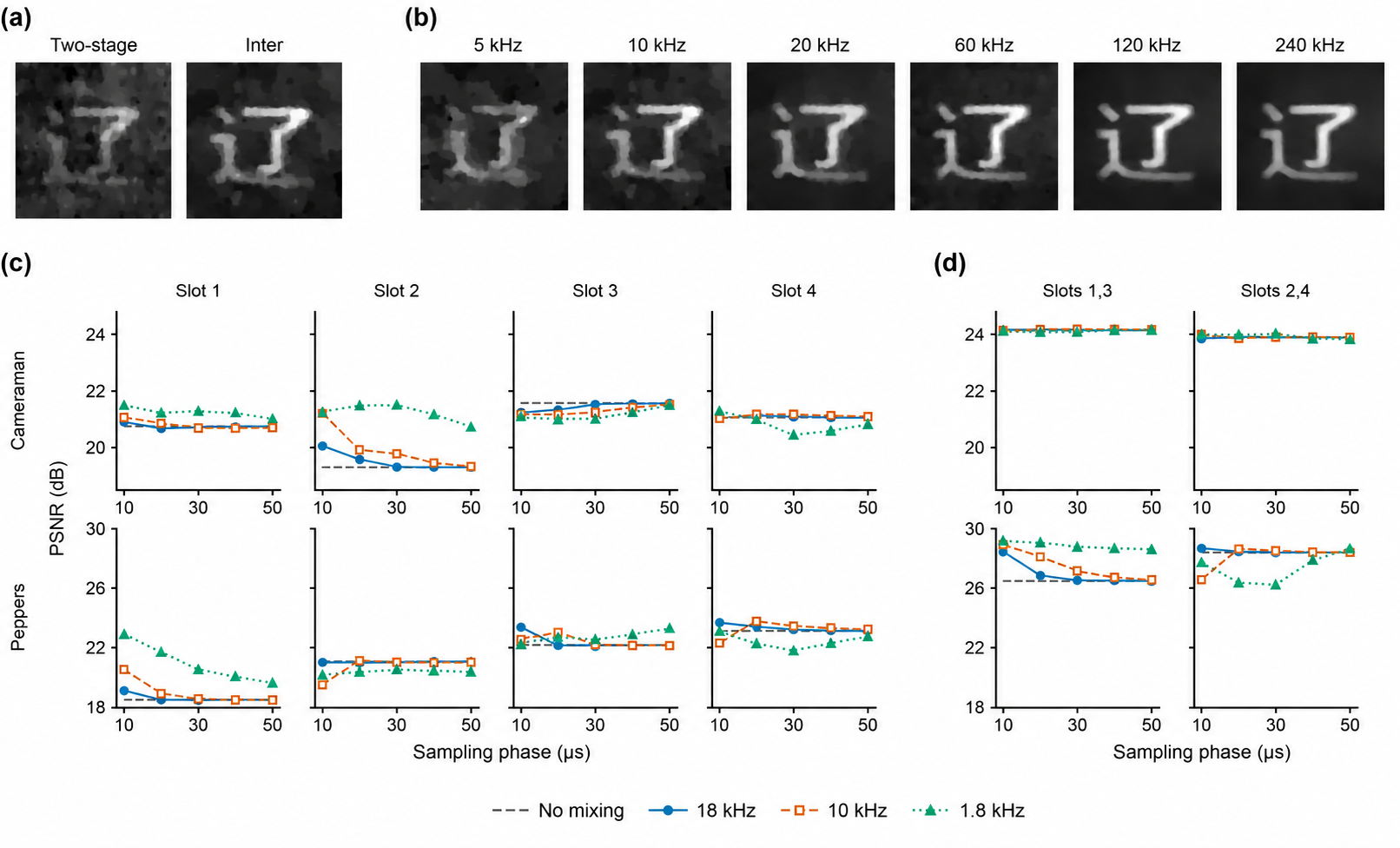}
\caption{Direct common-target recovery and historical encoding under limited ADC acquisition. (a) Two-stage pattern-projection recovery versus direct Inter reconstruction from the same 10~kHz ADC record. (b) Measured character reconstructions at ADC rates of 5, 10, 20, 60, 120, and 240~kHz; Inter is used for the first three and Joint for the latter three. (c) Noiseless PSNR comparison between memoryless sensing and first-order temporal responses at 5~kHz ADC for four sampled-slot positions and five sampling phases. (d) Corresponding comparison at 10~kHz ADC for the two sampled-slot pairs. Rows of (c) and (d) show Cameraman and Peppers; dashed lines indicate the memoryless references.}
\label{fig:supp_low_adc_history}
\end{figure}

Figure~S14(a) compares two recovery procedures on the same measured character-target record from the 1.8~kHz receiving chain at an ADC rate of 10~kHz. In the two-stage procedure, the matched temporal-history matrix is first inverted by a minimum-norm fit to estimate separate projections for the physical pattern slots. Complementary differences of these estimated projections are then used for TV image recovery. Inter instead combines the temporal history and spatial pattern sequence in one matched operator and directly reconstructs their common target. The same ADC observations are used by both procedures. Inter retains clearer strokes and a more coherent background, illustrating the advantage of imposing the shared spatial target constraint without first estimating an underdetermined set of individual pattern projections, as described in Section~S2.3.

Figure~S14(b) shows further measured character reconstructions using the same 1.8~kHz receiving chain at ADC rates of 5, 10, 20, 60, 120, and 240~kHz. Inter is used at 5, 10, and 20~kHz, and Joint at 60, 120, and 240~kHz. Even at 5~kHz, when several consecutive DMD patterns can pass without a direct ADC sample, the principal character remains recognizable. These acquisitions contain actual observation disturbances and ADC--DMD timing variation, with the sampling-to-pattern correspondence maintained using the procedures in Sections~S1.2 and S3.3. They demonstrate practical low-ADC recovery, while the following simulations isolate the historical-encoding mechanism at a fixed digital observation budget.

Figures~S14(c) and (d) compare ideal memoryless sensing against calibrated first-order temporal responses with bandwidths of 18, 10, and 1.8~kHz in a noiseless setting. Both Cameraman and Peppers use the same C44 physical pattern sequence, total acquisition duration, and digital observation count. At 5~kHz, one ADC sample is retained per four physical slots, with the sampled slot varied among the four positions. At 10~kHz, one sample is retained per two slots, comparing positions 1 and 3 against positions 2 and 4. Within each choice, sampling phases of 10, 20, 30, 40, and 50~$\mu$s are tested. The prescribed phases are all retained in this noiseless control, independently of the observation windows used for measured imaging. TV parameters are selected under the common simulation protocol for each configuration.

Two comparisons distinguish the influence of temporal mixing from sampling-position selection. First, with the ADC rate, sampled slots, and phases fixed identically, mixed measurements can substantially outperform the memoryless case. For Peppers at 5~kHz using slot 1 and a 10~$\mu$s phase, PSNR rises from 18.51 to 22.89~dB with the 1.8~kHz response, a gain of 4.38~dB. At 10~kHz using slots 1 and 3 and the same phase, the corresponding values are 26.48 and 29.17~dB, a gain of 2.69~dB. Cameraman at 5~kHz using slot 2 and a 30~$\mu$s phase similarly increases from 19.30 to 21.49~dB, a gain of 2.20~dB. Because the sampling schedule and observation count are identical in each pair, these differences demonstrate the effect of historical spatial coding within the acquired samples.

Second, when each sensing configuration is allowed to select its best slot and phase from the prescribed set, the advantages are more dependent on target and timing. For Peppers, the best memoryless results at 5 and 10~kHz are 23.10 and 28.38~dB; the best mixed configurations reach 23.78 and 29.17~dB, gains of 0.68 and 0.79~dB, respectively. For Cameraman, the corresponding best results are nearly identical: approximately 21.56~dB in both configurations at 5~kHz and 24.14 versus 24.19~dB at 10~kHz. The fixed-schedule comparison and the best-over-schedules comparison therefore answer different questions: mixing can improve a particular digital sampling design, while optimizing timing for each response can reduce or retain the advantage depending on the target. Since these controls contain no noise, the gains arise from the effective spatial constraints of the mixed observations rather than noise attenuation.

Across Sections~S4.2--S4.5, modulation rate, pattern ordering, temporal response, sampling schedule, and composite readout jointly shape the effective operator defined in Eq.~(4) of the main text. Calibrated temporal memory changes how the complete physical pattern history enters the finite ADC observations; the composite readout determines which of those target-dependent variations are used for common-target recovery. Their joint design, rather than the isolated minimization of temporal mixing, determines the achievable imaging benefit under the tested acquisition and disturbance conditions. Complete acquisition, simulation, and parameter-selection records for this subsection are provided in Supplementary Data 1.

%% file: section_s5.tex
\section{Direct spatiotemporal video recovery and real-time dynamic imaging under temporal mixing}
\label{sec:supp_s5}

Sections~S1--S4 establish and validate the calibrated complete-history framework $\widetilde{\mathbf{A}}=\mathbf{R}\mathbf{W}\mathbf{P}$ for static scenes. Extending single-pixel imaging to continuous dynamic scenes introduces a more demanding regime in which nonstationary object motion, severe per-frame spatial undersampling, and motion- or illumination-induced flux variations occur simultaneously. Complementing Section~S1.5 and Section~3.4 of the main text, this section establishes full-sampling static benchmarks for diverse targets on the dynamic receiving chain in Section~S5.1, formulates the active advantages of temporal mixing for dynamic sensing together with the direct video-recovery paradigm and causal two-frame hybrid solver in Section~S5.2, validates dual-branch and cross-frame historical constraints quantitatively on held-out real dynamic observations in Section~S5.3, and presents real-time runtime benchmarks and extended multi-scenario video reconstructions in Section~S5.4.

\subsection{Full-sampling benchmarks for diverse targets on the dynamic receiving chain}
\label{sec:supp_s5_1}

Prior to introducing severe per-frame undersampling and continuous motion, Fig.~S15 establishes the baseline spatial fidelity of the $1.8~\mathrm{kHz}$ receiving chain at $64\times 64$ spatial resolution ($N=4,096$) under $20~\mathrm{kHz}$ DMD modulation and $240~\mathrm{kHz}$ ADC sampling. Each acquisition probes all 4,095 non-DC Walsh--Hadamard directions via 8,190 complementary binary masks arranged in grouped C44 order and reconstructs the target using Joint under the variational protocol of Section~S4.1.

\begin{figure}[!htb]
\centering
\includegraphics[width=\linewidth,height=0.57\textheight,keepaspectratio]{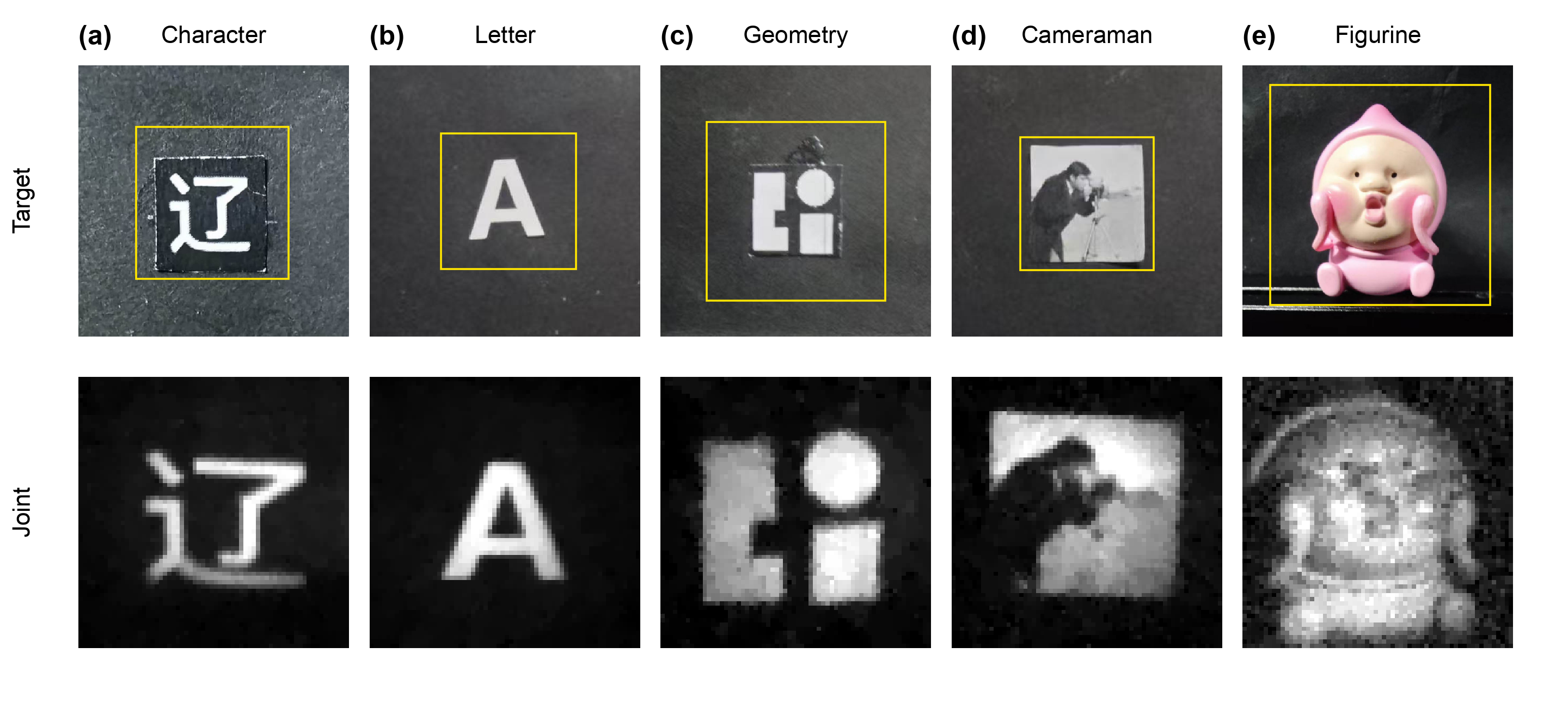}
\caption{Full-sampling $64\times64$ Joint benchmarks for diverse targets on the 1.8~kHz receiving chain at $20~\mathrm{kHz}$ DMD modulation and $240~\mathrm{kHz}$ ADC sampling. The upper row shows target photographs with yellow boxes marking the imaged regions, and the lower row shows Joint reconstructions across all 4,095 non-DC Hadamard directions for (a) Character, (b) Letter, (c) Geometry, (d) Cameraman, and (e) Figurine.}
\label{fig:supp_64_benchmarks}
\end{figure}

Across the five targets spanning diverse spatial structures and reflectance profiles in Fig.~S15(a)--(e), namely the character, letter, geometry, Cameraman, and figurine targets, Joint reconstructions cleanly resolve fine stroke intersections, sharp polygonal and circular boundaries, continuous photographic grayscale gradients, and curved diffuse shading despite modulating more than an order of magnitude above the $1.8~\mathrm{kHz}$ receiver bandwidth. These full-sampling reconstructions confirm that calibrated complete-history Joint readout preserves high spatial fidelity on the 1.8 kHz hardware platform across diverse targets, providing a static reference benchmark for the dynamic experiments that follow.

\subsection{Temporal mixing advantages for dynamic sensing and direct video-sequence recovery}
\label{sec:supp_s5_2}

\paragraph{Active advantages of temporal mixing in dynamic imaging.}

Conventional dynamic single-pixel imaging faces an acute hardware conflict between per-slot steady-state settling and frame-rate temporal resolution. Enforcing per-slot settling on a high-sensitivity, narrowband photodetector forces slow pattern modulation, stretching the frame duration $T_{\mathrm{frame}}$ and introducing severe intra-frame motion blur; conversely, accelerating modulation beyond the receiver bandwidth under a memoryless frame-wise model corrupts reconstructions through both intra-frame inter-pattern crosstalk and persistent cross-frame tails at frame boundaries. Embedding calibrated electro-optical memory into the dynamic forward model of Section~S1.5 resolves this conflict and converts temporal mixing into three constructive advantages for dynamic sensing.

First, removing the per-slot settling requirement decouples modulation speed from receiver bandwidth, allowing our cost-effective $1.8~\mathrm{kHz}$ receiving chain to operate at $20~\mathrm{kHz}$ DMD modulation and compress 784 physical slots into a $39.2~\mathrm{ms}$ frame window while retaining high transimpedance gain and front-end suppression of broadband upstream noise. Second, under the discrete-frame complete-history relation in Eq.~(S11) of Section~S1.5, impulse tails from masks displayed near the end of frame $n-1$ naturally persist into early ADC readings of frame $n$. Rather than discarding boundary slots at the expense of duty cycle or misattributing cross-boundary tails to $\mathbf{x}_n$, explicit historical modeling turns cross-frame persistence into physical coupling constraints linking consecutive scene states without imposing heuristic optical-flow priors. Third, unconstrained object translation, in-plane rotation, and illumination changes continuously perturb total backscattered flux and baseline levels; combining inter-slot mean differences with zero-sum intra-slot transient slopes harvests within-slot phase evolution while rejecting slot-constant offsets and suppressing slow flux drift as analyzed in Section~S2.2.

\paragraph{Elevating direct recovery from static images to spatiotemporal video sequences.}

In high-speed dynamic regimes with aggressive per-frame spatial undersampling ($9.57\%$ at $64\times 64$ and $0.60\%$ at $256\times 256$), attempting to deconvolve recorded waveforms into unmixed steady-state slot projections prior to frame reconstruction introduces an intermediate inverse problem before video recovery. At the reconstruction level, estimating slot projections as unconstrained temporal unknowns leaves their common-scene spatial consistency unenforced and can amplify recorded-domain noise; physically, faithfully restoring the pre-response optical waveform re-introduces upstream illumination and motion-induced flux fluctuations that the low-pass receiver had attenuated, extending the waveform-fidelity conflict established in Section~S2.3 to time-varying streams. Computationally, a separate temporal inversion stage adds processing steps before spatial reconstruction, whereas direct recovery concentrates computation on the video sequence itself.

Our framework avoids this intermediate stage by elevating the direct recovery principle from a single static scene $\widehat{\mathbf{x}}$ to the spatiotemporal video sequence $\{\widehat{\mathbf{x}}_n\}$ itself, bypassing steady-state waveform restoration and isolated single-frame bucket estimation altogether. Because the calibrated response support $\tau_H=3.20~\mathrm{ms}$ on the $1.8~\mathrm{kHz}$ chain is much shorter than the $39.2~\mathrm{ms}$ frame interval, cross-frame persistence in Eq.~(S11) of Section~S1.5 couples each frame $\mathbf{x}_n\in\mathbb{R}^{N}$ exclusively with its immediate predecessor $\mathbf{x}_{n-1}\in\mathbb{R}^{N}$. Stacking consecutive scene states as $\mathbf{X}_n=[\mathbf{x}_{n-1}^{\mathsf T},\mathbf{x}_n^{\mathsf T}]^{\mathsf T}\in\mathbb{R}^{2N}$ over a two-frame window yields the coupled composite observation model
\begin{equation}
\mathbf{z}_n
=
\begin{bmatrix}
\mathbf{z}_{\mathrm{out},n}\\[3pt]
\mathbf{z}_{\mathrm{in},n}
\end{bmatrix}
=
\begin{bmatrix}
\widetilde{\mathbf{A}}_{\mathrm{out},n}^{(n-1)} & \widetilde{\mathbf{A}}_{\mathrm{out},n}^{(n)}\\[4pt]
\widetilde{\mathbf{A}}_{\mathrm{in},n}^{(n-1)} & \widetilde{\mathbf{A}}_{\mathrm{in},n}^{(n)}
\end{bmatrix}
\begin{bmatrix}
\mathbf{x}_{n-1}\\[3pt]
\mathbf{x}_n
\end{bmatrix}
+
\boldsymbol{\eta}_n
=
\widetilde{\mathbf{A}}_n\mathbf{X}_n+\boldsymbol{\eta}_n,
\end{equation}
where $\mathbf{z}_{\mathrm{out},n}$ and $\mathbf{z}_{\mathrm{in},n}$ collect the valid inter-slot mean differences and intra-slot slopes within the window up to the physical completion of frame $n$, $\widetilde{\mathbf{A}}_n$ retains the full binary pattern histories across both frames so that inter-frame scene-brightness changes are physically represented, and $\boldsymbol{\eta}_n$ aggregates measurement noise together with minor calibration and within-frame motion residuals. Formulating video sequence estimation directly against Eq.~(S20) exemplifies a broader task-direct paradigm: whenever a downstream spatiotemporal state or physical parameter would conventionally be extracted by post-processing steady-state waveform projections, embedding calibrated receiver memory into the forward operator allows the ultimate task output to be recovered directly from mixed observations, laying the foundation for the generalized extensions discussed in Section~4 of the main text.

\paragraph{Synchronization-free loop timing and causal physics-guided hybrid recovery.}

Continuous dynamic acquisitions on the $1.8~\mathrm{kHz}$ chain operate at $20~\mathrm{kHz}$ DMD modulation and $240~\mathrm{kHz}$ asynchronous ADC sampling in repeating 25-frame cycles, where each cycle begins with one 168-slot spatially uniform binary timing reference followed by 25 consecutive scene frames of 784 C44 pattern slots each (probing the 392 lowest-order non-DC Cake-cutting Hadamard directions~\cite{ref8,ref9} over a $39.2~\mathrm{ms}$ nominal frame interval). Online timing registration matches each embedded reference against a template synthesized from the frozen $1.8~\mathrm{kHz}$ response $\widehat{\mathbf{h}}$ via normalized correlation as in Section~S1.3. Qualified references causally update slot-phase alignment, whereas references obscured during object entry, exit, or transient saturation retain the preceding valid timing model through causal holdover. Valid imaging samples exclude the first $10~\mu\mathrm{s}$ of each slot and require complete lag support within known $50\%$-open window masks.

Upon reaching the physical endpoint of frame $n$, $\widetilde{\mathbf{A}}_n$ and $\mathbf{z}_n$ in Eq.~(S20) are assembled from the causally arrived ADC stream, balanced across the inter- and intra-slot branches, and spectrally normalized. To achieve real-time throughput under strong undersampling, $\widehat{\mathbf{X}}_n$ is recovered by a physics-guided hybrid neural solver~\cite{ref10} that initializes from a regularized linear physical estimate on Eq.~(S20) and alternates measurement-consistency gradient updates using $\widetilde{\mathbf{A}}_n$ with learned spatial residual corrections from a shared-weight U-Net regularizer~\cite{ref11}, emitting $\widehat{\mathbf{x}}_n$ immediately without modifying previously displayed frames. Crucially, because $\widetilde{\mathbf{A}}_n$ explicitly accounts for the calibrated impulse response $\widehat{\mathbf{h}}$ and online timing $\widehat{q}$, the network need not learn hardware-specific temporal deconvolution and acts purely as a spatial image prior. Network parameters are therefore trained offline exclusively on synthetic two-frame sequences rendered through the frozen $1.8~\mathrm{kHz}$ forward model with shared-sample simulated ADC noise passed through both $\mathbf{R}_{\mathrm{out}}$ and $\mathbf{R}_{\mathrm{in}}$ to preserve the cross-branch noise covariance of Eq.~(S16), requiring zero experimental ground-truth video recordings. Dynamic acquisition and timing settings are summarized in Supplementary Data~1.

\subsection{Quantitative validation on held-out dynamic observations}
\label{sec:supp_s5_3}

Because freely moving hand-held objects lack pixelwise ground-truth video references, we quantitatively evaluate whether dual-branch Joint readout and cross-frame historical modeling improve physical scene estimation on real dynamic streams via held-out observation prediction across all four $64\times 64$ dynamic recordings, denoted Records~1--4 and provided as Supplementary Videos~1--4. Within each qualified two-frame window, a disjoint subset of inter-slot complementary observations and their underlying ADC samples are withheld entirely from reconstruction. The hybrid solver reconstructs $\widehat{\mathbf{X}}_n$ from the remaining observations, and $\widehat{\mathbf{X}}_n$ is projected through the calibrated forward rows of the withheld subset to compute the held-out prediction RMSE in ADC counts, summarized by the median across windows for each recording.

First, comparing inter-slot-only dynamic recovery against dual-branch Joint recovery tests the physical contribution of intra-slot transient slopes. Across Records~1--4, inter-slot-only recovery yields median held-out prediction RMSE values of 37.11, 52.67, 54.43, and 35.52~ADC counts, whereas incorporating the intra-slot branch in Joint reduces these errors to 30.37, 40.36, 40.97, and 29.91~ADC counts, corresponding to reductions of 18.16\%, 23.38\%, 24.74\%, and 15.80\%, respectively. Because the evaluation metric is computed strictly on unseen inter-slot observations rather than intra-slot residuals, these consistent reductions across all four sequences confirm that within-slot phase evolution supplies genuine complementary scene information during dynamic motion.

Second, to isolate the effect of cross-frame historical coupling in Eq.~(S20), full complete-history Joint recovery is compared against an ablated Joint variant that deletes readout rows whose lag support crosses the boundary between frames $n-1$ and $n$. Retaining cross-frame mixed rows lowers median held-out RMSE from 48.77 to 40.36~ADC counts on Record~2 (a 17.24\% reduction), from 41.50 to 40.97~ADC counts on Record~3 (1.29\%), and from 31.02 to 29.91~ADC counts on Record~4 (3.59\%), while increasing from 28.63 to 30.37 ADC counts on Record 1 (an increase of 6.07\%). This degradation may reflect motion-induced mismatch in the piecewise-constant frame approximation described in Section S1.5, although its cause was not independently established. Across the four records, retaining cross-frame observations reduced the held-out prediction error in three cases, indicating that the benefit was not uniform across the tested dynamic scenes.

\subsection{Real-time runtime performance and multi-scenario dynamic demonstrations}
\label{sec:supp_s5_4}

Table~S2 summarizes the online per-frame computation times measured on an Apple M4 platform across the four $64\times 64$ dynamic recordings (CPU execution) and the $256\times 256$ dynamic recording (CPU operator processing paired with Metal Performance Shaders GPU neural inference), complementing Section~3.4 of the main text. Reported runtimes average qualified two-frame reconstructions over three sequential online stages, namely per-frame operator and readout assembly, operator scaling and solver preparation, and hybrid reconstruction solving, excluding serial acquisition waiting, file I/O, and display rendering. At $64\times 64$ resolution, per-frame processing averages 4.35--4.53~ms for assembly, 5.71--5.98~ms for preparation, and 9.39--10.40~ms for solving, yielding total per-frame times of 19.48--20.91~ms across Records~1--4. Scaling to $256\times 256$ resolution yields 4.61~ms for assembly, 2.90~ms for preparation, and 22.87~ms for solving, for a total of 30.38~ms per frame. All mean per-frame processing times remain comfortably below the $39.2~\mathrm{ms}$ nominal frame playback interval, supporting continuous synchronization-free operation at approximately $25~\mathrm{frames/s}$.

\begin{table}[!htb]
\centering
\caption{Mean online per-frame computation times (ms) on an Apple M4 platform across dynamic imaging recordings, excluding acquisition waiting, file I/O, and display refresh.}
\label{tab:supp_dynamic_runtime}
\small
\setlength{\tabcolsep}{5pt}
\begin{tabular}{llcccc}
\hline
Dynamic recording & Compute backend & Assembly & Preparation & Solving & Total \\
\hline
$64\times64$ Record~1 & Apple M4 CPU & 4.36 & 5.73 & 9.39 & 19.48 \\
$64\times64$ Record~2 & Apple M4 CPU & 4.35 & 5.71 & 10.03 & 20.09 \\
$64\times64$ Record~3 & Apple M4 CPU & 4.53 & 5.98 & 10.40 & 20.91 \\
$64\times64$ Record~4 & Apple M4 CPU & 4.42 & 5.79 & 10.25 & 20.46 \\
$256\times256$ Record & Apple M4 CPU + MPS & 4.61 & 2.90 & 22.87 & 30.38 \\
\hline
\end{tabular}
\end{table}

Figure~S16 presents extended chronological frame sequences from these real-time dynamic experiments, expanding upon the representative frames in Fig.~7(c) and (d) of the main text. Across the four $64\times 64$ sequences in Fig.~S16(a) ($9.57\%$ spatial sampling per frame; Supplementary Videos~1--4), the online Joint reconstructor tracks unconstrained hand-held translation, large-angle in-plane rotation, object entry into and exit from the field of view, and pronounced illumination-induced frame-to-frame brightness shifts across letter, Cameraman, character, and geometric targets while preserving stroke topology, grayscale contrast, sharp corners, and clean backgrounds. In the $256\times 256$ moving-letter demonstration in Fig.~S16(b) ($0.60\%$ spatial sampling per frame using the same 392 Cake-cutting directions per frame; Supplementary Video~5), the moving letter L remains clearly resolved with smooth frame-to-frame trajectory continuity at approximately $25~\mathrm{frames/s}$.

\begin{figure}[!htb]
\centering
\includegraphics[width=\linewidth,height=0.50\textheight,keepaspectratio]{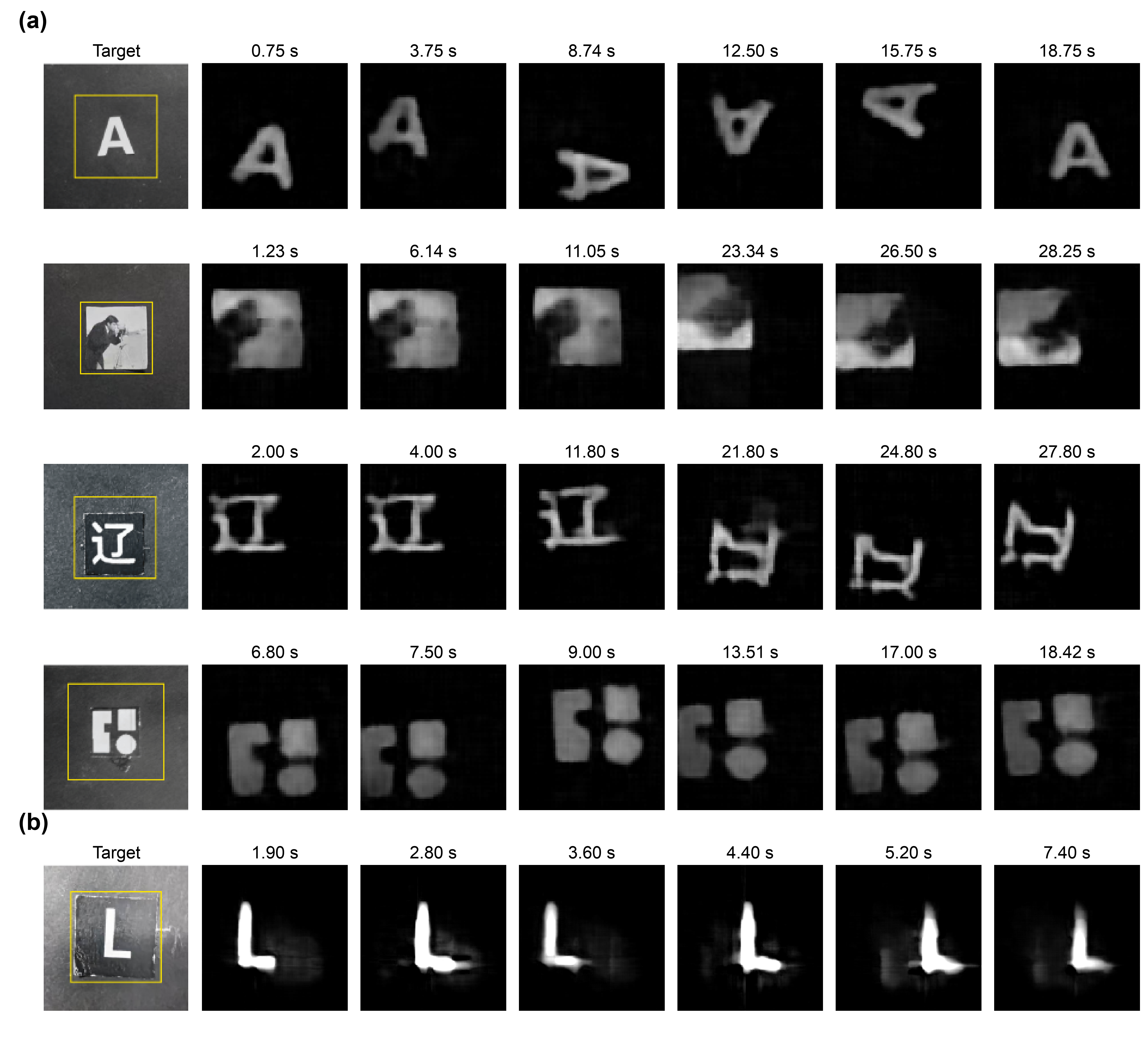}
\caption{Extended real-time dynamic reconstruction sequences on the 1.8~kHz receiving chain at $20~\mathrm{kHz}$ DMD modulation, $240~\mathrm{kHz}$ ADC sampling, and approximately $25~\mathrm{frames/s}$. (a) Chronological frames from the four $64\times64$ dynamic recordings capturing continuous translation, large-angle in-plane rotation, field-of-view entry and exit, and illumination-induced brightness variations across letter, Cameraman, character, and geometric targets. (b) Chronological frames from the $256\times256$ moving-letter sequence.}
\label{fig:supp_dynamic_sequences}
\end{figure}

Together, Sections~S1--S5 establish an end-to-end theoretical, algorithmic, and experimental framework for exploiting temporal mixing across static and dynamic single-pixel imaging. By integrating data-driven complete-history calibration, reference-based asynchronous timing registration, and composite inter- and intra-slot readouts around direct recovery of the target image or video sequence, hardware temporal memory is transformed from a speed-limiting bottleneck into a transferable, programmable measurement resource. Held-out dynamic evaluation summaries and mean per-frame runtime records are provided in Supplementary Data~1.

\paragraph{Access to supplementary data and videos}
Supplementary Data~1 and Supplementary Videos~1--5 are available from Jing-Sen Li (\texttt{luzi23887@163.com}) upon reasonable request.